\documentclass[
reprint,
amsmath,amssymb,
aps
]{revtex4-2}
\usepackage{dcolumn}
\usepackage{subfigure}
\usepackage{graphicx}
\usepackage{mathtools}
\usepackage{times,bm}
\usepackage{amsmath} 
\usepackage{amsfonts}
\usepackage{amssymb}
\usepackage{mathrsfs}  
\usepackage{longtable}
\usepackage{color}
\usepackage{xcolor}

\begin{document}
\title{Creases and cusps in growing soft matter\\}% Force line breaks with \\
\thanks{ Instability}%

\author{Martine Ben Amar}

 \affiliation{Laboratoire de Physique de l'Ecole normale sup\'erieure, ENS, Universit\'e PSL, CNRS, Sorbonne
Universit\'e, Universit\'e  Paris-Cit\'e, F-75005 Paris, France}
   \affiliation{Institut Universitaire de Canc\'erologie, Facult\'e de m\'edecine, Sorbonne Universit\'e, 91 Bd de l'H\^opital,  75013 Paris, France}           %  
\date{today}   

\begin{abstract}
The buckling of a soft elastic sample under growth or swelling has highlighted a new interest in materials science, morphogenesis, and biology or physiology. Indeed, the change of mass or volume is a common fact of any living species, and on a scale larger than the cell size, a macroscopic view can help to explain many features of common observation. Many morphologies of soft materials result from the accumulation of elastic compressive stress due to growth, and thus from the minimization of a nonlinear elastic energy. The similarity between growth and compression of a piece of rubber has revived the instability formalism  of nonlinear elastic samples under compression, and in particular Biot's instability. Here we present a modern treatment of this instability in the light of complex analysis and demonstrate the richness of possible profiles that an interface can present under buckling, even if one restricts oneself to the two spatial dimensions. Special attention is given to wrinkles, folds and cusps, a surprising observation in swelling gels or clays. The standard techniques of complex analysis, nonlinear bifurcation theory and path-independent integrals are revisited to highlight the role of physical parameters at the origin of the observed patterns below and above the Biot threshold. 
\end{abstract}

%Use showkeys class option if keyword
                              %display desired
\maketitle
\tableofcontents
\section{Introduction}
	The buckling of the outer surface of a living tissue during growth \cite{nelson2013buckling,volokh2003tissue,legoff2016mechanical}  and the corrugation of the surface of a swelling gel \cite{tanaka1987mechanical,li1990kinetics,hong2009inhomogeneous} are often observed in nature or in the laboratory. In the last three decades, a large number of studies have been devoted to such patterns in order to explain complex geometries in embryogenesis \cite{amar2013anisotropic, amar2017mimicking,bayly2013cortical}, botanical morphogenesis \cite{mendelson1989bending,gordon2011possible,nelson2016buckling}, but also in tumorogenesis \cite{buzney1984scleral,dervaux2011buckling,zhang2020morphomechanics}  and organ pathologies (e.g. wound healing \cite{wu2015growth,bowden2016morphoelastic,koppenol2017biomedical}). These shape instabilities affect thick samples that experience large volume variations in a non-isotropic manner. Obviously, in a free environment, the constant growth of a homogeneous sample does not generate stress, but if there is a constraint, such as a substrate, or if there is a material or growth inhomogeneity, then  the stress is generated that changes the shape of the body. It can buckle, but only if there is enough growth. This suggests a shape change once the relative volume increase exceeds a threshold, about  2 times the original. The origin of the observed patterns at free surfaces results from the compressive stress generated by growth coupled with the hyperelastic properties of soft tissues. These tissues exhibit large deformations even at low stress values, and classical linear elasticity cannot explain the observed shapes. Focusing on the simplest case of a gel layer of constant thickness $H_0$ placed on a substrate, the growth process
occurs mainly in the vertical direction and leads to a thickening of the layer with: 
$H = J_t H_0$, where $J_t$ is  the relative growth per unit 
 volume at  a time $t$ in this simple geometry. When $J_t$ is increased to a critical value, the top surface begins to wrinkle. For neo-Hookean elasticity, this value $J_B$ of order $3.38$ can be related to the critical value found by Biot for samples under compression. Of course, this instability is common and not limited to the ideal gel layer. The threshold for wrinkling depends on the nonlinear elasticity model \cite{amar2010swelling,pandurangi2022nucleation}, or on the initial geometry of the sample \cite{goriely2017mathematics,wu2015growth}, or possibly on the growth anisotropy \cite{ben2017mimicking}, but the order of magnitude of this number seems quite robust.

 The mechanical interpretation of a material under compression was first given by M.A. Biot in a seminal paper "Surface instability of rubber in compression" \cite{biot1963surface}. Surface instability means that the instability is more visible at the surface of the sample, but actually  occurs throughout the volume, as opposed to the Azaro-Tiller-Grenfield instability \cite{asaro1972interface,grinfeld1993stress}, which results from surface diffusion.  This instabilty is also different  from wrinkles formed by a two-layer system where the top layer is thin and stiff and plays the role of a hard skin \cite{box2020cloaking}. In this case, the surface topography can be realized in a very controlled way and is  of enormous importance in industrial
 and biomedical applications \cite{dimmock2020biomedical}. Biot's instability was first demonstrated for a compressed neo-Hookean hyperelastic sample with a free surface in infinite geometry. It describes a two-dimensional infinite periodic pattern that occurs above a characteristic threshold for the compression level, but when the material geometry is more complex, such as bilayers  \cite{ackermann2022modeling,pandurangi2022nucleation}, or when the compression results from anisotropic or inhomogeneous growth, the interface buckling is recovered experimentally, but the analysis can be less straightforward.
However, if smooth surface undulations can also be considered \cite{li2012mechanics}, the experimental patterns quickly evolve to nonlinear mode coupling \cite{javili2015computational,hutchinson2017nonlinear,crawford1991introduction,cross1993pattern,charru2011hydrodynamic} and even to wrinkles, which are less understood, although they are easily and commonly observed in experiments and are also noted in the physiology of the brain cortex, for example  \cite{libero2019longitudinal}. 

An even more puzzling observation  concerns more cusped interfaces as shown in Fig.(\ref{cuspexp}) (A1) to (A6). In one dimension, a cusp  is a special point of a curve where the radius of curvature vanishes (or the curvature is infinite), while a "wrinkle" represents a more or less deep local folding of the interface. Other different interpretations of surface wrinkles concern singular points at the origin of a self-contacting interface, which of course indicates a much more singular interface deformation, see Fig. (\ref{cuspexp}) (A9) and  \cite{hong2009formation,jin2011creases,karpitschka2017cusp,ciarletta2018matched, ciarletta2019soft}. Do they result from a highly nonlinear coupling of modes occurring after the bifurcation, or do they belong to another class of solutions? In the latter case, they can appear below the Biot threshold $J_B$ and even inhibit the classical instability  \cite{chen2012surface,chen2014controlled}. More recently,
the idea that there can be new families of solutions below the Biot threshold has been supported by matched asymptotic analysis \cite{hong2009formation,jin2011creases,karpitschka2017cusp,ciarletta2018matched, ciarletta2019soft} or by the nucleation of new solutions in more complex elasticity models and geometries \cite{pandurangi2020stable,pandurangi2022nucleation}. Some experimental evidence realized on rubber in compression or on swelling gels also seems to favor the second hypothesis  \cite{ hong2009formation,jin2011creases,weiss2013creases}.  Of course, numerical evidence is always difficult in the case of spatial singularities, but we must mention the finite element numerical investigation of  \cite{hohlfeld2011unfolding,hohlfeld2012scale} in favor of a subcritical (or discontinuous bifurcation) before $J_B $ which becomes supercritical (or continuous) at $J_B$ with an important sensitivity of the results to the conditions imposed on the substrate.  
Another way to study the cusp formation experimentally and theoretically \cite{karpitschka2017cusp} is to create a localized defect in a controlled experiment, mimicking in some way experiments in viscous fluids where the defect is realized by contra-rotating cylinders  \cite{jeong1992free}. It should  be noted that localized singular structures easily occur in tubes but here the geometry helps the appearance of singular deformations \cite{emery2021post,fu2021necking}.  

Despite the similarity that exists between compressive forcing and homogeneous growth in the neo-Hookean approach, this review article focuses on volumetric growth, which is ubiquitous in life. Most of our organs exhibit  Biot's instability, which explains our fingerprints, the convolutions of our brains, the villi and the mucosa of the intestines. All  these structures appear after a certain time after fertilization in foetal life.  They are present in most mammals, except for small rodents. These two observations support an interpretation in terms of morpho-elasticity: the shape of the organ is a determinant factor, as is the volumetric growth, which increases with time from $J = 1 $ (no growth expansion) up to critical values.

Before giving mathematical proofs concerning wrinkles, our presentation will begin with a selection of experiments (section \ref{experiment}) and a brief introduction to the principles of nonlinear elasticity. In this field of study,  positive quantities called invariants $I_J$ are introduced to evaluate the elastic energy density.
Since they are specific to finite elasticity, they will be introduced in detail in section \ref{intro}. In addition, the local growth per unit volume creates an external field that does not obey physical rules and is imposed a priori inside the sample. It is not fully comparable to an externally applied compressive dead load, see Sec. IV. We first revisit the original model of Biot for neo-Hookean elasticity in the incompressibility limit and in semi-infinite geometry \cite{biot1939xliii,biot1963surface}, but for the threshold determination $J_B$ and for nonlinear buckling and wrinkling,
we follow a different strategy based on variational principles.  Euler-Lagrange equations derived by incremental perturbation techniques are at the origin of the periodic modes and also of $J_B$, the threshold. We then apply the
nonlinear techniques of bifurcations, combined with complex analysis,  which greatly simplifies the intermediate algebra. The results of Biot are recovered in a much simpler way and nonlinearities  are  treated above and below the threshold without difficulty. First, subcritical bifurcations,  as indicated  by \cite{breid2012controlling,jia2013theoretical,jin2015bifurcation}, are  demonstrated by nonlinear sinusoidal mode coupling. Second, wrinkles above and below the Biot threshold are analytically justified by introducing singularities either inside or above the elastic sample.

This notion can be rather abstract, but has been successfully introduced for interfacial flows such as viscous fingering \cite{cummings1999two,combescot1988shape,shraiman1986velocity,hong1986analytic}, for bubbles in Laplacian and Stokes flows \cite{cummings1999two,crowdy2009multiple}, for vortices \cite{crowdy1999class,crowdy2010new}, and for diffusive growth \cite{amar1986theory,cummings1999evolution}. In fluids, singularities outside the physical plane are used to select the length scale of the interface patterns, but they can be physically introduced into the flow in the experimental setup, leading to a complete change of the interface shape. For example, a coin or a bubble in front of a viscous finger completely changes the shape into a dendritic one \cite{thome1990controlling}, and a theoretical interpretation has been given in terms of a dipole. 
\\
This idea  of a dipole was
taken up later \cite{snoeijer2015interface}  in fluids and in linear elastic solids. Also, when vortices are created in viscous fluids, they generate cusps at the interface \cite{richardson1968two,howison1995cusp}  (in the mathematical sense), which are transformed into sharp wrinkles when a weak surface tension is included \cite{jeong1992free,kelly1997numerical}. Following a similar strategy, we will consider singularities outside and inside the physical domain, with the aim of discovering the main physical ingredients necessary to predict the observed wrinkles.\\

In conclusion, the existence of wrinkles in growing soft materials benefits from many theoretical analyses carried out in the last decades on viscous flows (interfacial and vortex flows) and from treatments of singularities in elasticity based on the Noether theorem and path independent integrals, see section \ref{path}. These classical but not easy techniques are presented in the following. We limit ourselves to a very simple modeling of hyperelasticity, being convinced that, once established, it will be possible to extend the mathematics to arbitrary geometries and complex structures of soft materials.
After the presentation of some experimental examples in  section \ref{experiment}  and a reminder of the foundations of nonlinear or finite elasticity (sections \ref{intro} to \ref{pressuregrowth}), we focus on a variational energy method, section \ref{main}, where buckling modes are treated at the linear, (section \ref{linear}),  and nonlinear, (section \ref{energetic}), levels. We then study the possibility of stress focusing \cite{witten2007stress} inside the material just below the interface, which can induce interfacial wrinkles, in section \ref{escape}. If these zones can be perfectly characterized in morphoelastic growth, (section \ref{evidence}),  there is no clear threshold for their observation as demonstrated by the technique of path independent integrals, (section \ref{path}). Finally, we come back to the buckling of thin films of finite thickness comparable to the wavelength in section \ref{Finitesize}.

\section{Selection of creases in experiments}
\label{experiment}
\begin{figure*}	
\includegraphics[width=1.\textwidth]{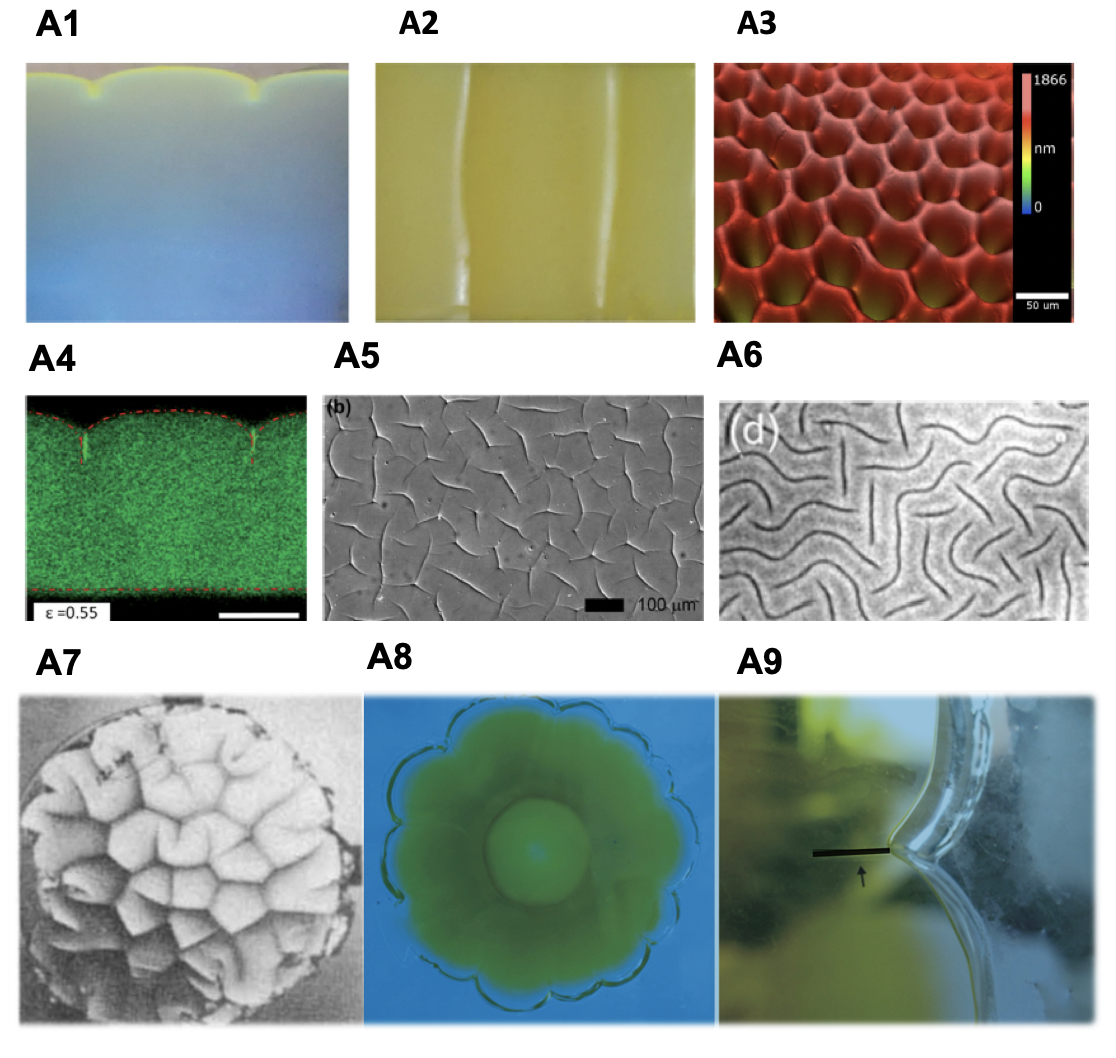} \\
\caption{In (A1) and (A2), compression of a parallelepiped specimen: micrographs of a pair of wrinkles from the front and  from the top view at a strain level of $45\%$, above the Biot threshold.
\cite{tang2017dimension}. The critical strain for wrinkling is $37.2\%$. In (A3) experimental herringbone array of a PDMS swelling gel sample, courtesy of Derek Breid \cite{breid2012controlling,jia2013theoretical}.\\
In (A4) Confocal microscopy for elastomer surfaces under compressive strain with an initial thickness of  $23 \mu$m \cite{cai2012creasing}. In (A5) and (A6) two optical micrographs of wrinkle growth for a gel containing  $15$
mol  $\%$ NaAc,  obtained by cooling, with an initial thickness of $15\mu$m  \cite{yoon2010nucleation}, in (A5) from to $33.2 ^\circ  C$ C to $31.7 ^\circ  C$ in
(A6) up to to $25 ^\circ  C$.\\
In (A7) Creases in circular geometry: a pioneering experiment
by T. Tanaka {\it  et al.} on the swelling of an ionized acrylamide gel in water. In (A8) A ring of charged polyacrylamide gel (yellow) around a hard disk of neutral polyacrylamide gel (transparent) viewed from above: initial diameter: $50$ mm and imposed thickness of $1$  mm. The outer ring swells by immersion in distilled water; the swelling is highly inhomogeneous in this geometry. The inner disk acts as a constraint, and after the appearance of smooth wrinkles, wrinkles develop at the outer boundary above a certain threshold of volume variation \cite{dervaux2011buckling}. In (A9) the same experimental setup as in (b) with a focus on a single cuspoidal point \cite{dervaux2011shape}. For clarity, the attached line of fracture or refolding has been underlined in black. Note that it may appear as a self-contacting interface or as a fracture in compression \cite{karpitschka2017cusp} .}
\label{cuspexp}
\end{figure*}

The formation of wrinkles and creases  in samples of elastomers or swelling gels has fascinated physicists for decades and probably still does. Examples of compressed elastomers are given in Fig.(\ref{cuspexp}) panels $A1,A2,A4$, and all the other panels concern swelling gels in different experimental setups. In fact, the nucleation of wrinkles in materials known to be highly deformable without plasticity is quite astonishing. It contrasts with the difficulty of nucleating a fracture in a 3D brittle sample under tensile loading: in this case, an initial notch or slit must be deliberately made \cite{pomeau1992brisure,bonn1998delayed}. Experimentally, it is difficult to elucidate the threshold for the appearance of these wrinkles. Indeed, the homogeneous volumetric growth of a material is equivalent to a compression, but the linear instability threshold discovered by Biot has not been precisely verified experimentally. As for wrinkles, it seems even worse, although there is a tendency to detect them above the Biot threshold. It is true that the geometry of the experimental setup has its importance on the threshold, as well as the fact that the material is attached to a solid substrate or to another elastic sample. Another important point concerns the size of the experimental setup compared to the instability wavelength and the fact that the neo-Hookean model (or any hyperelastic model) is not really adapted to swelling. The poroelastic model is more appropriate  in this case \cite{hong2008theory,hong2009formation,dervaux2011buckling}. Independently, R. Hayward and collaborators \cite{trujillo2008creasing,chen2014controlled,jin2015bifurcation}  point out in a series of articles that the bifurcation around the Biot threshold is probably subcritical, which makes a precise experimental determination difficult. However, singular profiles certainly exist, and the last panel (A9) shows the strong stress concentration that leads to the ejection of material pieces from the outer ring \cite{dervaux2010morphogenese,dervaux2011shape} during the course of the experiment. Our main concern in the following will be the prediction of patterns around the Biot threshold or below.
Nevertheless, let us recall the theory of finite elasticity with or without growth. It will be a way to introduce the main principles of application as well as the mathematical tools. A short presentation of the theory of swelling gels is also included to emphasize the difference between swelling and volumetric growth.

\section{A basic introduction to nonlinear elasticity}
\label{intro}
\subsection{ A brief reminder of the principles of linear elasticity}
 Linear elasticity is limited to weak to moderate deformations corresponding to small strains, estimated by the ratio: deformation over a typical length of the deformed specimen. These deformations often occur under external loads, possibly under external fields such as temperature changes. Unlike other heuristic models, such as the Canham-Helfrich \cite{canham1970minimum,helfrich1973elastic}  models for lipid membranes, elasticity requires knowledge of the initial shape of the body, which is assumed to be free of stress, and focuses on its deformation. Until recently, the goal was to explain and quantify the deformations of stiff materials: steel, wood, concrete, paper, nylon, etc., and their stiffness is usually given by the Young's modulus $E$
in Pascals. For these materials, the value 
of $E$ is on the order of  $10^9$ to $10^{12}$  Pascals, which immediately indicates that it will be very difficult to stretch a cuboid by human force. Nevertheless, the field of linear elasticity remains very active: Being closely related to geometry, any peculiarity leads to a strong interest and curiosity, such as the crumpling of paper \cite{lobkovsky1995scaling,kramer1997stress,ben1997crumpled,blair2005geometry}, the formation of folds \cite{mahadevan1999periodic,conti2008confining,thiria2011relaxation}, or the science of origami \cite{mahadevan2005self,brunck2016elastic}. The linearity of the relationship between displacement and load does not automatically imply that the equilibrium equations are linear, as demonstrated by the Foppl-Von Karman equation, where the Hooke formalism is applied but the deformation is extended  to the third order \cite{audoly2010elasticity}. In particular, origami  and paper crumpling studies introduce geometric singularities that can be treated with linear elasticity \cite{witten2007stress}, while folding involves nonlinear elasticity. The linearity of the Hooke's law does not automatically imply simplicity of the theoretical treatment  when the initial shape is complex. In fact,  the formalism leads to partial differential equations, and this geometric complexity is also recovered in nonlinear elasticity. Thus, the main question is when nonlinear elasticity is required a priori.

\subsection{The principles of nonlinear elasticity}
\label{Invariant}

Once a material is soft, even very soft, with a Young's modulus $E$ not greater than $10^5$, the displacement of any point of the sample under load can be of the order of the original shape. Then, a precise description for the internal stresses
and also for the geometry of the deformations are required. Not all nonlinear descriptions of the elastic energy density W are possible because  they must satisfy strong mathematical properties dictated by the laws of mechanics, such as objectivity and convexity.
 Objectivity means that the elastic energy remains invariant under rigid rotation or translation. Convexity means that for small displacements $u$, $\delta W\sim \alpha u^2$ with $\alpha>0$.  We consider an undeformed body with no  internal stresses,  where  each point $M$ is  represented by the capital letters $M (X, Y, Z)$ (for simplicity, Cartesian coordinates are chosen and maintained throughout the manuscript).  Then there exists a vectorial mapping function $\chi$  such that relates the new coordinates of the displaced point $m$ to the  coordinates of the original point such that $\vec Om=\vec OM+\vec u$.  $\vec u$ is the vector displacement according to  the  same definition as in linear elasticity. One of the most important mathematical tools is the  deformation gradient tensor which reads:
 \begin{equation}
\label{map}
{\bf F}=\nabla \chi\quad {\mbox{where}}\quad  F_{ij}=\frac{\partial  x_i}{\partial X_j}=\delta_{ij}+\frac{\partial  u_i}{\partial X_j}\,. 
\end{equation}
The hyperelastic energy density $W$ must respect spatial isotropy (if there is no preferred direction in the structure of the body) and be invariant for any change in the coordinate system.
Consequently, it must be represented by the trace or determinant of tensors constructed with ${\bf F}$. We start with the simplest invariants, the most common one being defined with the Cauchy right tensor $\bf C=\bf F^T \bf F$  to satisfy the objectivity requirement.
\begin{equation}
\label{invariant}
I_1=\rm{Tr}({\bf C}) \quad I_2=\frac{1}{2}\left \{(Tr({\bf C}))^2-Tr({\bf C}^2)\right\},
\end{equation}
$I_1$ can be written as $I_1=F_{ij}\cdot F_{ij}$  where summation on repeated indices holds. The third invariant $I_3=$Det$({\bf F})$ is related to the local volume variation and must be a positive number.
 Homogeneous hyperelastic energy densities are basically functions of these $3$ invariants, but can also be restricted to two of them, generally $I_1$ and $I_3$ as for  the neo-Hookean energy density,
 while the linear combination of $I_1$, $I_2$ and $I_3$ is called the Mooney-Rivlin model. One may wonder how to recover the weakly nonlinear energy density described by the Lam\'e coefficients.
 The simplest way is to  define   $\bf H=F-I$\, first,
and then the elastic energy density W as
\begin{equation}
W=\frac{1}{2}
\left\{ \mu_L (\mbox{Tr }({\bf H^T}\cdot {\bf H})+\mbox{Tr}({\bf H})^2) + \lambda_L \mbox{Tr}({\bf H}^2)\right\}.
\end{equation}
Note that such a formulation is not suitable for incompressible materials, since the coefficient  $\lambda_L$  diverges. In fact, for incompressible materials, $I_3 = 1$, a limit corresponding to a Poisson ratio $\sigma = 0.5$ in linear elasticity. If a preferred direction is present in the materials, as is often the case in organs
such as heart, arteries, and skeletal muscles, more invariants are needed indicating an increase in stiffness. These invariants will depend on ${\bf C}$ and on the orientation of a unit vector ${\vec e_0}$ which indicates the direction of the fibers, assuming that this direction is unique.
 The Helmoltz free energy
for an incompressible sample is then
\begin{equation}
\label{energiew}
    {\cal E}=\int \!\!\!\int \!\!\! \int_ {\Omega} d V\,\, W(I_1,I_2,I_4,I_5)-Q\,\,(I_3-1)\,,
    \end{equation}
where dV is the volume element in the reference configuration and $Q$ is a Lagrange multiplier that fixes the physical property of incompressibility. The energy density $W$ is a positive scalar that vanishes for ${\bf C = I}$.  
If a material is anisotropic only in a single direction, defined by the unit vector ${\vec e_0}$  in the reference configuration, then two invariants must be added, such as $I_4$ and $I_5$, given by $I_4=\vec e_0.( {\bf C} \vec e_0)$ and $I_5=\vec e_0.( {\bf C^2} \vec e_0)$ \cite{holzapfel2019fibre}. In the biological
context, materials can have other directions of anisotropy, in which case other invariants are introduced with a new vector ${\vec e_1}$.  For compressible materials, the energy is composed of two terms: a volumetric term, which is a function of $I_3$: $\Psi (I_3)$, and a strain energy function, where all components of the strains are divided by  $I_3^{1/3}$  in $3D$ so $\bar I_1=I_1/I_3^{2/3}$ and $\bar I_2=I_2/I_3^{4/3}$  :
 \begin{equation}
\label{energiecomp}
    {\cal E}=\int \!\!\!\int \!\!\! \int_ {\Omega} d V\,\, W(\bar I_1,\bar I_2) +\Psi(I_3) \,.
    \end{equation}
  Note that in $2$D, the new strains are divided by $\sqrt{I_3}$.  Compressible elasticity leads to much more complex calculations in practice and  different simpler models can be found in the literature \cite{holzapfel2000nonlinear} as the compressible Mooney-Rivlin model \cite{ciarlet1982constitutive}:
\begin{equation}
\label{mooneycomp}
\begin{cases}
 W_{MR}&= c_1(I_1-3)+c_2(I_2-3)+c(I_3-1)^2\\
 &-2 (c_1+2 c_2) Log(I_3)\,.
 \end{cases}
 \end{equation}
    Finally, if an external mechanical load $\vec B$  is applied onto the  system  and/or on its  surface $\vec T$, the work they exert on the sample must be added to eq.(\ref{energiew}) or to eq.(\ref{mooneycomp}) according to:
      \begin{equation}
\label{addition} 
{\cal E}_{add}=-\int \!\!\!\int \!\!\! \int_ {\Omega} d V\,\, \vec B \cdot\vec x - \int \!\!\! \int_ {\partial\Omega} d{\cal A} \,\,\vec T \cdot \vec x.
\end{equation}
 Let us now derive the so-called constitutive equations, which are the counterpart of the Hooke's law of the linear elasticity theory. 
    
  \subsection{Constitutive equations in finite elasticity and definition of the elastic stresses}
   \label{defstress}
The constitutive equation is the relation between the stress tensor ${\bf S}$ and the gradient  of the deformation tensor ${\bf F}$ which can be obtained from the variation of the elastic energy. The Euler-Lagrange equation results from the extremum of ${\cal E}+{\cal E}_{add} $
with respect to the variation of the new position $\delta x$ and also of $Q$. Mathematically, it reads:
\begin{equation}
    \delta [{\cal E}+ {\cal  E}_{add}] (x,y,z;x_i)=0\quad \mbox{and}\quad   \delta {\cal E} (x,y,z;Q)=0\,,
\end{equation}
for arbitrary variation of $x_i$ and $Q$. As before $x_i$ means either $x$, or $y$, or $z$, which are the current coordinates of the displaced point  $m$, initially located at $M$. 
Then 
 \begin{equation}
 \label{firstvariation}
  \delta {\cal E}=\int \!\!\!\int \!\!\! \int_ {\Omega} d V \left(\frac{\partial W}{\partial {\bf F}}-Q\,{\bf F^{-T}}\right) \delta{\bf F},
 \end{equation}
 where we have used the tensorial relation for an arbitrary tensor ${\bf A}$, which is  $\partial$  Det( ${\bf A})/\partial {\bf A}=$ Det(${\bf A})\,{\bf A^{-T}}$. Then we derive the Piola stress tensor ${\bf S}$ for an incompressible material:
\begin{equation}
\label{stressdef}
{\bf S}= \frac{\partial W}{\partial {\bf F}}-Q\,{\bf F^{-T}}\,.
\end{equation}
Note that the Piola stress tensor, also called the first Piola-Kirchfoff stress tensor \cite{holzapfel2000nonlinear} is the transpose of the nominal stress tensor \cite{ogden1997non}.  Once $W$  is selected, this relation represents the constitutive relation of the material. Since we must perform the variation with respect to the current position $\vec x$  in the 
 coordinate  system of the reference configuration ${\vec X}$, an integration by part leads for $\delta {\cal E}+ \delta {\cal  E}_{add} $:
\begin{equation}
\begin{cases}
    \delta {\cal E}+\delta {\cal  E}_{add} =\int \!\!\! \int_ {\partial\Omega} d{\cal A} \,\,(-\vec T+{\bf S}\cdot\vec N)\cdot \vec \delta x\\
    -\int \!\!\!\int \!\!\! \int_ {\Omega} d V \, ({\rm Div} \, {\bf S} +\vec B)\cdot \delta \vec x=0\,.
\end{cases}
\end{equation}
When the equilibrium is reached:
\begin{equation}
\label{diverg}
     {\rm Div}{\bf S}+ \vec B =0,\quad  {\bf S}\cdot {\vec N}={\vec T}\,.
\end{equation}
The Piola stress tensor ${\bf S}$  is not the only stress that can be defined in finite elasticity. In fact, by definition, a stress is the ratio between a force and a surface, and the value is not the same in the reference or in the current configuration where the Cauchy stress is evaluated according to:
\begin{equation}
\int \!\!\! \int_ {\partial\Omega} d {\cal A}\,\,({\bf S}.\vec N)=\int \!\!\! \int_ {\partial\Omega} d\,a\,\,({\bm \sigma}.\vec n)\,.
    \end{equation}
Using Nanson's formula: $ d a\,\vec n= d {\cal A}\vec N ($Det$({\bf F}) {\bf F^{-T}})$, we obtain the Cauchy stress ${\bm \sigma}$:
\begin{equation}
\label{stressdefbis}
{\bm \sigma}={\rm Det} ({\bf F})^{-1} {\bf S F^T}\quad \mbox{and}\quad  { \bf S F^T} ={\bf F S^T}\,.
    \end{equation}
The Cauchy stress is imposed to be symmetric unlike  ${\bf S}$ and  the last equality results  for  the Piola stress tensor $\bf S$ which is not symmetric. Note that although in this section the determinant of ${\bf F}$ is equal to one, we keep this notation which will change when growth is considered. 
In the literature and in classical textbooks (see \cite{ogden1997non,holzapfel2000nonlinear,goriely2017mathematics} for instance)
there are other alternative stress tensors, all of which are related to the Piola stress tensor, as opposed to linear elasticity. Relations between them can be  established as soon as ${\bf F}$ is known.

\subsection{Simple geometry and stretches}
\label{stretch}
\begin{figure}	
        \includegraphics[width=0.45\textwidth]{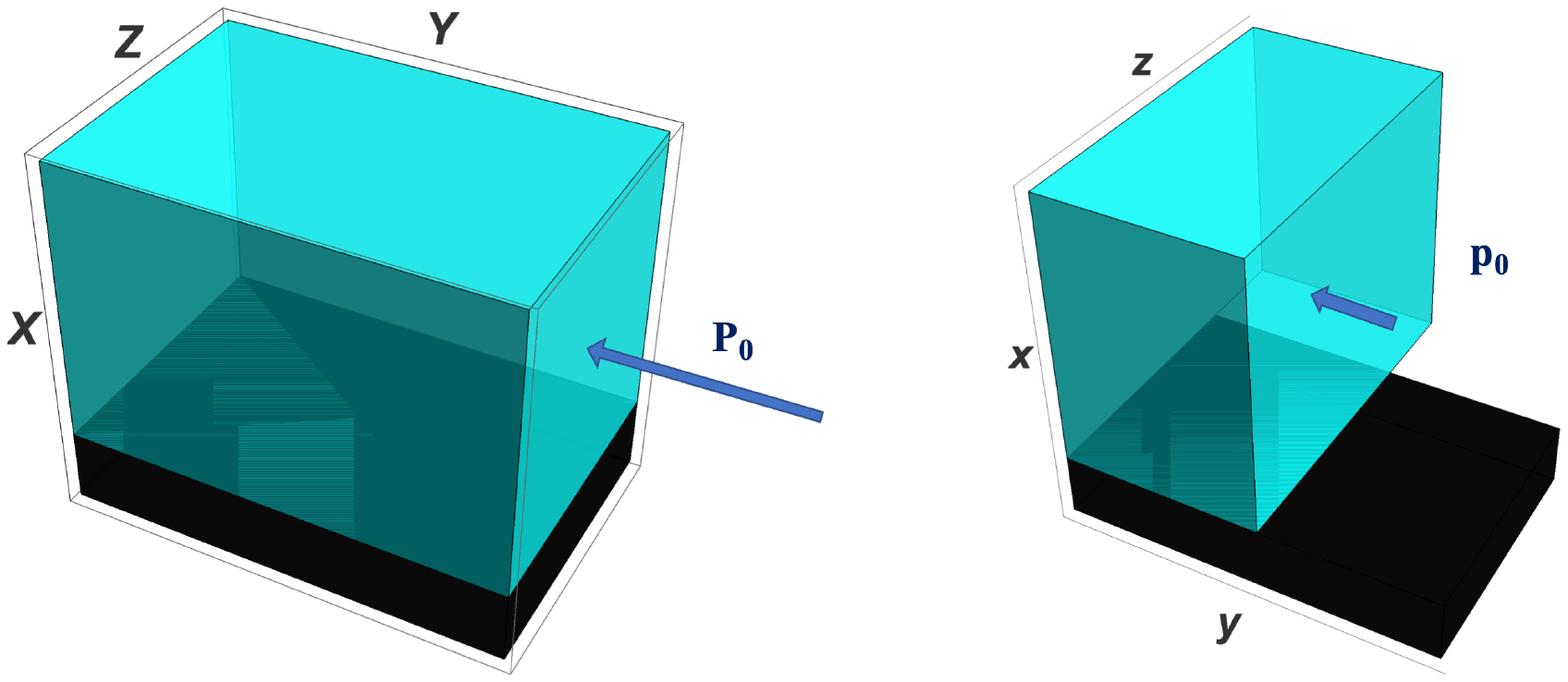}
        \caption{On the left a schematic representation of a soft material in blue, in the initial configuration, on the right the same sample in the current configuration.  A normal pressure  $P_0$ is applied to both surfaces  (X,Z)  on the left, which becomes  $p_0$ on the right. For clarity, only one side is shown. To emphasize  the deformation, a non-deformable substrate  is shown in black  and it is assumed that the sample slides  on this substrate without friction. Note that the pressure in the reference and current configurations are different due to the expansion of the lateral surfaces.}
\label{cuboid}
\end{figure}

When the specimen geometry is simple such as the cube, the cylinder and the sphere, the deformation gradient tensor can  be diagonal in the corresponding coordinate system  and the equations of elasticity become  simpler if the deformations follow the same symmetry.
Let us start  with a parallelepiped  with coordinates ${{0<X<L_X},{0<Y<L_Y},{0<Z<L_Z}}$, subjected to a compressive force on the two opposite faces  normal to $\vec e_Y$ (see Fig.(\ref{cuboid})).
In this case, we expect a simple deformation $x=\lambda_1 X$,\,$y=\lambda_2 Y$ and $z=\lambda_3 Z$ and the   diagonal tensors ${\bf F}$ and ${\bf S}$ are easily obtained:
\begin{equation}\begin{cases}
{\bf F} =\rm{Diag}(\lambda_1,\lambda_2,\lambda_3),\\
\mbox{and}\\
{\bf S} =\rm{Diag}(\frac{\partial W}{\partial \lambda_1}-\frac{Q}{\lambda_1},\frac{\partial W}{\partial \lambda_2}-\frac{Q}{\lambda_2},\frac{\partial W}{\partial \lambda_3}-\frac{Q}{\lambda_3}) \,.
\end{cases}
\end{equation}
where ${\bf S}$ follows the definition of eq.(\ref{stressdef}). In this simple geometry and for constant values of $\lambda_i$,  ${\bf S}$ is diagonal with constant  components, so it  automatically satisfies the equilibrium equation eq.(\ref{diverg}) in the absence of internal mechanical load $\vec B$.  The eigenvalues of ${\bf F} $ are called stretches. Since there is no force acting  on the surfaces perpendicular  to $\vec e_X$ and $\vec e_Z$, the Lagrange parameter $Q$ is then 
\begin{equation}
 Q=\lambda_1\frac{\partial W}{\partial \lambda_1}\quad \mbox{and}\quad  Q=\lambda_3\frac{\partial W}{\partial \lambda_3}\,.
 \end{equation}
 For an isotropic sample, $W$ is a symmetric function of the stretches $\lambda_i$, and there is no reason to distinguish  between both directions, here $1$ and $3$ so $\lambda_1=\lambda_3=1/\sqrt{\lambda_2}$ due to the assumption of incompressibility. After applying   a compressive load, we finally get:
 \begin{equation}
     \frac{\partial W}{\partial \lambda_2}-\frac{\lambda_1}{\lambda_2}\frac{\partial W}{\partial \lambda_1}=-P_0\,.
 \end{equation}
Assuming a neo-Hookean material with a shear modulus $\mu$ chosen as the unit of stress, then the energy density is  $W=1/2(I_1-3)$ and  the stretch $\lambda_2$ is the solution of the cubic equation:
 \begin{equation}
 \label{lam1}
     \lambda_2^3+P_0 \lambda_2^2-1=0,
     \end{equation}
 which has a unique real root: $\lambda_2\sim 1- P_0/3$ for small $P_0$ and for large compression, the stretch is close to zero so  $\lambda_2\sim 1/ \sqrt{P_0}$. Note  the simplicity of the derivation of such a solution which, however, implies  that the points at the bottom of the cube can slide freely, without friction.

\section{Competition between elasticity and continuous fields}
\label{volumetric}
Independent of local forces applied to the surface, the shape of a body can change due to different  external  fields applied and elasticity can be only one cause   of the deformation among others. The nonlinear elastic formalism  explained above concerns only a part of the global visible deformation and in practice  it is not so easy to separate  the elastic part from the overall shape. In the case of  volumetric growth, each small piece of the sample which  initially has a  volume $\delta \Omega$  becomes $\delta \omega$ after a growth or  drying   process that results in a change  in the total volume but also in a change in shape or morphology.  In the following, the word growth will be used to refer  either an increase or a decrease in volume.  Furthermore, growth can refer to  the  cell proliferation, as in embryos,  or to the swelling of gels, as already shown  in the experiments mentioned in section \ref{experiment}. It can also refer to drying or volume decrease. 

To separate the growth from the elastic deformation, we keep the definition of the mapping $\chi$ between the initial state and  the observed state  at time $t$ as it is defined in eq.(\ref{map}). This mapping 
 gives only a geometric information and we split the tensor ${\bf F}$ into two components:  a tensor  mimicking the growth $\bf G$ and the  tensor ${\bf F_e}$ for the elasticity, so that :
\begin{equation}
\label{rodriguez}
    \bf{F= F_e G}\,\,\,\mbox{so}\quad   \bf{F_e}= F G^{-1}.
\end{equation}
This relation, inspired by plasticity modeling and proposed in biomechanics  by Rodriguez et al. \cite{rodriguez1994stress} is local, and   ${\bf G}$ is the growth tensor at a point $M$ of the sample, obtained  after a period $t$. This law is cumulative in time meaning that the determinant of ${\bf G}$ gives the local amount of growth variation between the initial time and  the time of observation. This approach assumes that transient states are quickly eliminated to make room for a slowly  adiabatic dependent growth state  where the time is an index.  Although not intuitive, this formalism actually allows to quantitatively translate some aspects of biological growth, such as inhomogeneity, but also  anisotropy of growth: ${\bf G} $ is a tensor, so it simultaneously represents $3$ directions and $3 $ eigenvalues, each of them associated with a direction. 

A question that immediately comes to mind is the order of the two tensors  ${\bf{ F_e}}$ and ${\bf G}$ when they do not commute.   This and other questions  have been discussed, see \cite{naghdi1990critical}. A physicist would argue that, since the stresses are due to the growth, then the position of ${\bf G}$ is obviously on the right side. Another difficult problem arises simply  from the fact that growth is often associated with a process defined per unit time and may be  better represented in an  Eulerian description while here we are faced with a Lagrangian formulation that relates an initial  state to a  current state   at time $t$. This approach  more or less intuitively assumes that the time scale of growth is extremely long  compared to any time scale at the origin of the dissipation, reorganization, or remodeling of the samples \cite{ambrosi2019growth}. Despite its apparent conceptual simplicity, this formalism has generated significant contributions in  embryogenesis, morphogenesis and also in the description of various pathologies such as wound healing, fibrosis and tumorogenesis. As suggested by eq.(\ref{rodriguez}), growth induces stresses so not only a change in  volume but also  a change in its shape and one may wonder if this is always the case. In the next section, we will examine the origin of the stresses induced by growth. 
\subsection{The origin of the elastic stresses} 
\subsubsection {Growth without stress generation}
  Materials can grow without stress  if they can  follow and adapt themselves  to the imposed growth tensor. This is possible if there are no boundary conditions restricting  the growth. Homogeneous growth ${\bf G}=G_0 {\bf I}$ of a spherical object (without weight) does not generate any stress in  the  material.  If the growth tensor is more complex, e.g. inhomogeneous and anisotropic, the shape of the body will change, as it  grows.  The question of a stress-free process has  recently been  explored \cite{chen2020stress,chen2022generating} and examples from living systems have been given. If the deformation of the body can exactly follow the growth process, then  ${\bf F=G}$ and is independent of the material properties of the body. Such a relation allows to obtain the tensor ${\bf G}$ and thus the properties of the growth which are mostly  unknown for macroscopic samples. This process requires the absence of constraints from boundaries and  external  forces such as  gravity. The best example can be given by fresh planar leaves \cite{alim2016leaf}. To verify  such a hypothesis, one possible test is to cut  the material at right angles. If there is no crack opening, then the material is considered as stress-free. When the leaves have in-plane residual stresses due to growth, they cannot remain planar, as shown in \cite{dervaux2011shape}, and they buckle. Recently, a general proof of stress-free growth by conformal mapping  was given \cite{dai2022minimizing}. 
\subsubsection{Constrained growth process} 
Obviously, the main source  of stress  comes from boundary conditions, especially from  rigid walls. Imagine a parallelepiped  where only one side is  rigidly attached to a substrate and then it cannot evolve freely. This is the case with gels, where the chains of polymers adhere to the substrate, and then mimic clamped conditions. But it is also  the case of  parallel layers with different elastic properties, that are attached to each other, and grow  according to  their own rules. The best example concerns growing  epithelia, always connected to their ECM (extracellular matrix), such as the imaginal disc of the Drosophila wing \cite{ackermann2022modeling},  the skin layers of the epidermis \cite{ciarletta2012papillary,kucken2005fingerprint}, and  also the cortex of the brain connected to the white matter \cite{ben2017mimicking,holland2018symmetry}, in  the embryonic period.\\
Finally, it is known that life is compatible with elastic stresses, which is the basis of the criterion of  homeostasis for mammals: Compressive  stress above the homeostatic pressure reduces the cell proliferation, while tensile stress favors the proliferatin.
\subsection{Volumetric Growth and elasticity}
The elasticity invariants  defined in eq.(\ref{invariant}) refer to the elastic tensor ${\bf F_e}$ and not to the deformation gradient ${\bf F}$. ${\cal E}$ must now  take into account the growth per unit volume of the sample, which is represented by Det $({\bf F})=$ Det  $({\bf G})=J$  for an incompressible material, and the elastic energy becomes
\begin{equation}
\label{energiegrowth}
    {\cal E}=\int \!\!\!\int \!\!\! \int_ {\Omega} d V\,\,J 
    \left\{ W(I_1,I_2,I_4,I_5)-Q\,\,(I_3-1)\right \}.
    \end{equation}

The invariants are given by eq.(\ref{invariant}) where ${\bf F}$ is replaced by ${\bf F_e}$. In eq.(\ref{energiegrowth}) the   growth appears  explicitly by  the factor $J$ which indicates that the material volume has changed  and implicitly in the substitution of ${\bf F}$ by ${\bf F_e}$ in all the invariants. If we also consider  this substitution in the definition of ${\bf S}$ in eq.(\ref{stressdef})  and of ${\bf \sigma}$  in  eq.(\ref{stressdefbis}), we have
\begin{equation}
\label{stressgrowth}
{\bf S}= J {\bf G^{-1}} \left \{\frac{\partial W}{\partial {\bf F_e}}-Q\,{\bf F_e^{-T}}\right\}; \,\,
{\bm \sigma}={\bf F_e }\frac{\partial W}{\partial {\bf F_e}}-Q\,{\bf I}\,.
\end{equation}
In contrast to the Piola stress tensor, ${\bf S}$, the Cauchy stress ${\bm \sigma}$  shows no signature of the growth, which can be surprising. At this stage, it is important to emphasize that, first, these two tensors are not defined in the same coordinate basis, and second,  only forces acting on a given surface are invariant quantities,  as will be shown later.   
To illustrate this paragraph, we consider a growth process that is anisotropic and the example of section \ref{stretch}. There is no change in the elastic stretches  except for the compressive loading $P_0$ which becomes $P_G=P_0(g_1g_3)$, if we want to keep the same stress level. The stretches do not change and $\lambda_1$ is the solution of eq.(\ref{lam1}) with $P_0$. However, due to the growth, the new coordinates will be: $x_i=\lambda_i g_i X_i$.

Now consider  the case where the bottom surface of the cuboid is attached to a rigid substrate, assuming anisotropic growth but no applied external compressive stress.  Then for $X=0$, the points of this surface cannot move  and $y=Y$ and  $z=Z$. If no displacement is  possible in the $Y$ and $Z$ directions, the simplest choice is to choose the same rules  $y=Y$ and $z=Z$, everywhere in the sample, and only the allowed displacements are in the $X$ direction so  that $x=J X$.  The elastic stretches are then:
\begin{equation}
 \lambda_2=\frac{1}{g_2};\quad \lambda_3=\frac{1}{g_3};\quad \lambda_1= \frac{J}{g_1}=g_2 g_3.
\end{equation}
According to eq.(\ref{stressgrowth}) the Piola stress tensor at the top in the neo-Hookean approach becomes:
\begin{equation}
    S_1=g_2g_3 \left (g_2g_3-Q\frac{1}{g_2g_3} \right )=0;\,\,  Q=g_2^2 g_3^2.
\end{equation}
 In  both horizontal directions, we have: 
\begin{equation}
    S_2=g_1g_3 \left(\frac{1}{g_2}-g_2^3g_3^2\right);S_3=g_1g_2\left(\frac{1}{g_3}-g_2^2g_3^3\right).
    \end{equation}
Note that the horizontal stresses are compressive, which means  $g_i>1$, indicating that compressive stresses must be applied to the vertical faces at $\pm L_Y$ and at $\pm L_Z$ to maintain such deformation. Another possibility  is an infinite sample in the $Y$ and $Z$ directions. 
However, growth can also induce  a buckling instability which will be studied in  detail in the following. When buckling occurs, this  simple  choice of deformations must be modified, but the main deformation remains for  low  stress levels above the buckling threshold. \\
In conclusion, a  substrate that prohibits any displacement at  the bottom of the parallelepiped is an obstacle  to free growth at the origin of compressive stresses, leading eventually to a shape bifurcation.

\section{Swelling of gels}
Swelling hydrogels have the advantage of mimicking the mechanical behavior of growing soft tissue while being precisely controllable. They consist of reticulated networks of  polymer chains with a high
proportion of small solvent molecules. A phase transition can be induced in the sample 
when it comes into contact with a  reservoir of solvent, resulting in an amazing increase in volume. Although they are perfect candidates for mimicking growing  tissues, growth and swelling have different microscopic origins. A swollen hydrogel is a system in both mechanical and thermodynamic equilibrium, and the swelling does not produce any new polymeric components, which constitute  the only elastic phase and become increasingly  dilute during the swelling process. In addition, the solvent has no reason to be uniformly distributed in the sample. For this reason, different poroelastic models have been proposed for the swelling \cite{quesada2011gel,barros2012surface,hu2023thermo} but also for plant or animal tissues \cite{yang1991possible,humphrey2002constrained,leng2021using,amar2020tip,esposito2022symmetry}. Here, we choose a presentation by Hong et al. \cite{hong2008theory,hong2009inhomogeneous} slightly modified to be as close as possible to the section \ref{volumetric}. 

In fact, at equilibrium, the minimization concerns the grand potential $\hat {\cal W}= \cal W({\bf F},C)-\mu C$  where $C$ is the solvent concentration and $\mu$ is the chemical potential: $\mu=\partial {\cal W({\bf F},C)}/\partial C$. If the gel is in contact with a reservoir full of solvent, then at the interface between the reservoir and the swelling gel, the two chemical potentials in both phases are  equal: $\mu=\mu_s$. If incompressibility is assumed and can be considered as the sum of its incompressible components, then $C$ is  related to Det$({\bf F})$ by the relation: 
Det$({\bf F})=1+\nu C$, where $\nu C$  is simply the ratio between the volume  of the solvent molecules and that   of the dry matrix.  Obviously, although the experiments on swelling gels are easier to perform and show interesting patterns similar to those  observed in nature, we are still faced with two coupled fields: the elastic and the chemical one. Let us consider  the variation of the free energy density:
\begin{equation}
 \delta \hat {\cal W}= \delta W({\bf F}, C)-\mu \delta C=\frac{\partial W}{\partial {\bf F}} \delta {\bf F}+\left(\frac{\partial W}{\partial C}-\mu \right)\delta C\,,
 \end{equation}
 where $\delta C$ is replaced by $\delta$ Det(${\bf F})/\nu$=Det$({\bf F})\, {\bf F^{-T}}\delta{\bf  F}/\nu$. Then,  the corresponding stress becomes:
\begin{equation}
     {\bf S}=\frac{\partial W}{\partial \bf{F}}+\frac{1}{\nu}\left(\frac{\partial W}{\partial C}-\mu \right)\text{Det}({\bf F}) {\bf F^{-T}} \,.
 \end{equation}
 The free energy density $W({\bf F},C)$  is often represented by the addition of two components: $W_e({\bf F})$ and $ W _c(C)$, where the first represents the elastic energy of the polymer matrix, the second, the contribution $W _c(C)$which  depends  only on $C$. For $W_e({\bf F})$,   a classical formulation due to Flory and Rehner \cite{flory1943statistical} leads to :
  \begin{equation}
   W_e({\bf F}) =\frac{1}{2} N kT \left(I_1-3-2 Log(\lambda_1 \lambda_2 \lambda_3)\right),
   \end{equation}
   for a compressible polymer matrix that satisfies the neo-Hookean elasticity, $N$ is the number of the polymer chains,  while for $W_c(C)$ we have:
   \begin{equation}
   W_c(C)=-\frac{k T}{\nu} \left(\nu C Log[\frac{(1 +\nu C)}{\nu C}]+\frac{\Upsilon}{1+\nu C}\right).
 \end{equation}
If we consider  the case of a cuboid with clamped conditions at the bottom, then we can again imagine a diagonal strain and stress tensors with $\lambda_2=\lambda_3$ and  ${\bf F}_{11}=\lambda_1$\,,  so that
\begin{equation}
  S_{1}=N k T\left \{\lambda_1-\frac{1}{\lambda_1} -\frac{1}{N\nu} \lambda_2^2 \left(w'+ \frac{\mu}{kT} \right)\right\} =0\,,
  \end{equation}
  \begin{equation}
  S_{2}=NkT \left \{ \lambda_2  -\frac{1}{\lambda_2}-\frac{1}{N\nu}\lambda_1  \lambda_2 \left (w'+ \frac{\mu}{kT} \right)  \right\},
  \end{equation}
 with  
 \begin{equation}
  w'=-\left(Log(\frac{\lambda_1-1}{\lambda_1}) +\frac{1}{\lambda_1} +\frac{\Upsilon}{\lambda_1^2} \right),
 \end{equation}
 and a similar result for $S_3$, which is equal to  $S_2$.
  The relative increase of the height $\lambda_1$  in the vertical direction leads to a compressive stress in the horizontal directions,  at the origin of the buckling of the sample.
Here the control parameter is $\mu/\nu$ at the origin of the swelling/deswelling. Although there is  an  analogy between volumetric growth and swelling, the theoretical approach will be  more uncertain in the second case and also more dependent on the experimental conditions. Therefore, for our purposes, and in the following, we will restrict ourselves to the simplest initial geometry and suggest  how we can interpret the experiments shown in section \ref{experiment}.
\section{ Biot's theory applied to rubber in compression versus volumetric growth}
\label{pressuregrowth}
\subsection{Compression and critical instability threshold}
Thick samples can buckle under compression. This volumetric instability occurs when the compressive stresses due to  load reach a threshold value. In fact, as mentioned in section \ref{experiment}, experimentalists often characterize  buckling by the compressive strain $\lambda_2$ rather than  by the  compressive load. In fact, strain, which is  the ratio of the length of the speciment to the initial length, is more easily evaluated.  Biot has studied  this buckling instability in detail, in particular  for the neo-Hookean  and Mooney-Rivlin models for a semi-infinite sample representing  a free surface, subjected  to a lateral compression, which we will call  $P_0$. This simple geometry allows a diagonal representation of the strains and stresses  before the bifurcation, and this instability is often called a surface instability because it is more easily observed  at the surface. His proof concerns a  simple plane strain instability  controlled by a  parameter $\xi$, above which the simple diagonal representation  ceases to be valid. $\xi$ and $\xi_B$ are  given by: 

\begin{equation}
 \xi=\frac{\lambda_1^2-\lambda_2^2}{\lambda_1^2+\lambda_2^2}\quad \mbox{and}\quad  \xi_B=0.839287.
\end{equation}
For the neo-Hookean model, Biot \cite{biot1963surface} has established the following relation for  $\xi_B$:
\begin{equation}
\label{eqBiot}
{\cal Q}_B=\xi_B^3+2 \xi_B^2-2=0\,.
\end{equation}

We will consider three different cases, the first two were considered in \cite{biot1965mechanics}. The stresses  are defined in the current configuration and   ${\bm \sigma}$ represents the Cauchy stress. In the  following three cases there  is no stress on the top  free surface, which leads to :  $\sigma_1=\lambda_1^2-Q=0$ when the shear modulus is chosen as unity: $\mu=1$.  It  gives $Q=\lambda_1^2$. Remember that in this case $\sigma_1-\sigma_i=-\sigma_i=\lambda_1^2-\lambda_i^2$\, for $i=2$ or $3$.
\subsubsection{Case one} 
\label{caseone}
We assume that there is no strain in the $Z$ direction and $\lambda_3=1$
    \begin{equation}
       {\bf F} =\rm{Diag} (\lambda_1,\lambda_2,1); \,\,{\bm\sigma} =\rm{Diag} (0,\lambda_2^2-\lambda_1^2,1-\lambda_1^2).
    \end{equation}
With this choice, incompressibility imposes: $\lambda_2=1/\lambda_1$  and the parameter $\xi$ becomes:
\begin{equation}
    \xi=\frac{\lambda_1^2-1/\lambda_1^2}{\lambda_1^2+1/\lambda_1^2}\quad {\mbox{so}}\quad  \lambda_1=\left(\frac{1+\xi}{1-\xi}\right)^{1/4}.
\end{equation}
At the threshold of stability, the value of  the stretches are then given by $\xi=\xi_B$, and   $\lambda_1=1.839287$ so $\lambda_2=0.543689$, and  compressive stresses occur in both directions for $Y$ with  $\sigma_2=-3.0873$ and for $Z$\, with  $\sigma_3=-2.38298$.
\subsubsection{ Case two}
\label{casetwo}
Choosing now $\lambda_1=\lambda_3$ 
\begin{equation}
\label{F2}
       {\bf F}=\rm{Diag} (\lambda_1,\lambda_2,\lambda_1) \quad {\bm \sigma}=\rm{Diag} (0,\lambda_2^2-\lambda_1^2,0) \,.
    \end{equation}
With this choice, the incompressibility imposes:  $\lambda_2=1/\lambda_1^2$  and the parameter $\xi$ and $\lambda_1$  become: 
     \begin{equation}
    \xi=\frac{\lambda_1^2-1/\lambda_1^4}{\lambda_1^2+1/\lambda_1^4}\quad {\mbox{so}}\quad  \lambda_1=\left(\frac{1+\xi}{1-\xi}\right )^{1/6}\,,
\end{equation}
which gives the instability when $\lambda_1=1.50118$ and $\lambda_2=0.443746$. The compressive stress occurs only in the $Y$ direction with $\sigma_2=-2.05663$.\\

\subsubsection{Case three}
\label{casethree}
Finally for the third case, we assume that the compressive loads act similarly in both directions: $Y$ and $Z$.
\begin{equation}
\label{F3}
       {\bf F}=\rm{Diag} (\lambda_1,\lambda_2,\lambda_2);\,\,{\bm \sigma} =\rm{Diag} (0,\lambda_2^2-\lambda_1^2,\lambda_2^2-\lambda_1^2).
    \end{equation}
With this choice, incompressibility imposes $\lambda_2=1/\sqrt{\lambda_1}$  and the parameter $\xi$ and $\lambda_1$ become:
 \begin{equation}
    \xi=\frac{\lambda_1^2-1/\lambda_1}{\lambda_1^2+1/\lambda_1}\quad {\mbox{so}}\quad  \lambda_1=\left(\frac{1+\xi}{1-\xi}\right )^{1/3},
\end{equation}
which gives the instability when $\lambda_1=2.25354$ and $\lambda_2=0.666142$ and a compressive stress equal in $Y$ and $Z$ direction: $\sigma_2=\sigma_3= -4.6347$. Note that this last case is not  considered by Biot.

\subsection{Semi-infinite samples under volumetric growth}
As shown earlier, the Biot instability is mostly controlled by the strains that are directly observable for a solid under compression. There is no difference between the elastic and the geometric strains as opposed to  growth. Assuming that the previous analysis remains valid, we will try to apply the Biot approach to volumetric growth. To do so, we will reconsider the three cases defined above.\\
\subsubsection{ Case one}
\label{caseonebis}
This case  concerns  $\lambda_3=1$, which means that  in this direction the displacement is  equal  to growth. Then the critical elastic strains evaluated in section \ref{caseone} are equal to 
$\lambda_1\sim 1.839$ and $\lambda_2=0.543689$. There are  several  cases  depending on how 
the growth is organized in the sample. For isotropic growth without displacement in the $Y$ direction, we have $x=J X$, $y=Y$ and $z=g Z$ with $\lambda_1=J/g$, $\lambda_2=1/g$ and $J=g^2$. So the expansion in the $X$ and $Z$ direction  at criticality is $g=1.839$. These values were determined directly in \cite{amar2010swelling} and are  recovered in a different way in section \ref{main}. The compressive stresses in the $Y$ and $Z$ directions become: $\sigma_2 =-3.0874$ and $\sigma_3 =-2.383$. $J$ can be evaluated by noting that $\xi=(J^2-1)/(J^2+1)$ which  once introduced into  eq.(\ref{eqBiot}) leads to the polynomial for $J_B$:
\begin{equation}
 {\cal Q}_J=J_B^3-3 J_B^2-J_B-1=0; \,\,  J_B=3.38298\,.
\end{equation}
This configuration  will be examined in detail in all the following sections.
\subsubsection{Case two}
\label{casetwobis}
This case concerns the growth of a sample with $2$ sides\, without stress. Assuming $x=J_1 X$, $y=J_2 Y$ and $Z=J_1 X$, then at the threshold   $ \lambda_1=J_1/g=\lambda_3=1.5012$ and  $\lambda_2=J_2/g=0.4437$ with $g$ defined as $g^3=J_1^2J_2$. There is  only a compressive stress in the $Y$ direction with the same value as in section \ref{casetwo}: $\sigma_2=-2.0567$

\subsubsection{ Case three}
In this case it is assumed that  $x=J_1 X$, $y=J_2 Y$ and $z=J_2 Z$. If  the displacement is forbidden along the $Y$ and $Z$ directions, then $J_2=J_3=1$ and $J=J_1=g^3$. 
 \begin{equation}
 \begin{cases}
{\bf G}=\rm{Diag} ({g,g,g)};\\
 {\bf F}=\rm{Diag} (\frac{J_1}{g},1,1);\\
 {\bf F_e}=\rm{Diag} (\frac{J_1}{g},\frac{1}{g},\frac{1}{g});\\
  {\bm\sigma}=\rm{Diag} (0,\frac{1}{g^2}-g^4,\frac{1}{g^2}-g^4) \,.
 \end{cases} 
 \end{equation}
This unidirectional growth process produces  lateral compressive stresses when $g$ and $J_1$ are  greater  than one. In the opposite case $J_1<1$, the stresses are tensile. This case is similar to eq.(\ref{F3}) and 
\begin{equation}
\xi_B=\frac{g^4-1/g^2}{g^4+1/g^2}=
\frac{J_1^2-1}{J_1^2+1} \,.
\end{equation}
At the threshold, replacing $\xi_B$ by $J_B$ in eq.(\ref{eqBiot}) we obtain the critical threshold for such growth process given by:
\begin{equation}
{\cal Q}_J=J_B^3-3 J_B^2-J_B-1=0\,.
\end{equation}
The solution for $J_B$ is then $J_B = 3.38298$, the critical strain is then $\lambda_1=2.25354$ and $\lambda_2=0.666142$. Note that we recover the same threshold for the growth parameter as for section \ref{caseonebis}.\\
Growth anisotropy increases the space of possible instability parameters. Here we limit ourselves to   three cases and restrict ourselves to homogeneous growth. The Biot instability is generic,  but depending on the situation, the thresholds  can be different and must be evaluated each time.  In the following, we will consider only one case  with a different theoretical approach, without  using the  Biot's method, which imposes  a critical parameter $\xi_B$  \cite{biot1963surface}. We prefer a presentation in terms of variational analysis.
 
\section{Growth of a semi-infinite sample}
\label{main}
It is  impossible  to list all the publications on  volumetric growth in soft matter. If growing layers, multilayers, shells, disks, spheres are the most frequently chosen geometries \cite{goriely2017mathematics}, numerical treatments with advanced finite elements methods softwares allow to represent a variety of shapes closer to reality \cite{wang2022}. Our purpose is different since we want to give exact results with the simplest and standard hyperelastic model, that is the neo-Hookean model \cite{ogden1997non,holzapfel2000nonlinear,goriely2017mathematics} for incompressible materials. In addition, instead of considering  all possible growth processes that  can be found in nature, anisotropic \cite{amar2013anisotropic} or space dependent \cite{chen2020stress,chen2022generating}, we focus on a spatially constant growth that  evolves on a rather long time scale in order to neglect any transient dynamics. Since elasticity goes  hand in hand with  geometry \cite{audoly2010elasticity}, we   start with the geometry of the sample  to fix the notations used in the following.

\subsection{The geometry}
We consider a semi-infinite growing sample bounded by the plane  $X=0$, infinite in the positive $X$ direction and extending laterally between $-\infty, \infty$ in the $Y$ and $Z$ directions.  We assume $\lambda_3=1$, so that no elastic strain exits in the third direction. The growth is assumed to be isotropic and homogeneous with a  constant  relative volume expansion $J=g^3$. Due to the Biot instability (see the previous section), periodic patterns will appear on top of the sample with a spatial periodicity  $\Lambda$ chosen as the length unit. This geometry orients the growth  mostly in the  $X$ direction and the  new position for an arbitrary material point inside the sample  leads to compressive stresses in the $Y$ direction, as described  before in section \ref{caseonebis}. Thus, defining a Cartesian coordinate system  $X,Y$ in the initial configuration, the position of each point after growth and the elastic deformation  becomes $x\sim J X$ and $y\sim Y$, in leading order and  $J_{2D}=J=g^2$. Since an adiabatic approach to the growth process is assumed, i.e. transient deformations are quickly eliminated, a free energy describes the possible  patterns  resulting from a symmetry breaking. Our approach, which is  poorly followed in the mechanics community, will be based on energy variation and  will avoid tensorial algebra. 
\subsection{The variational method based on the free energy minimization}
\label{variational}
\subsubsection{The free energy: elasticity and capillarity}
The Euler-Lagrange  equations or the equilibrium equations result from the extremum of the free energy, the sum of the elastic and possibly  surface energy. Assuming a perfect periodicity of the patterns, we make a virtual partition of the initial domain into stripes of unity width and focus on ${\cal P}$\, the domain between $-1/2<Y<1/2$, see the blue domains in Fig.(\ref{riemann}). The  neo-Hookean model  depends on only  two invariants: $ I_1$, for the elastic  deformations and $ I_3$ for the relative volume change due to elastic stresses, which  we renormalize into the geometric invariants: $\tilde  I_1= J I_1$ and $\tilde I_3=J I_3$ :
\begin{equation}
\label{Ione}
\tilde I_1=x_X^2+ x_Y^2+y_X^2 + y_Y^2-2J; \tilde I_3=x_X y_Y- y_X x_Y-J\,,
\end{equation}
where the subscript $X$ (resp.\,$Y$) denotes the  partial derivative of any function  with respect to the variable $X$ (resp.\,$Y$).\\
The invariants $I_1$ and $I_3$ have already been defined in section \ref{Invariant}. The energy unit is chosen as the product: $\mu\cdot (\Lambda^2 t_3)$.  $t_3$  is the thickness of the sample in the orthogonal direction which is  irrelevant for plane strain deformations and we have for the elastic energy of a single  strip:
\begin{equation}
\label{energy}
{\cal E}_e=\frac{1}{2} \iint_{{\cal P}}dS\left (\tilde I_1  -2 Q\tilde I_3 \right ).
\end{equation}
 The Lagrange parameter $Q$ is  also a function of $X$ and $Y$  fixing the incompressibility constraint $I_3=1$ or $\tilde I_3=0$ and $dS=dXdY$. The capillary energy is often written in  Eulerian coordinates:
\begin{equation}
\label{eulcap}
{\cal  \tilde E}_c= \gamma_0
 \int_{\partial {\cal P}} dy 
  \sqrt{1+ x_y^2}\,.
 \end{equation}
Considering the upper boundary $\partial {\cal P}$:
$$   X=0;\quad Y \in [-1/2,1/2],$$
where  the capillary energy  is defined,  the following relations hold : 
\begin{equation}
dy=\frac{\partial y}{\partial Y} \vert_{_{X=0}} \,dY \quad \mbox{and}\quad   dx=\frac{\partial x}{\partial Y}\vert_{_{X=0}}\, dY,
\end{equation}
then  eq.(\ref{eulcap}) is  transformed into: 
\begin{equation}
\label{energycap}
{\cal E}_c= \gamma_0
 \int_{\partial {\cal P}} dY 
 ( \sqrt{x_Y^2+y_Y^2}-1)\,,
\end{equation}
where $\gamma_0$ is the  rescaled capillarity coefficient and  is equal to  $\gamma_0=\gamma/(\mu \Lambda)$ ( $\gamma$ is the surface tension). 
Capillarity represents the average energy difference  between the microscopic components of  the sample (as atoms, molecules) located in the bulk or at the interface.   It is positive when the interface separates a dense material from a more dilute phase. 
 In practice, the capillary coefficient $\gamma_0$  is very weak for ordinary gels and  plays a significant role only when  the sample size  is of the order of  $0.1$mm and for extremely soft gels \cite{mora2010capillarity}.\, However a skin effect can occur on top of elastic samples due to inhomogeneity of the shear modulus or to the growth process itself. This is especially true for the swelling of gels. Despite the weakness of this energy, it plays a crucial role   in the determination of  the wavelength and  in the local regularization of  singular profiles. 
 \subsubsection{The Euler-Lagrange equations}
 They  simply result from the first variational  derivative of  the functional ${\cal E}_e$ with respect to  the small variation of $x$ and $y$:
\begin{equation}
\label{euler}
\begin{cases}
x_{XX}+x_{YY}= Q_ X\, y_Y- Q_Y \, y_ X=\{Q,y\} \,,\\
y_{XX}+y_{YY}=- Q_ X \,x_Y+ Q_Y\, x_ X=-\{Q,x\}\,.
\end{cases}
\end{equation}
The left-hand side of the equation (\ref{euler}) represents the Laplacian $\Delta$ in Cartesian coordinates, and $\{P,x_i\}$ is the Poisson bracket of $P$ and $x_i$. This mathematical symbol has important properties in mechanics \cite{landau1969mecanique}.  The zero-order solution : $x=J X$ and $y=Y$ verify these equations when the Lagrange parameter is a constant, so $Q=Q_0$. 
Boundary conditions are also derived from the first variational derivative of   
 ${\cal E}_e$ and ${\cal E}_c$ with respect to the elementary variation of $x$ and $y$, a process which allows the cancellation of the normal components $S_{11} $ and $S_{21}$ of the Piola stress tensor $\bf S$ \cite{ogden1997non,holzapfel2000nonlinear},
 at the free boundary $\partial {\cal P}$:
%%%%%%%%%%%%
\begin{equation}
\label{borderstress}
S_{11} =x_X-Q\, y_Y \quad \mbox{and} \quad  S_{21} =y_X+Q\, x_Y\,.
\end{equation}
On top, for $X=0$,  the cancellation of  $S_{11}$  gives $Q_0=J$ while $S_{21}=0$ is automatically obtained for the zero order solution.  Capillarity appears for buckled solutions and is responsible for  the  normal $\Gamma_{11}$ and tangential $\Gamma_{21}$ components:
\begin{equation}
\label{capstressone}
\Gamma_{11}=  \gamma_ 0 \frac{\partial }{\partial Y} \frac{ x_Y}{(x_Y^2+y_Y^2)^{1/2}}\,,
\end{equation}
and
\begin{equation}
\label{capstresstwo}
\Gamma_{21} = \gamma_0 \frac{\partial}{\partial  Y} \frac{ y_Y}{(x_Y^2+y_Y^2)^{1/2}} \,,
\end{equation}
which must be added to the normal stresses at  $X=0$.
Note   the strong nonlinearities  in the surface energy. However, since $\gamma_0$  is in practice a very small parameter,  the role of the capillary stresses is probably negligible for  smooth patterns, but may  become  important in the case of creases.
For completeness, the  other two components of the stresses are also given:
\begin{equation}
\label{borderstressno}
S_{12} =x_Y+Q\, y_X \quad \mbox{and} \quad  S_{22} =y_Y-Q\, x_X\,.
\end{equation}
So far, it  is assumed that the interface is regular and admits  a regular curvature everywhere. Self-contacting interfaces are not considered, although in the last panel  (A9)of Fig(\ref{cuspexp}) on the right, such a property can explain the highly singular pattern obtained in the radial geometry. Assuming that it happens  at a position $Y=0$, then two additive stress boundary conditions must be imposed locally \cite{karpitschka2017cusp, ciarletta2018matched,ciarletta2019soft},
\begin{equation}
    S_{22}\vert_{Y=0^{+}}=S_{22}\vert_{Y=0^{-}}\,\, {\mbox{and}}\,\, S_{12}\vert_{Y=0^{+}}=S_{12}\vert_{Y=0^{-}},
    \end{equation}
the second condition indicates the absence of friction on the singular line.\\
Finally, it is easy to show that the Euler-Lagrange equations, eq.(\ref{euler}), are equivalent to the cancellation of the divergence of the Piola stress tensor,  see also section \ref{defstress} and eq.(\ref{diverg}). In Cartesian coordinates, Div$({\bf S})_i=\partial Sij/\partial X_j$.

%Figure obtenue par CUSPFINALAUGUST21%
\subsection{Incremental approach and solution of the Euler-Lagrange equations}
The classical way to detect a bifurcation in the elasticity  is to  expand the general solution by adding a small perturbation scaled by a small parameter $\epsilon$.  The following results are obtained for $x$ and $y$ and $Q$ :
\begin{equation}
\label{eps}
\begin{cases}
Q=J+\epsilon q(X,Y) \,,\\
    x=J X+\epsilon U(X,Y)\quad \mbox{with} \quad   \Delta U=q_X\,,\\
    y=Y+\epsilon V(X,Y)\quad \mbox{with} \quad  \Delta V=J q_Y\,.
    \end{cases}
\end{equation}
The incompressibility condition at $\epsilon$ order  imposes the following constraint $U_X+J V_Y=0$ and the elimination of $q$  is easy by cross-derivation  of the previous equations, eq.( \ref{eps}): 
$J \partial_Y \Delta U-\partial_X \Delta  V=0$ which can be derivated a second time to isolate $U$ from $V$. Defining  $\Delta_J=\partial^2_{XX}+J^2\partial^2_{YY}$: 
\begin{equation}
\label{sol}
    \Delta_J (\Delta U)=\Delta_J (\Delta V)=0.
\end{equation}
and   $\Delta_J q=0$. The fourth order operator $\Delta_J \Delta=\Delta \Delta_J$ accepts as possible solutions $\Re \,(\Phi_1(Z)+\Phi_2(Z_1))$ where both functions are holomorphic functions of $Z=X+I Y$ and $Z_1=J X+I Y$. Nevertheless, due to the boundary conditions for $X=0$ (on  the top of the strip), $\Phi_1$ and $\Phi_2$ are related and we finally get for $x$ and $y$:
\begin{figure*} 	
    \centering 
    \begin{minipage}[b]{1\textwidth}{} 
      \subfigure{\,}{\includegraphics[width=.9\textwidth]{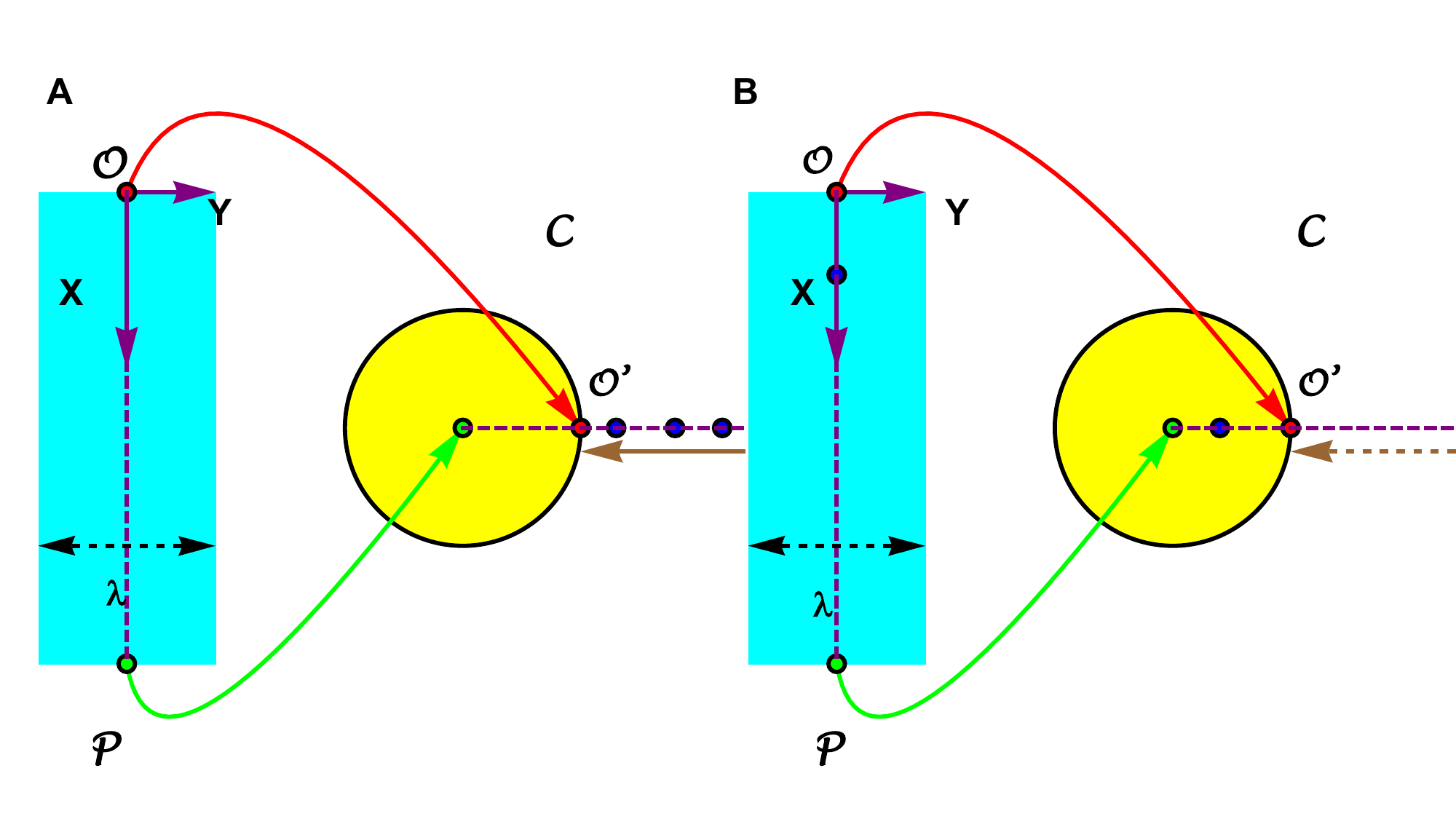} }
      \hspace{2in} 
      \subfigure[\,\bf{Profile for an outer pole}]{ 
            \includegraphics[width=0.33\textwidth]{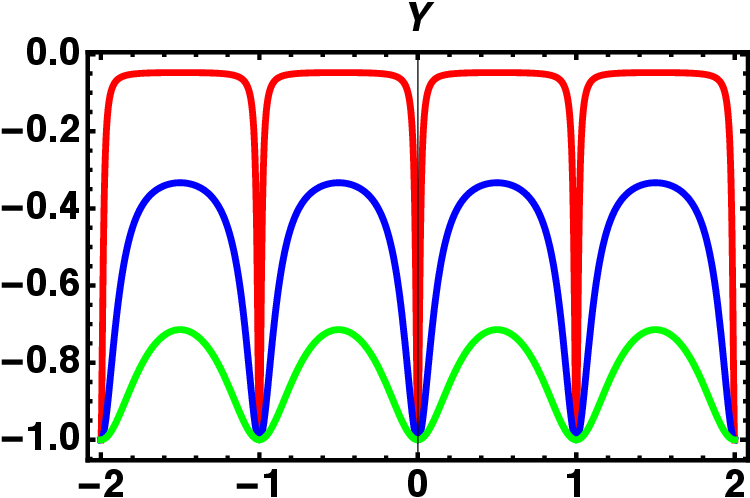} 
        } 
        \hspace{0in} 
        \subfigure[\,\bf{Real Part} ]{ 
            \includegraphics[width=0.3\textwidth]{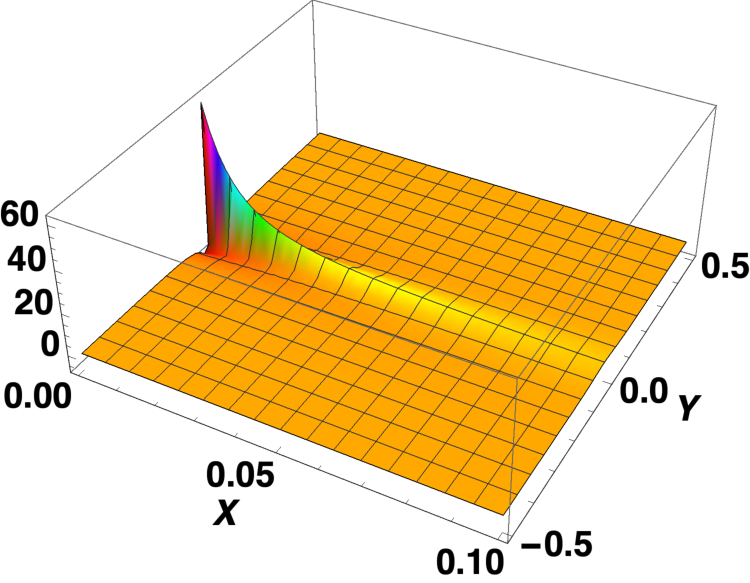} 
        } 
        \hspace{0in} 
        \subfigure[\,\bf{Imaginary part}]{ 
            \includegraphics[width=0.3\textwidth]{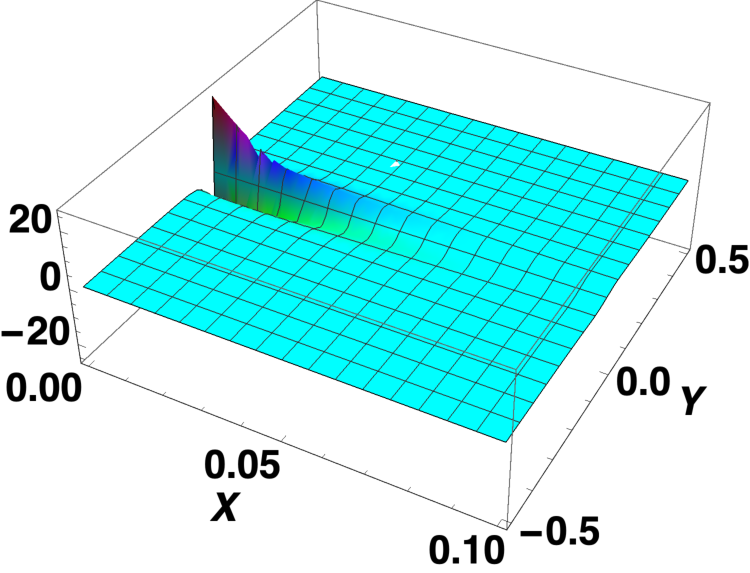} 
        } 
    \end{minipage} 
    \caption{At the top, panels A and B, the  physical domain ${\cal P}$, corresponding to one wavelength in cyan, is restricted to $-1/2 \leq Y \leq 1/2$, $X \in [0, +\infty]$.  Only one wavelength is plotted. To the  right of each panel, is the ${\cal C}$ plane with the unit Riemann disk in yellow. The red arrow indicates the correspondence between the point $\mathcal{O}$ ( $X=0,Y=0$ ) of the physical space and the point  $\mathcal{O'}$ (${1,0}$)
    of the ${\cal C}$ plane.
 The region $X-> +\infty$  is contracted and associated with the center of the Riemann disk (green arrow). Only the interior and the contour of the Riemann disk mimic the physical domain. The blue dots represent the possible singularities, which  are not singular for the physical domain in panel (A),  but can generate sharp interface variations. Note that the dashed purple lines are images of each other due to  the conformal mapping. In (B), the same plot except that a  singular point enters both  the Riemann disk and  the physical plane {\cal P}. Below, in (b), the interface profile for an outer pole, see   eq.(\ref{singular}), with the parameter $a=0.1$ for the red curve, $a=1$ for the blue curve, $a=5$ for the green curve. The profile appears  quite singular for $a=0.1$, but it remains finite. In (c) the real part of the derivative which contributes to the stresses, see $S_{11}$, according to  eq.(\ref{stress}). One wavelength, ${\mathcal{P}}$, is shown and the rainbow colors indicate the sharp variations near  the origin $\mathcal{O}$. Note also that the scale for $X$ varies between $0$ and $0.1$. Although quite singular near $\mathcal{O}$, the stress remains finite and decreases rapidly. Similarly,  (d) shows the opposite of the imaginary part present in  $S_{21}$ of eq.(\ref{stress}).} 
   \label{riemann}
    \end{figure*}

\begin{equation}
\label{expansion}
\begin{cases} 
x &=J X+ J \epsilon \Re \, [\Phi (Z)+  \tau_1 \Phi (Z_1)] \,,
 \\
y &=Y-\epsilon  \Im \, [\Phi(Z)+J \tau_1 \Phi (Z_1)]\,,
\\
Q &=J+\epsilon \tau_0 \tau_1 \Re \, [\Phi_{Z_1}];\,\,  \tau_0=J^2-1 \,,
\end{cases}
\end{equation}
where $\Re \,$ and $\Im \,$ are  the real and imaginary parts of the holomorphic function $\Phi$.  The notation $\tau_0=J^2-1$ is introduced for convenience and $\tau_1$ is  free at this stage, and  will be determined later. A priori $\Phi$ is arbitrary with the restriction that it must vanish for $X \rightarrow \infty$ which  automatically implies that $\Phi$ is singular.  Any singularity  occurring  outside of the domain ${\cal P}$ (for $X<0$) is physically appropriate  while singularities within the physical domain ${\cal P}$ must be considered with caution. 
The balance of the elastic and capillary stresses  at the surface $\partial {\cal P}$ gives the value of $\tau_1$ as well as  the threshold $J_B$ for the buckling instability. Let us first evaluate the stresses in linear order in $\epsilon$:
 The calculation is not difficult  and can be  easily done  using the Mathematica software as an example (see also the Appendix, section \ref{appendexp}):
\begin{equation}
\label{stress}
\begin{cases}
S_{11} &= \epsilon   \Re \,[ 2 J\Phi_Z+(1+J^2) \tau_1 \Phi_{Z_1}]\,,\\
S_{21} &= -\epsilon   \Im \,[ (1+J^2)\Phi_Z + 2 J^2 \tau_1 \Phi_{Z_1}] \,,\\
S_{12} &=-\epsilon  J \Im \,[ 2\Phi_Z +(1+J^2) \tau_1 \Phi_{Z_1}]\,,\\ 
S_{22}&=1-J^2-\epsilon \Re \,[(1+J^2) \Phi_Z+2 J^3 \tau_1 \Phi_{Z_1}]\,.
\end{cases}
\end{equation}
Only $S_{22}$ shows a zero order  contribution in  $(-\tau_0)=1-J^2$, which is negative for a growing material since $J>1$. This compressive stress  explains the buckling instability and is associated with an  elastic strain along $Y$.
\subsection{The boundary conditions at the top of the sample, the Biot threshold and the harmonic modes}
%Biot Solution and Biot threshold}
 To derive the condition of a buckling instability, the quantities of interest are the normal $S_{11}$ and shear $S_{21}$ stresses at the top,  which must include the capillary contribution. Only the normal capillary stress $\Gamma_{11}$ is of order $\epsilon$ while $\Gamma_{21} $ is of order $O(\epsilon^2)$ and can be discarded so it reads, for $X=0$ and for $S_{11}$:  
\begin{equation}
\label{disper}
\epsilon\left( 2 J +(1+J^2)\tau_1\right )\cdot \Re \,[\Phi_{Z}]
-\epsilon \gamma_ 0 J(1+\tau_1)  \Re \,[\Phi_{ZZ}]\,,\\
\end{equation}
where $\tau_1$ is not modified.
We first neglect the surface tension. Then,  the cancellation of $S_{21}$ gives  the value of  $\tau_1$: $\tau_1=-(1+J^2)/(2 J^2)$.
%For an harmonic mode $\Phi(Z)=e^{-2 n\pi Z}$ and a  small capillarity effect:
%\begin{equation}
% \tau_1=-\frac{1+J^2}{2 J^2} +n \pi\gamma_0 \frac{(J-1)^2}{2J^3}  
% \end{equation}.
 Once this $\tau_1$ value is introduced into  $S_{11}$, there are two possibilities:
\begin{itemize}
\item Cancellation of ${\cal Q}(J)$, leading to the determination of $J_B$ such that  :
\begin{equation}
\label{charact}
{\cal Q}(J_B)=(J_B^3-3 J_B^2-J_B-1)=0 \,,
\end{equation}
for any profile function $\Phi(Z)$. This value was also found by Biot, see section  \ref{caseonebis} where another demonstration by Biot is proposed \,\cite{biot1965mechanics} .\\
\item
  $\Re \,[\Phi_{Z}]=0$  which defines a family of suitable profiles but not a threshold  value for observing  the interface buckling. It requires that $\Phi$ is an even function of $Z$.
\end{itemize}
The second case does not imply any specific value of  $J$, but selects  shape profiles, unlike the first case which occurs  above $J_B$ for any profile. It suggests and explains the diversity of experimental observations for the layer buckling: Indeed, the absence of mode selection at the Biot threshold  automatically induces a spontaneous coupling of harmonic modes. 
The only real root $J_B$ of ${\cal Q}(J)$ is
\begin{equation}
\label{biot}
J_B=\frac{1}{3}(3 +  6^{1/3}\{(9 - \sqrt{33})^{1/3} +(9 + \sqrt{33})^{1/3}\})\,,
\end{equation}
$J_B\sim 3.38$. But, as mentioned above, all holomorphic periodic functions of $Z$ that  vanish for $X  \rightarrow  +\infty$  are possible eigenmodes of  deformations that occur for the same threshold value. In the original papers, Biot only focused  on the harmonic modes: $\Phi_B= e^{-2 \pi n Z}$, which appear for a compressed  rubber sample. The polynomial that gives the threshold is not always the same depending on the experiment. Any modification in the physics of the problem such as more sophisticated hyperelasticity (Mooney-Rivlin, Ogden model, Fung-model \cite{ogden1997non,holzapfel2000nonlinear}, anisotropy of the material \cite{holzapfel2019fibre} or of the growth \cite{amar2013anisotropic,amar2017mimicking}, possibly   external loading, will modify  the incremental result eq.(\ref{expansion}) and the critical polynomial $\cal{Q}$, but not the fundamental property of instability.\\
\\
However, this model does not provide a choice of  wavelength at the threshold, unlike the similar instabilities of fluids such as Rayleigh-B\'enard or B\'enard Marangoni \cite{charru2011hydrodynamic}. Above the threshold, the determination of the wavelength for periodic fluid instabilities remains  a difficult theoretical challenge
\cite{guyon2001hydrodynamique,charru2011hydrodynamic,meron2015nonlinear} giving rise to an important literature and sometimes controversies  as for the Rayleigh-B\'enard convection or  the diffusion patterns of directional solidification. In \cite{amar2010swelling}, a surface tension selection mechanism is proposed for a layer of finite height  $d$. 
It  induces a shift of the threshold $J_B$ and a selection of the  ratio: wavelength over  the height  $d$ of the sample, this ratio being of order one.  Here the sample height is infinite and the wavelength is chosen as length unit,  which means that the selection must in principle provide the value of the critical threshold $J_C$. A discussion of finite size effects is deferred to the last section \ref{Finitesize}.  When capillarity is introduced, the normal stress $S_{11}$ and the shear stress $S_{21}$, given by  eq.(\ref{disper}) are modified by  the capillary contribution.
Only the  periodic function $\Phi=e^{-2 n \pi Z}$ (where n is an integer) gives  a simple solution with a shift of the bifurcation threshold due to  $\Gamma_{11}$:
 \begin{equation}
 \label{capillarity}
 \delta J=J_C-J_B= n \pi\gamma_0 J_B\frac{(1 + J_B)}{(3 J_B^2-6 J_B-1)}\,.
 \end{equation}
 It is possible to recover this threshold  directly minimizing the the total energy: elastic and capillary energy. In the next section, we give examples of such an evaluation
 which takes advantage of the expansion given in section \ref{develop}.
\section{Periodic profiles and the Riemann theorem}
\label{linear}
\subsection{Construction of periodic profiles}
The choice of the  periodic  functions $\Phi(Z)$ follows from  the Riemann theorem, which states that there exists a bijective mapping between a non-empty and simply connected domain  and the unit disk  \, ${\cal C}$. In our case the domain is  ${\cal P}$, which covers one period, see  Fig.(\ref{riemann}). Introducing the complex variable  $\zeta= e^{-2 \pi Z}$,  Fig.(\ref{riemann}) shows the points of correspondence, in particular the upper boundary ($X=0$) of $\partial {\cal P}$, which is mapped onto  the outer circle $\partial {\cal C}$, and the zone at infinity of ${\cal P}$  ($X\rightarrow + \infty$) concentrated in the center of the unit disk. The exterior of the unit disk corresponds to the non-physical half-plane where $X<0$. The central  vertical  axis of the strip $X>0,Y=0$, (dashed purple line in Fig.(\ref{riemann})) is associated with the horizontal axis of the ${\cal C}$ plane (purple dashed lines) which we extend into the non-physical domain. Every holomorphic function, except constants, $\Phi(Z)$ or $\Psi(\zeta)$, has singularities. If they are located in the non-physical plane, these solutions  are physically relevant, since they contribute  to a finite elastic energy density.  But this is not the case when they are located inside ${\cal P}$ or ${\cal C}$, where they require special attention. When they are near  the boundary or  of the Riemann circle $\partial {\cal C}$, they become good candidates for generating creases. We will consider  the regular profiles first.   

\subsection{Regular patterns above the Biot threshold.}
\label{quasisinguar}
The patterns of interest are of course the harmonic modes  $\zeta_k=e^{-2 \pi k Z}$ and their superposition:\, $\Phi(Z)=a_k \zeta^k$  where the Einstein notation on double indices is assumed and with $k$ a positive integer. The Biot solution is  simply  $\zeta=e^{-2 \pi  Z}$. All these modes without specific parity in the change $Z \curvearrowright -Z $  occur strictly at the Biot threshold and  can easily overlap. However, when focusing on folds occurring at the interface, a more appropriate choice must be made  with singularities located near the interface. The word creases is chosen to describe a sharp and localized variation of  the interface shape $\partial {\cal P}$, which is mathematically represented  by a continuous function, such  that the profile  $x(Y)$  remains differentiable, at least $2$ times.  Another definition can be that the elastic and/or capillary energy density remains locally finite. A fancy representation in complex analysis has been given by the  viscous interfacial flow  experts. \cite{tanveer1995time,vasconcelos1993exact,cummings1999two,crowdy2019analytical}.  The creases are then simply  generated at the threshold by using the conformal mapping technique  \cite{crowdy2011analytical,crowdy2013analytical}.
Defining the neighborhood of the central line $ \zeta_a= \zeta-1-a$, with $a>0$, possible solutions with a pole, a logarithm or a square root  can be good representations of quasi-singular profiles  in the neighborhood near the center of the strip $\mathcal{O}$ or near $ \mathcal{O'}$:
\begin{equation}
\label{singular}
\Phi=\frac{a}{\zeta_a};\,\,\Phi=-\frac{Log(-\zeta_a)}{Log(a)};\,\, \Phi=(-\zeta_a)^{1/2}\,,
\end{equation}
 $a$  decreases as one approaches the point $ \mathcal{O'}$, or the interface near $ \mathcal{O}$ (see Fig.(\ref{riemann})). The amplitude of the singular profile is normalized in the definition given by eq.(\ref{singular}). Fig.(\ref{riemann}) shows different profile solutions for $a=0.1$ (red curve), $a=1$ (blue curve) and $a=5$ (green curve)  for a pole singularity  (first choice of eq.(\ref{singular}) on the left)  corresponding to a distance $d_a$ from the point $ \cal {O} $ of the physical plane with    $d_a=-0.0152,-0.110,-0.285$ respectively.  Since $\Phi_Z$ goes directly in the  stress definition, its value gives  information about the stresses, see eq.(\ref{stress}) and Fig.(\ref{riemann}), panels (c)\,  and (d). Plotted on a single wavelength  $-0.5<Y<0.5$ (with $a=0.1$), the real and imaginary parts of $\Phi$ show a strong  localization near the interface for $a=0.1$ so $d_a=0.0152$ and they  quickly disappear with increasing values of $X$. However, even if the stresses at the interface are large, the solution is not singular and the linear expansion remains valid for sufficiently  small values of $\epsilon$.  For the logarithm or square root  choices presented in eq.(\ref{singular}), see the Appendix (\ref{appenprofile}).

 \subsection{The case of even holomorphic function of  $Z$}
 The second way to satisfy the cancellation of the stress $S_{11}$ for $X=0$ is to choose an even function of $Z$, which means that $\Phi(Z)= a_k (\zeta^k +\zeta^{-k})$ which will automatically   diverge in the center of the Riemann disk or at infinity of ${\cal P}$. The only way to satisfy the convergence at infinity is to  introduce a singularity inside the Riemann disk. The choice of such a singularity is huge, but $2D$ elasticity allows only logarithm and square root singularities for  elastic energy convergence. In linear elasticity, it is well known that  square roots c correspond to fractures and logarithms 
 to wedge dislocations, see \cite{landau1967theorie}. Before proceeding  further in this direction,
 let us  start with the  nonlinear  coupling of regular modes. 
  
\section{Nonlinear bifurcation analysis via  energy minimisation and  mode coupling }
\label{energetic}
 All modes emerge at the Biot threshold and the mechanisms of selection are never a simple matter in the physics  of continuous media. For  example in diffusion-limited growth, the selection of the size and velocity of the needle-crystal remained a puzzle  for five decades. In fact,  Ivantsov \cite{ivantsov1947temperature,horvay1961dendritic,langer1980instabilities,gollub1999pattern}
     has demonstrated the relationship between the tip radius of the crystal times its  velocity as a function of the undercooling since $1947$ but  disentangling  both quantities by the appropriate  correct physical process has generated  many conflicting hypotheses and discussions. The role of the surface tension, much debated due to mathematical difficulties, is now understood by including the  surface tension anisotropy \, \cite{amar1986theory,brener1991pattern}. In the same class of problems, the displacement of a  viscous finger in a channel remains unsolved for  about thirty years \cite{bensimon1986viscous} and it has been  again demonstrated the role of the surface tension which chooses  a discrete set of solutions among a continuum  \cite{shraiman1986velocity,hong1986analytic,combescot1988shape}. When an energy emerges, as in our formulation of volumetric growth, solutions can be selected by an energy minimization, which was not the case for the two examples mentioned above. However our continuum here is a continuum of possible functions, not just a selection of a pure number characteristic of the pattern.  Using an expansion of the deformation at  $\epsilon$ order, the energy density can be expanded in $E=  \tau_0/2+\delta E$, as follows:
\begin{equation}
\label{energyexp}
\delta {\cal  E}=\epsilon  \int_{{\cal P}} dS ( E_1 + \epsilon E_2 +\epsilon^2 E_3)= \epsilon {\cal E}_1+ \epsilon^2 {\cal E}_2 + \epsilon^3 { \cal E}_3\,,
\end{equation}
where  each order is given in the Appendix (\ref{develop}).

If  the system is governed  by an energy, it is a possible  to analyze  the bifurcation in more detail, to deduce its characteristics and finally to obtain  the amplitude $\epsilon$ of the selected mode  by expanding the energy.
 To prepare such a calculation, which can be really tedious and even impossible for an arbitrary choice of $\Phi$,  we take advantage of the complex formulation.
 \subsection{Elastic energy evaluation, order by order}
 \label{Ordertwo}
Such an evaluation requires surface integrals  covering   the entire domain of interest ${\cal P}$ (Fig.(\ref{3modester}), top  left panel) which can be obtained   in two ways: either in $X,Y$ coordinates, or in $Z,\bar Z$ coordinates. The latter choice makes the calculus much more easier for the first and second order, as soon as holomorphic  functions are chosen in the  rectangular geometry. First, we define  these surface integrals defined on  ${\cal P}$ as:
% Indeed according to  \cite{davis1972double}, it reads for 
\begin{equation}
    \begin{cases}
K^{(1)}(f,\bar g)=\frac{1}{2 I}\iint_{{\cal P}} dZ d\bar Z f_Z  \bar g_{\bar Z}\,,\\
 K^{(2)}(f_1, \bar g_1)=\frac{1}{2 J I}\iint_{{\cal P}} d Z_1 d\bar Z_1 f_{Z_1}  \bar g_{\bar Z_1}\,,\\
 K^{(3)}(f\bar g_1)=\frac{1}{ I(J+1)}\iint_{{\cal P}} dZ d\bar Z_1 f_Z  \bar g_{\bar Z_1}\,,\\
  K^{(4)}(f, g_1)=\frac{1}{I(J-1)J }\iint_{{\cal P}} d Z d Z_1 f_{Z}  g_{ Z_1}\,.
  \end{cases}
 \end{equation} 
According to  \cite{davis1972double}, these integrals can be transformed  into contour integrals such that:
\begin{equation}
 \label{modelK2} 
 \begin{cases}
K^{(1)}(f,\bar g) &=\frac{1}{2 I}\oint_{\partial {\cal P}}  dZ  f_Z  \bar g(\bar Z)\,,\\ K^{(2)}(f_1, \bar g_1) 
 &=\frac{1}{2 I J}\oint_{\partial {\cal P}}  d Z_1  f_{Z_1}  \bar g(\bar Z_1)\,,
\end{cases}
\end{equation}
and for $K^{(3)}$ and $K^{(4)}$ which  mix $Z$ and $Z_1$  using :
\begin{equation}
   Z_1=Z\frac{1+J}{2}+ \bar Z\frac{J-1}{2}\,,
\end{equation}
it comes:
\begin{equation}
\label{modelb}
\begin{cases}
K^{(3)}(f,\bar g_1)&=\frac{1}{ I(J+1)}\oint_{\partial {\cal P}}  dZ  f_Z  \bar g(\bar Z_1)\,,\\
 K^{(4)}(f, g_1) &= \frac{1}{ I (J-1)}\oint_{\partial {\cal P}}  d Z  f_{Z}  g( Z_1)\,.
\end{cases}
\end{equation}

The first order corresponds to
\begin{equation}
\label{Eonetwo}
\begin{cases}
{ \cal E}_1 &= \tau_0 \int_{{\cal P}} dS \Re \,(\Phi_Z+J \tau_1 \Phi_{Z_1})\\
&=\tau_0 \Re \,[K^{(1)}(\Phi,1)]+J \tau_1 K^{(2)}(\Phi,1)]\,.
 \end{cases}
 \end{equation}
 Since $\Phi$ has no singularity inside the sample, the contour integral for $\Phi $ vanishes and ${ \cal E}_1=0$.
  ${\cal E}_2$ and ${ \cal E}_3$ can be found in  section \ref{develop}, eq.(\ref{densityel}).
 Using eq.(\ref{densityel}), expansion of ${\cal E}$ at second order gives for ${\cal E}_2$, :
\begin{equation}
\label{Eonethree}
\begin{cases}
 {\cal E}_2&=\frac{1}{2} (1+3 J^2)(K_1+J^2\tau_1^2 K_2) \\
 &+\frac{J \tau_1}{2} ((J+1)^3 K_3-(J-1)^3 K_4)\,,
\end{cases}
\end{equation}
with $K_1=K^{(1)}(\Phi,\bar \Phi);\, K_2= K^{(2)}(\Phi_1,\bar \Phi_1);\,K_3=K^{(3)}(\Phi, \bar \Phi_1);\,\quad K_4=K^{(4)}(\Phi,\Phi_1)$.
All these quantities are reduced to contour integrals obtained
 along  $\partial {\cal P}$, see  Fig.(\ref{3modester}(A) on top). We divide the  outer contour into horizontal lines  and  vertical lines  travelled in the negative sense. Because of the periodicity, the two vertical  contour integrals cancel  each other out, (blue lines of  Fig.\ref{3modester}) above.  At infinity  $\Phi_Z$ vanishes so only  the integral along  ${\cal C}_0$ contributes to the energy at this order. This result is valid  since  there is {\it  no singularity} inside the physical domain ${\cal P}$. Finally, we get $K_1$:
\begin{equation}
\label{K1}
K_1=-\frac{1}{2}  \int_{-1/2}^{1/2} d Y \bar \Phi(-I Y) \Phi_Z (I Y)\,,
\end{equation}
and  $K_2=K_1/J; \quad K_3=2 K_1/(J+1);\quad K_4=0$. The energy density at second order simplifies:
\begin{equation}
\label{Esecond}
{\cal E}_2 =-{\cal Q}(J) \frac{(1+J)(1-J)^2}{8 J^3} K_1\,.
\end{equation}
Near the Biot  threshold, ${\cal E}_2$ behaves as  ${\cal E}_2 \sim -E_f(J-J_B)$. Defining first
\begin{equation}
\label{Qb}
  Q_B=\frac{d{\cal Q}}{dJ}\vert_{J=J_B}=(3 J_B^2-6 J_B-1).  
\end{equation}
$E_f$  reads: 
\begin{equation}
\label{calQ}
E_f=K_1 {\cal Q}_2;\,\mbox{where} \quad 
 {\cal Q}_2 =Q_B\frac{(J_B-1)^2(J_B+1)}{8 J^3}.
 \end{equation}
At this order of perturbation, we have recovered the linear stability result. It is possible to go one step further and specify the nature of the bifurcation that occurs near  $J_B$. For this we consider ${\cal E}_3$.
%We begin by a quasi-singular profile correspond to a logarithmic function. 
A third order, it reads:
\begin{equation}
\label{Eonefor}
 \frac{{\cal E}_3}{p_e} =L_1 +J^2 \tau_1^2 L_2+ \frac{\tau_1(J+1)^2}{2} L_3-\frac{\tau_1(J-1)^2}{2} L_4\,,
    \end{equation} 
with $p_e=J \tau_0\tau_1$ and:
\begin{equation}
\label{intLL} 
\begin{cases}
 L_1=<\Re \,[\Phi_Z \Phi_{\bar Z} \Phi_{Z_1}]> \,,\\
 L_2=<\Re \,[\Phi_{Z_1}  \Phi_{\bar Z_1} \Phi_{Z_1}]>\,,\\
 L_3=<\Re \,[\Phi_{Z_1}]\Re \,[ \Phi_{\bar Z_1} \Phi_Z]>\,,\\
 L_4=<\Re \,[\Phi_{Z_1}]\Re \,[ \Phi_{Z_1} \Phi_Z]>\,.
\end{cases}
\end{equation}
These formulas allow to calculate the third order for any profile function $\Phi$. The calculation is not always easy but can be done  as
demonstrated hereafter for the logarithmic function defined in eq.(\ref{singular}).

\subsection{Nonlinear coupling of quasi-singular profiles}
The purpose of this paragraph is to estimate the amplitude of the profile and the nature of the bifurcation near $J_B$.
 Since each case is  special, we limit ourselves  to  one of them, namely   the logarithmic mode, see  eq.(\ref{singular}): $\Phi=-Log(1+a-e^{-2\pi Z})/Log(a)$, with $a>0$ and shown  in Fig.(\ref{EC3EC4})(e). In this figure, only $\Re \,[\Phi]$ is shown for $X=0$, and the true profile function must  be multiplied by $\epsilon \tau_0/(2 J)$. Obviously, the  desired profile is chosen with a positive value of  $\epsilon$  to have  the sharp-pointed shape in the positive direction.  Such a solution appears a priori at the Biot threshold and  remains  a regular solution, even with  stresses accumulated at the interface. The corresponding elastic energy starts at the second order and the elastic energy expansion is  written as:
\begin{equation}
\label{energie}
{\cal E}={\cal E}_2 \epsilon^2+ {\cal E}_3 \epsilon^3=-E_f\left( \delta J \epsilon^2+e_3\epsilon^3\right)\,,
\end{equation}
where $\delta J=J-J_B$ and $e_3=-\frac{{\cal E}_3}{E_f}$. $E_f$, $K_1$ and  ${\cal Q}_2$ have been defined previously, see Eqs.(\ref{K1},\ref{calQ}). Thus, minimizing the energy with respect to $\epsilon$ leads to:
\begin{equation}
\label{exp1}
    \epsilon=-\frac{2}{3} \frac{\delta J}{ e_3};\quad \mbox{so} \quad   {\cal E}=-\frac{4}{27} E_f \frac{\delta J^3}{e_3^2}\,.
\end{equation}
To observe a bifurcation with such an expansion in $\epsilon$ requires a negative value of ${\cal E}$, so  $\delta J$ must be positive for positive values of $E_f$ and $K_1$.
$K_1$ depends on the logarithmic dependence of the profile,and  can be estimated as:
\begin{equation}
    K_1 \sim -\pi  \frac{Log(2 a)}{Log(a)^2} \quad \mbox{for}\quad  0<a<<1\,.
    \end{equation}
 The evaluation of the third order ${\cal E}_3$ is given in section \ref{quasisingular} and the corresponding  result in eq.(\ref{energquasi}). So  when $a$ is a small  quantity, we get for $e_3$:
 \begin{equation}
 \label{exp2}
     e_3=\frac{ 2J \pi  \tau_1 \Pi_1}{(J-1)^2 p_a Q_B};\,\, \mbox{where}\,\,  \,p_a= a Log(a) Log(2a).
 \end{equation}
 The numerical value of $e_3$ is then $e_3\simeq -11.71/p_a$ which decreases when $a$ increases.
Since $e_3<0$, $\delta J$ is positive to obtain $\epsilon>0$, which is required for  the profile  shown in Fig.(\ref{EC3EC4}). In this way, a bifurcation and a crease can be effectively observed. A negative sign  will be counterintuitive  with cusps oriented upward.  Nevertheless, the cusp amplitude will remain tiny approximately  given by $\left (2/3 p_a \delta J\right)/11.71\sim 0.01\delta J$ for $a=0.01$. This treatment does not include surface tension because of  obvious technical difficulties, \cite{jeong1992free}.  
\subsection{Nonlinear coupling of harmonic modes }
\label{nonlinearcoupling}
An efficient treatment of mode  coupling near a threshold  is to   multiply the main harmonic mode, here $\zeta$, by  a slowly varying amplitude satisfying the so-called amplitude equation derived from the separation of scales. This method is  easily  illustrated by the Euler Elastica \cite{euler1732solutio}. An explicit solution of the Elastica bending solution can be found in  \cite{landau1967theorie} section $19$ exercise $3$. Depending on the  boundary conditions applied at both ends, the threshold value of the force $F_c$ responsible for the bending is found,  and the nonlinearities  give  the amplitude of the bending profile as a function of the force above the threshold value. In this simple case, the bifurcation is supercritical since the amplitude varies as  $\pm \sqrt{ F-F_c}$, above the threshold. For  this simple example, there is  also  a free energy that includes   the elastic energy  of bending and  the work of the forcing. Then,  another way to study the bifurcation is also provided by the analysis of the free energy above the  threshold, this is the choice made in Appendix A of \cite{audoly2010elasticity}  that we will follow here. In fact, the treatment by the separation of scales is more tedious in our case for at least two reasons: first, three unknown functions $x,y,Q$ are coupled and second it requires an initial guess for the scaling dependence of the coupled functions,which is  not easy to find a priori.
The  energy analysis turns out to be much more efficient. and  is chosen  here.  We start with  the coupling of  $2$ harmonics and then $3$ harmonics. For the  linear order we have shown that all harmonic modes appear at  the same threshold value: $J_B$

\subsection{Intermediate algebra for the coupling of sinusoidal modes}

Consider the superposition of several modes where  $\Phi(Z)=\sum_k^{k_0} a_k \zeta^k$ with $k<k_0$, $k$ and $k_0$ being  positive integers \cite{budday2015period}.  Then  $K_1=\pi \sum_k k\vert a_k\vert^2 $ so  that $K_1$ is always positive and   ${\cal E}_2$ is negative above the Biot threshold. 
Unfortunately, at the third order, the calculus becomes much more tedious, even when sinusoidal modes  are imposed. Each integral involves  a triple series of the mode amplitudes $a_n$. 
\begin{equation}
\label{Limode}
\begin{cases}
\tilde L_1& = \sum_{p,q,r}\frac{ p q r a_p a_q a_r }{p+q+J r} (\delta_{p-q-r}+\delta_{p-q+r})\\
&=2  \sum_{0<p<q}   \frac{p q (q-p) a_p a_q a_{q-p}}{p(1-J)+q(1+J)} \,,\\
\\
\tilde L_2&= \sum_{p,q,r}\frac{ p q r a_p a_q a_r }{J (p+q+r)} (\delta_{p-q-r}+\delta_{p-q+r})\\
&= \frac{1}{J}  \sum_{0<p<q}  p  (q-p) a_p a_q a_{q-p}\,, \\
\\
\tilde L_3& =  \sum_{p,q,r} \frac{p q r a_p a_q a_r}{J(p+q)+r} (\delta_{p-q+r}+\delta_{p+q-r})\\
&=  \sum_{0<p\le q} (2-\delta_{p-q})  \frac{p q a_p a_q a_{q+p} }{J+1} +\tilde L_4\,, \\
\\
\tilde L_4 &=  \sum_{p,q,r} \frac{a_p a_q a_r p q r}{J(p+q)+r)} \delta_{p-q-r}\\
&= \sum_{0<p<q} a_p a_q\frac{ a_{q-p} p q (q-p) }{(J-1)p+(J+1) q}\,,
\end{cases}
\end{equation}
with $\tilde L_i=-2 \pi^2 L_i$. It is to be noted that  a non-vanishing third order in the energy exits if and only if modes are coupled. 
%\begin{itemize}
\subsubsection{Coupling two modes near the $J_B$ threshold }
\label{twomodes}
 In the case of two modes, $K_1=\pi(1+k \vert a_k\vert^2)$.
For the  third order in $\epsilon^3$, the only non-vanishing values  contributing to ${\cal  E}_3$, eq.(\ref{Eonefor}), are obtained for the exponents  $k=1$ and $k=2$. Thus, the two mode profile is limited to  $\zeta +a_2 \zeta^2$ where  $\vert a_2\vert$ is assumed  to be of order $1$, greater  than $\epsilon$ and $K_1=\pi(1+ 2 \vert a_2\vert^2)$.  Another scaling can be found below, in section \ref{supersub}. We have already found the second order of the energy ${\cal E}_2$, see eq.(\ref{Esecond}, \ref{calQ}).  Assuming $a_2$ is real, the results  for  the associated $L_i$, eq.(\ref{Limode}), are:
\begin{equation}
L_1=4 a_2/(3+J);\,\, L_2= a_2/J;\,\, L_4=2a_2/(3J+1)\,,
\end{equation}
and $L_3=L_4 +a_2/(1+J)$
which gives   ${\cal E}_3$ 
\begin{equation}
\begin{cases}
{\cal E}_3& =- \pi^2 a_2 {\cal Q}_3;
\\
{\cal Q}_3&= \frac{  (J-1)^4 (J+1) \left(J^2+1\right) \left(11 J^2+16 J+3\right)}{4 J^4 (J+3) (3 J+1)}\,,
\end{cases}
\end{equation}
and the generic results found in eq.(\ref{exp1}) and eq.(\ref{exp2})  apply and give:
\begin{equation}
\epsilon=-\frac{2}{3 }\frac{ {\cal Q}_2}{ {\cal Q}_3 }\delta J \frac{K_1}{\pi^2 a_2};\,\, 
{\cal E} =-\frac{4}{27} \frac{K_1^3 }{\pi^4 a_2^2} \frac{{\cal Q}_2^3}{{\cal Q}_3^2} \delta J^3\,.
\end{equation}
  We then deduce that the  two-mode profile is a minimizer of the elastic energy  above the  Biot threshold, $\delta J>0$. Such solution exists for every  finite value of $a_2$. The bifurcation occurs for $\epsilon a_2<0$ and is transcritical \cite{crawford1991introduction,meron2015nonlinear}. 
\subsubsection{Nonlinear three mode coupling in the vicinity of the $J_B$ threshold}
\label{threemodes}
We now consider the following shape deformation given by the three-mode coupling: $ \Phi(Z)=\zeta+a_2 \zeta^2+a_3 \zeta^3$. For simplicity, we choose real values for all the  coefficients $a_i$ and $K_1=\pi(1+2 a_2^2+3 a_3^2)$. Similarly, the expansion of the elastic energy up to the third order  reads: 
\begin{equation}
\label{3modes}
{\cal E} =- {\cal Q}_2 \delta J K_1\epsilon^2- \pi^2 a_2 {\cal Q}_3 (1+a_3  {\cal Q}_{33}) \epsilon^3 \,,
\end{equation}
where 
\begin{equation}
{\cal Q}_{33}=\frac{4 (3 + J) (1 + 3 J) {\cal \tilde Q}_{33}}{(2 + 
   J) (5 + J) (1 + 2 J) (1 + 5 J) (3 + 16 J + 11 J^2)}\,,
\nonumber  
\end{equation}
with ${\cal \tilde Q}_{33}=10 + 97 J + 254 J^2 + 196 J^3 + 37 J^4$.   The numerical value of   ${\cal Q}_{33}=3.8845$ for $J=J_B$. 
    The introduction of $ \zeta^3$  does not modify   the  result of section \ref{twomodes} unless   $a_3{\cal Q}_{33}<-1$. 
    The function $e_3$ which enters the eq.(\ref{exp1}) and eq.(\ref{exp2}) becomes 
 $e_3=\pi^2 a_2 {\cal Q}_3 (1+a_3  {\cal Q}_{33})$ and is shwon  numerically in density plots in Fig.(\ref{EC3EC4})(a). Again, the minimum non-trivial value of ${\cal E}$ is found for $\delta J>0$ with no possibility of obtaining a stable solution below the Biot threshold. Due to the complexity of the formula, we give here only the numerical value of the selected amplitude of the profile and the corresponding energy:
\begin{equation}
    \epsilon=-\beta_1 \frac{K_1 \delta J /\pi }{ a_2 (1 + \beta_2 a_3)};\,\,  {\cal E}=-\beta_3 \frac{(K_1 \delta J/\pi) ^3}{a_2^2 (1 + \beta_2 a_3)^2}\,,
\end{equation}
with $\beta_1= 0.02575$, $\beta_2=3.88446$, $\beta_3=0.0007271$.
\begin{figure*}	
 \centering 
    \begin{minipage}[b]{1\textwidth} 
    \subfigure[\,\,\,\bf{Domain of integration for the energy density}  ]{\includegraphics[width=.7\textwidth]{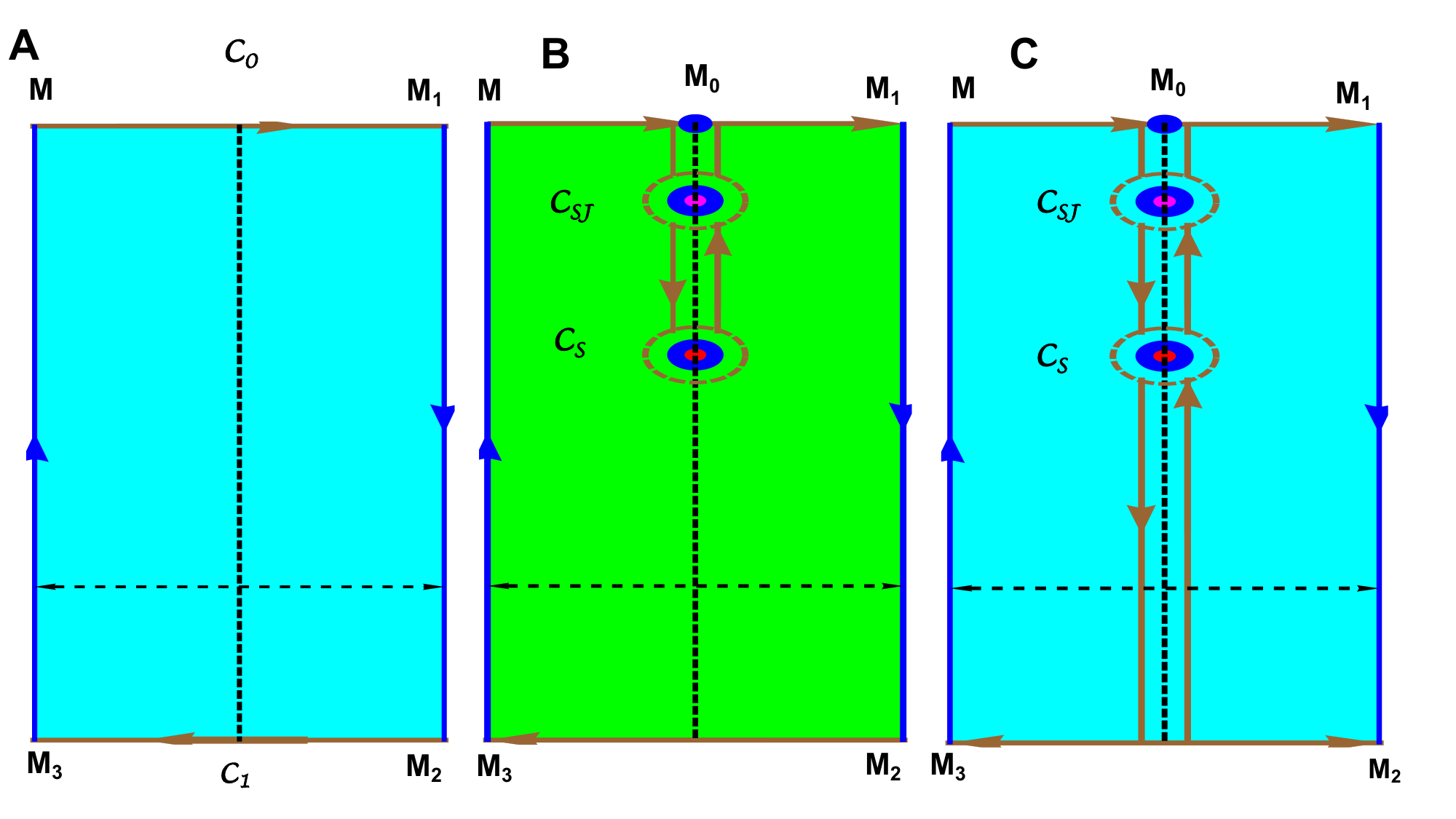}}   
        
           \hspace{0in} 
    \subfigure[\,\,\,\bf{Bifurcation diagram} ]{ 
         \includegraphics[width=.25\textwidth]{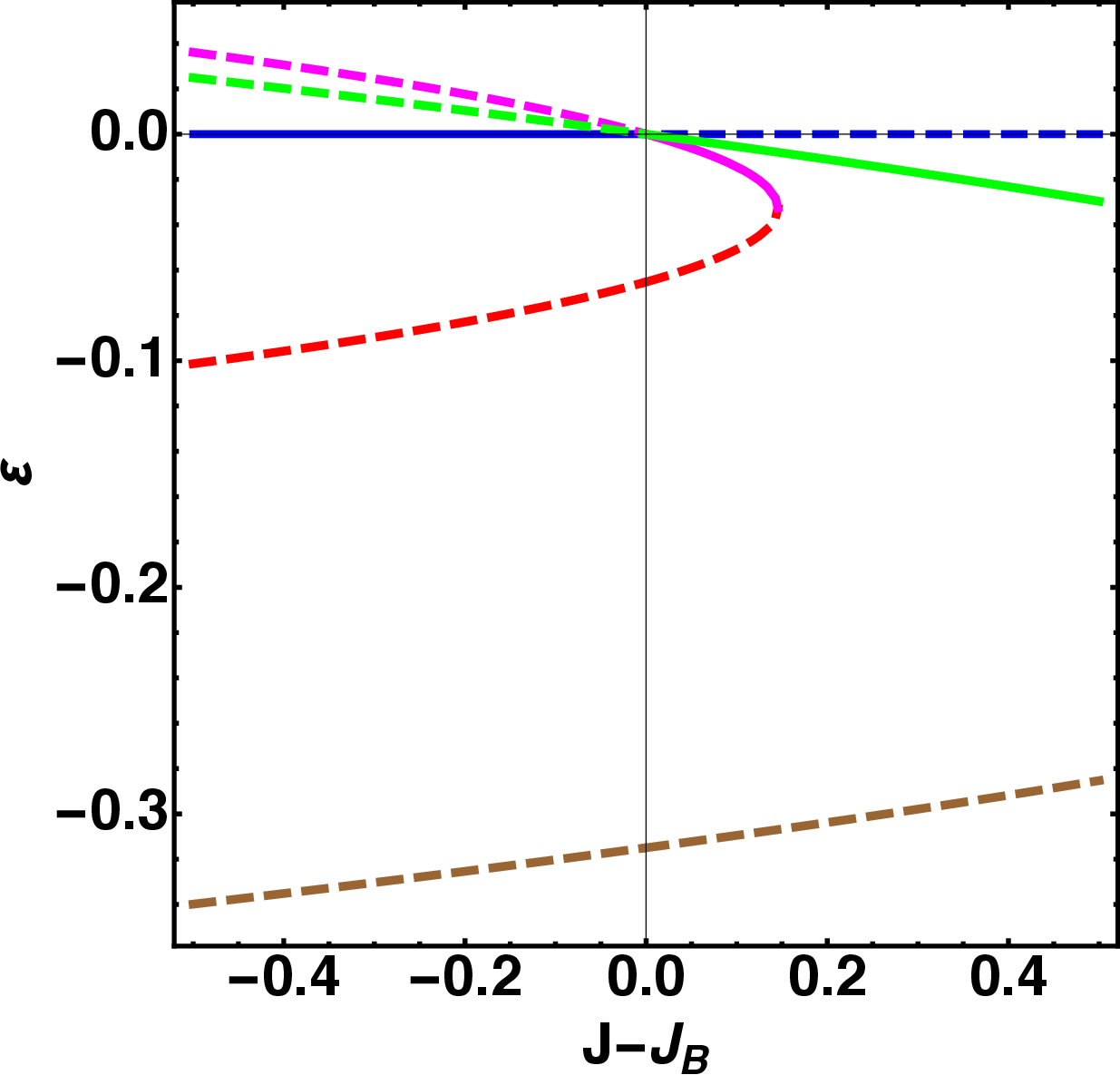} 
          } 
           \hspace{0in} 
       \subfigure[\,\,\,\bf{Bifurcation diagram}]{ 
         \includegraphics[width=.25\textwidth]{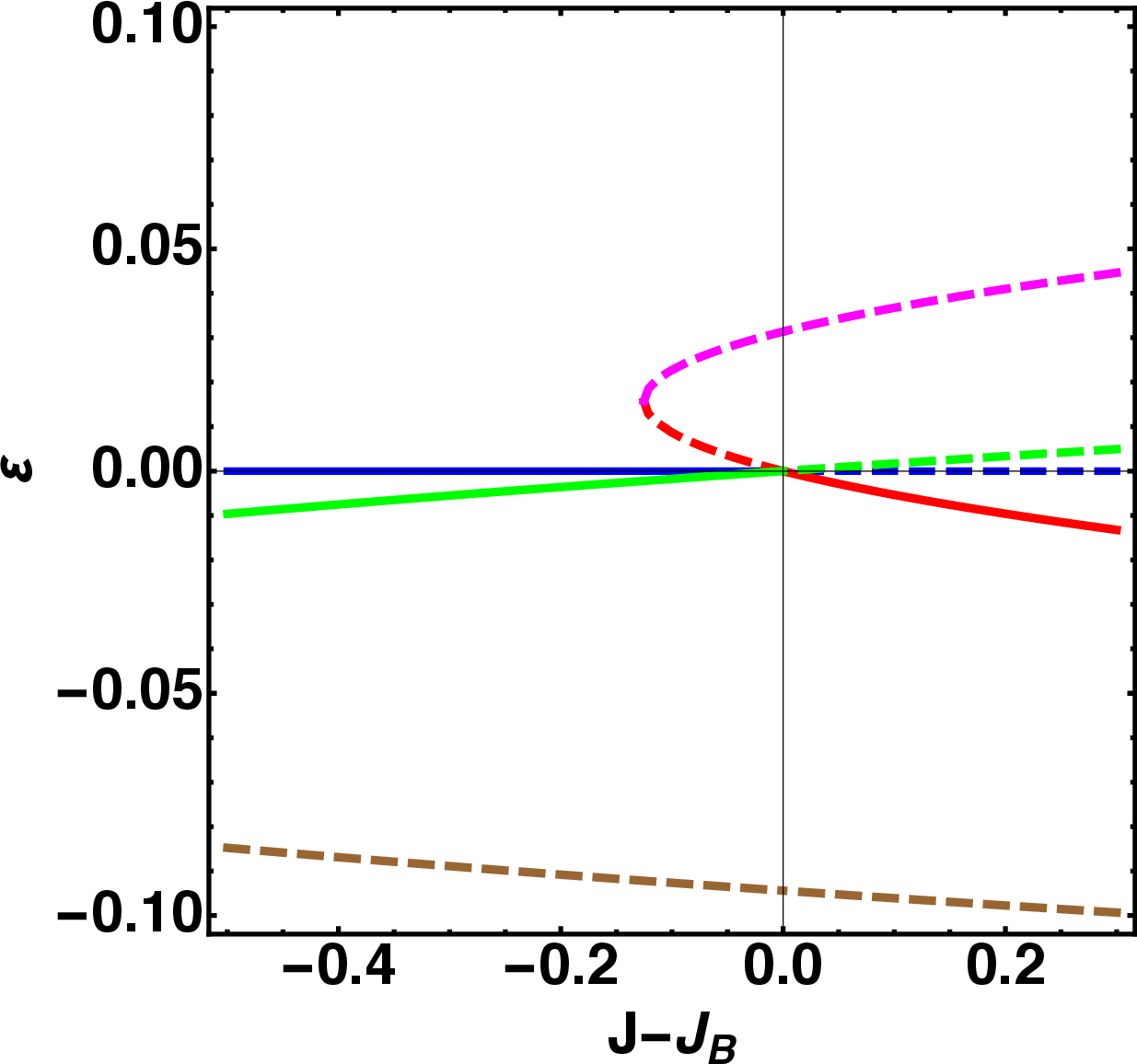} 
          } 
          \hspace{0in} 
        \subfigure[\,\,\, \bf{Upper interface profiles}]{
            \includegraphics[width=0.35\textwidth]{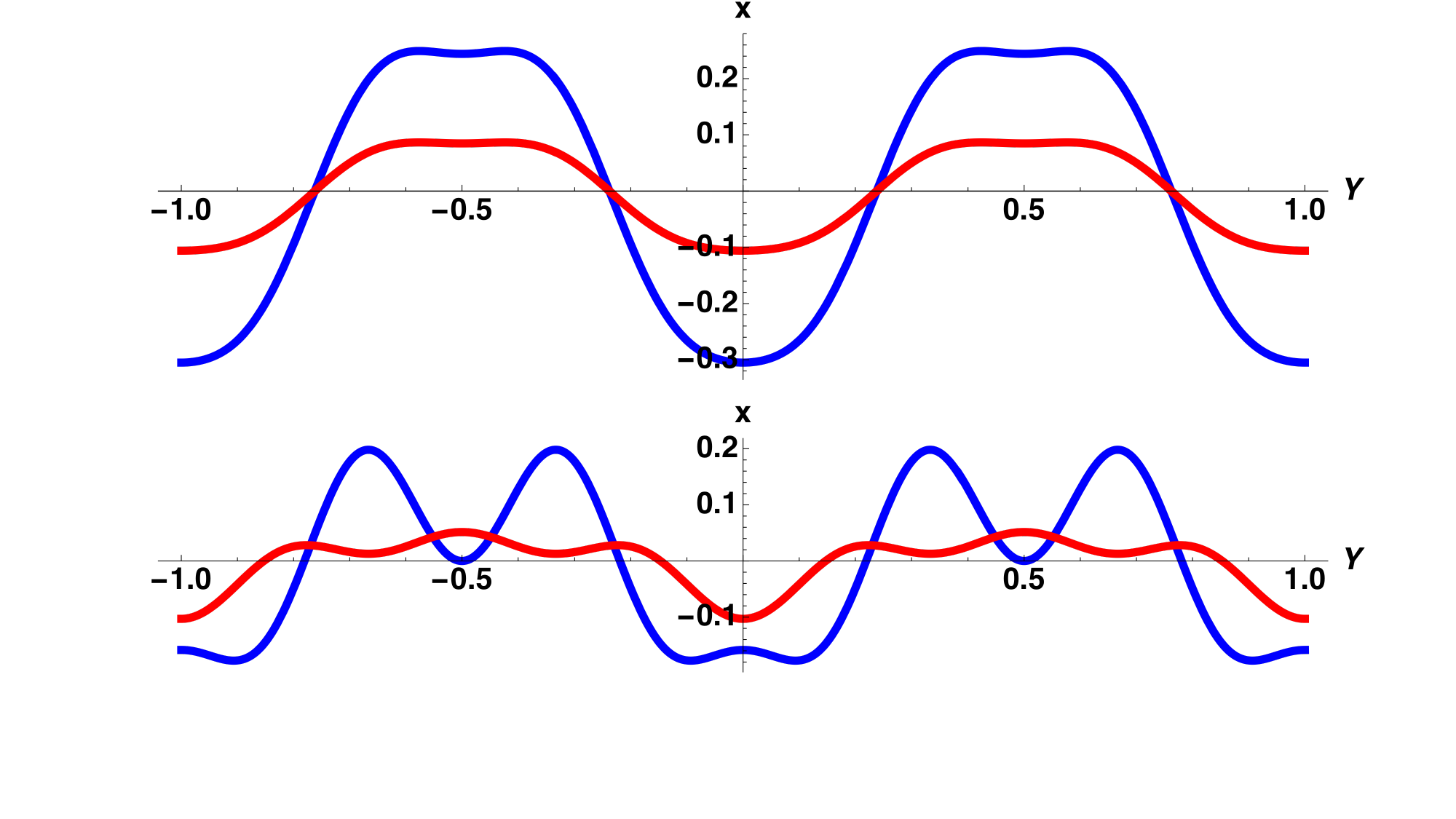} 
        } 
        \hspace{6in}
       \subfigure[\,\,\,\bf{Bifurcation diagram}]{ 
         \includegraphics[width=.4\textwidth]{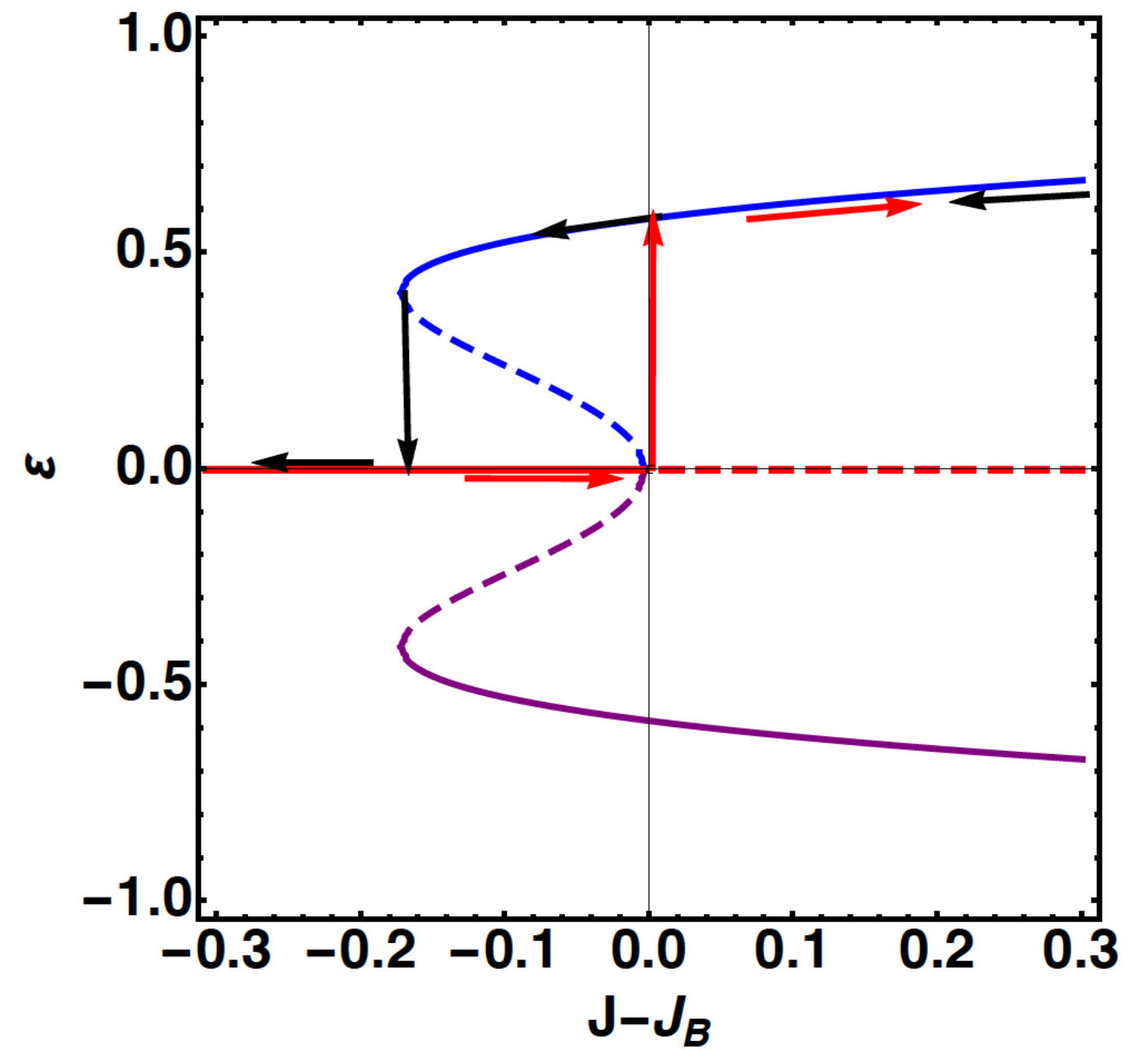} 
          } 
        \hspace{0in} 
        \subfigure[\,\,\, \bf{Elastic energy for} $\bm{J=J_B\pm 0.1}$]{
            \includegraphics[width=0.4\textwidth]{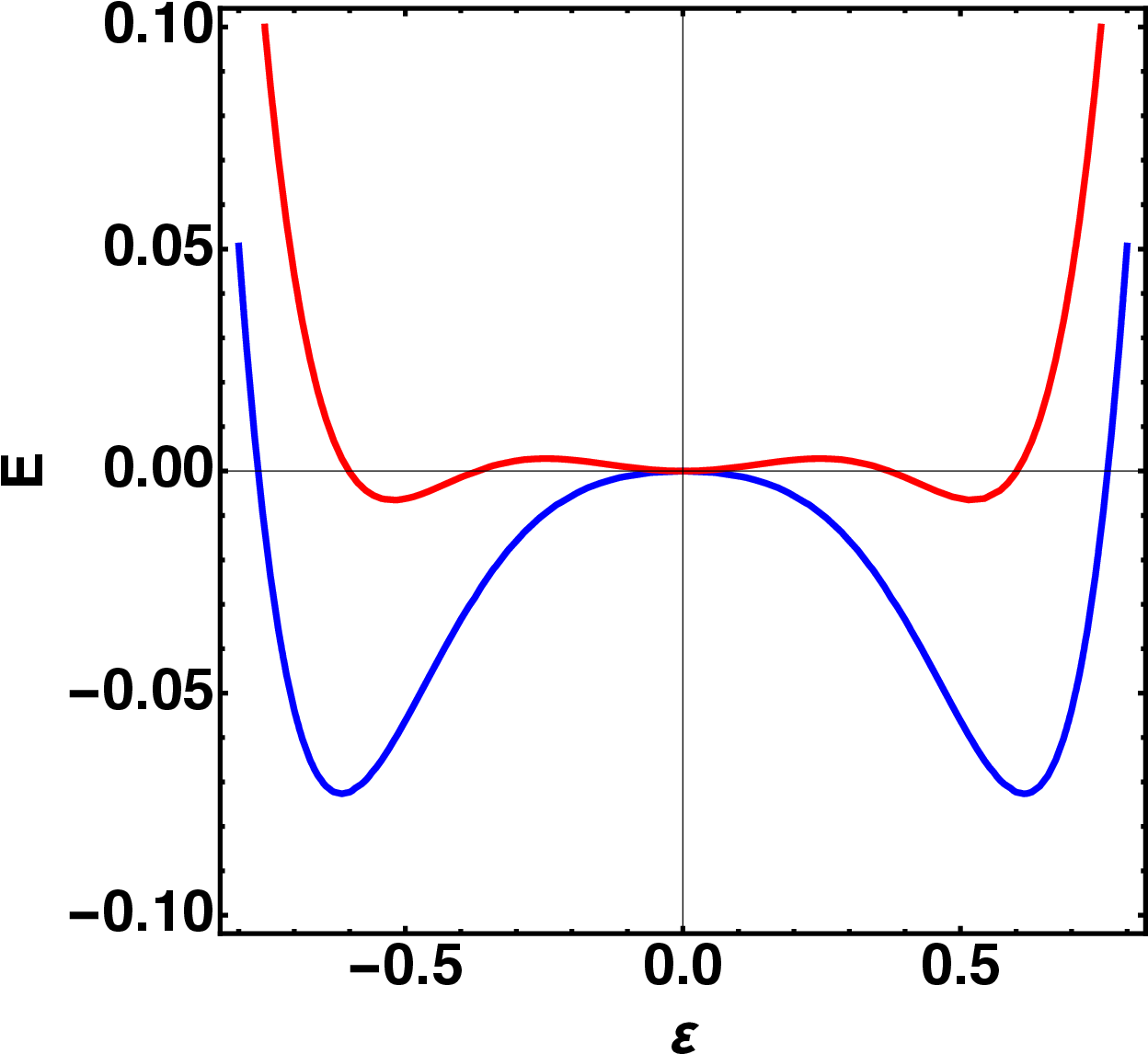} 
        }  
        \end{minipage}
    \caption{Domain ${\cal P}$ for integration of the energy density, restricted to  one wavelength $-1/2 \leq Y \leq 1/2$, $X \in [0, +\infty]$.  Only contours of interest are labeled, such as ${\cal C}_i$. Panel (A) is similar to panel (A) of Fig.(\ref{riemann}), only the contour ${\cal C}_0$ contributes to the elastic energy, as shown in section \ref{Ordertwo}. In (B) and (C) which is also panel (B) of Fig.(\ref{riemann}), the relevant contours are around each singular point $ {\cal S} $ and  ${\cal S}_J $.\\ In panels (b) and (c), the bifurcation diagram, $\epsilon$ versus $J-J_b$, for a triple mode coupling with surface tension: $\gamma_0=0.1$. In (b), $a_2=0.1$ and $a_3=-0.1$ (curves in red and magenta) and $a_2=0.1$ and $a_3=0.1$ (curves in brown  and green). Dashed curves indicate unstable solutions. In (c) $a_2=0.5$ and $a_3\pm 0.5$ with the same color code. For $a_3= 0.5$, the stable solution appears below the Biot threshold $J_B$ in contrast to the other cases. In (d) interface profiles (multiplied by 10 for both axes; $J-J_B=0.12$  for the upper case corresponding to the data of panel (b), $J-J_B=-0.2$ for the lower case. In blue $a_3=-a_2$, in red $a_3=a_2$ with the  exact values of $\epsilon$ chosen in each case. \\
    In (e) a subcritical bifurcation diagram for the amplitude $\epsilon$.  Continuous lines indicate locally stable solutions, dashed lines  unstable solutions, not observed experimentally. Red arrows indicate the trajectory for increasing values of $J$ while black arrows indicate  the trajectory for  decreasing values. Note the complete symmetry  between positive and negative values of $\epsilon$. The hysteresis cycle extends between the two  vertical arrows indicating the jump of the  $\epsilon$ amplitude at $J_C$ (see eq.(\ref{souscrit})) and at $J=J_B$. In  (f) the elastic energy density for $J=J_B+ 0.1$ in blue and $J=J_B -0.1$ in red as a function of  $\epsilon$ with $3$ stable solutions  below $J_B$,($3$ red minima)  and only $2$ (in blue) above.}
      \label{3modester}
    \end{figure*}
As shown in section {\ref{develop}}, the  surface tension creates  an  additive contribution in $\epsilon^4$ which can change the present result, giving two solutions for $\epsilon$ instead of one. An exhaustive study is really difficult due to the large number of degrees of freedom such as $a_2$, $a_3$ and $\gamma_0$, the dimensionless capillary number. However, this  number is rather weak and for the numerical study we choose $\gamma_0=0.1$. The amplitude $\epsilon$ which minimizes the elastic energy is a solution of a quadratic equation, so there are two solutions in addition to $\epsilon=0$. A first numerical investigation, for  coefficients $a_2=0.1$ and $0.5$ and $a_3=\pm a_2$ is shown in Fig.(\ref{3modester})(b,c) and demonstrates  nonlinear modes occurring  after or before the Biot threshold. Only stable solutions are considered, so only the continuous lines of  Fig.(\ref{3modester})(b,c). From these two examples, one can notice   that the $\epsilon$ values are rather weak, negative and less than $0.1$ in absolute value. The interface profiles are shown in Fig.(\ref{3modester})(d). At the  top we find  strongly distorted  profiles for a value of $J=3.50$ and $\epsilon=-0.019$\, ($a_3=-a_2$, $a_2=0.1$) and $\epsilon=-0.0065$ for $a_3=0.1$. Below, for $a_2=0.5=-a_3$ $\epsilon=-0.0096$ and $J=3.58$ and for $a_3=0.5$, $\epsilon=-0.003560$ and $J=3.18$. Fig.(\ref{3modester})(d) respects the scale but the height is magnified by a factor of $10$. In conclusion, this nonlinear treatment shows that nonlinear modes can occur before the Biot threshold, but for strong capillary numbers. As the amplitude of the coefficient $a_2$ increases, the mode becomes more and more distorted from the single sinusoidal solution but the amplitude of the interface remains small due to the small amplitude of  $\epsilon$.  To better understand the bifurcation plots, we choose a more appropriate  representation for the profile functions in the next section.

% voir code matematica triplemode et nonlinear24sept
\subsection{Super and subcritical bifurcations}
    \label{supersub}
   In the previous paragraph, we assumed that all the coupled harmonics are of  the same order of magnitude.  Now, we construct   profile functions where the harmonics slightly perturb the principal mode $\zeta$ such as : $ \Phi(Z)= \zeta -\epsilon A_2 \zeta^2(1+\alpha_3 \zeta)$, where  $A_2$ and $\alpha_3$ are constants  of order $1$. 
    In this case we get: 
\begin{equation}
\label{bifone}
{\cal E} =-  {\cal Q}_2 \delta J \pi \epsilon^2+ \pi^2 A_2 {\cal Q}_3  \epsilon^4 \,.
\end{equation}
For positive values of $A_2$, we recover the  classical supercritical bifurcation with $\epsilon \sim  \pm\sqrt{\delta J}$ above the Biot threshold. For the opposite sign of $A_2$ and  $\alpha_3=-\epsilon^2 B_3$ (which implies a very weak perturbation of the main mode $\zeta$), the selected  profile becomes $ \Phi(Z)= \zeta + \epsilon B_2 \zeta^2- \epsilon^2 B_3 \zeta^3 $, with $B_2$ and $B_3$ positive and in this case:
\begin{equation}
\label{biftwo}
{\cal E} =-  {\cal Q}_2 \delta J \pi \epsilon^2- \pi^2 B_2 {\cal Q}_3\epsilon^4   +\pi^2 B_3 B_2 {\cal Q}_3{\cal Q}_{33} \epsilon^6\,.
\end{equation}
The extrema of ${\cal E}$ are obtained for $\epsilon$ values given by:
\begin{equation}
\label{souscrit}
\epsilon(J)=\pm \frac{1}{\sqrt{3 B_3 {\cal Q}_{33}}} \left (1\pm \sqrt{1+\frac{3}{\pi}\frac{{\cal Q}_2 }{Q_3}\frac{B_3}{B_2}{\cal Q}_{33} \delta J}\right)^{1/2}\,.
\end{equation}
and \begin{equation}
    J_C=J_B-\frac{\pi}{3}\frac{B_2}{B_3}\frac{{\cal Q}_{3}}{{\cal Q}_{33}{\cal Q}_{2}}\,.
\end{equation}
Fig (\ref{3modester}) (e) and (f)  show the evolution of the profile amplitude $\epsilon$ as the volumetric growth coefficient $J$ increases and decreases in the vicinity of $J_B$. As $J$ increases, and  remains below $J_B$, the chosen value of $\epsilon$ remains zero (red solid curve), corresponding to the purely axial growth, but such a solution loses its stability at $J_B$, (red dashed curve). Then the value of $\epsilon$ makes a positive (or negative) jump
$\epsilon_G=\sqrt{2}/\sqrt{3 B_3 {\cal Q}_{33}}$ (resp. $-\epsilon_G$). Then  $\epsilon$ rises slightly above $J_B$ following the blue trajectory in the direction of the red arrow. If there is a decrease of $J$ from $J>J_B$, $\epsilon$ decreases along the blue line which is stable until $J=J_C$, where the blue trajectory loses its stability (blue dashed curve) and the flat pattern: $\epsilon=0$ is restored. At the transition there is also a jump in $\epsilon_D=1/\sqrt{3 B_3 {\cal Q}_{33}}$.
Note that  $\epsilon$ can be either positive or negative, only $\epsilon>0$ is shown for clarity but both signs are equivalent (see Fig.\ref{3modester})(f) which gives  the energy minima for two values of $(J-J_B)$. Only, E.Hohlfeld and Mahadevan  seem to have discovered this subcritical bifurcation by numerical means (finite elements, ABAQUS), while experimentally the hysteresis associated with  such a configuration was revealed  by J.Yoon, J. Kim  and R.C. Ryan Hayward in \cite{yoon2010nucleation}. This scheme nicely represents the hysteresis observed in experiments. Before closing the parenthesis on  nonlinear wrinkling patterns, studied with the classical techniques of bifurcation theory, let us outline a recent analysis performed with group theory methods concerning first the case of a compressed inextensible beam resting on a nonlinear foundation \cite{pandurangi2020stable},  second, the case of a thick compressed nonlinear sample \cite{pandurangi2022nucleation} with different  types of elasticity energy than the simple one considered here. Focusing on the first case, the very interesting point is that the authors succeed to capture localized patterns and one can wonder if it will not be possible to establish a nonlinear solitonic solution for the spatial modes detected here.
\subsection{Role of surface tension}
Surface tension is a weak parameter in processes controlled by elasticity. A typical order of magnitude is given by the dimensionless number $\gamma_0=\gamma/\mu H$, where $\gamma$ is in  Newtons per meter, the shear modulus $\mu$ is in Pascals and $H$ is a typical length, so in our case it will be the wavelength. An exact  value will depend  on the nature of the  elastic sample and possibly of  the fluid above the sample. Measurements based  on elasto-capillary waves \cite{mora2010capillarity,chakrabarti2013direct,shao2018extracting} made with extremely soft materials ($\mu\sim 100 Pa$)  give a value of  about $0.05 N/m$. Recently, the role of surface tension on creases has been considered and obviously,  surface tension plays an enormous role in the vicinity of quasi-singular profiles, as naively explained by the Laplace law of capillarity \cite{liu2019elastocapillary,masurel2019elastocapillary}. For small deformations, well represented by a few harmonics and  for ordinary elastic materials with a shear modulus around $10^4$Pa, the  surface tension may  be relevant only in the vicinity of  the bifurcation examined in the previous section  \ref{supersub}. We will first consider the case where the coupling with the first harmonic $\zeta$ is weak as in section \ref{supersub}. The expansion of the energy density, eq.(\ref{biftwo}),  must now include the capillary terms  order by order:
\begin{equation} 
\begin{cases}
{\cal E}_c ={\cal E}_{cs}+ \frac{\gamma_0}{2}   \left(\frac{\pi (J^2-1) \epsilon}{2 J}\right)^2 \\
\times \big( 4 B_2^2 +B_2\frac{(J-1)^2\pi}{J})\epsilon^2
 +e_{2c} \epsilon^4 \big)\,.
\end{cases}
\end{equation}
$E_{cs}$ is the capillary energy associated to the main mode $\zeta$. It is given by  eq.(\ref{coefec}) and eq.(\ref{onemode}) while  and $e_{2c}$ is given by eq.(\ref{B2cap}), in section \ref{evaluation}. $E_{cs}$ is the capillary energy associated with the main mode $\zeta$. Regarding the sign, the  fourth and sixth order terms can be positive or negative so they can change the nature of the bifurcation which can go from subcritical to supercritical if $\gamma_0$ is  strong enough.  One can now examine in more detail the case where the coupling of the $3$ modes has an equivalent weight, according  to the section \ref{threemodes}. In this parameter range, the surface tension becomes a small parameter for the standard range of values of  $\gamma_0$ and the capillarity plays a critical role at the fourth order. We rescale the free energy and rewrite the equation (\ref{exp1}) as follows:
\begin{equation}
\label{normalformbis}
{\cal E}_{t}= - E_f \epsilon^2 \left( \delta J +2\gamma_0 g_2)+(e_3+2\gamma_0 g_3)\epsilon+2 \gamma_0 g_4 \epsilon^2\right),
\nonumber
\end{equation}
where $E_f$ and $e_3$ have been defined in eq.(\ref{exp1}). Here we give only $g_2$
\begin{equation}
\label{coefcap}
 g_2=-\frac{\pi J (J+1) \left(1+4 a_2^2+9 a_3^2\right)}{(3 J^2-6 J-1) \left(1+2 a_2^2+3 a_3^2\right)}\,.
\end{equation} 
 Each coefficient  of the capillary energy is a function  of $J$, $a_2$ and $a_3$ and is listed in section \ref{evaluation}. The order of magnitude of these coefficients as   $a_2$ and $a_3$ vary can be found in Fig.(\ref{EC3EC4}) in section \ref{evaluation}. In fact, for normal values of the shear modulus, there is little chance that the capillary will alter the results given by eq.(\ref{exp1}). Since $g_2$ is negative, the bifurcation  threshold  is shifted to higher values by capillarity. This shift depends on the representation of the profile. 

 Post-buckling  creases  were studied extensively a decade ago \cite{li2012mechanics,cao2012wrinkles,weiss2013creases,jin2015bifurcation}. These studies suggest that creases can appear before the Biot threshold due to a subcritical bifurcation, as shown here in section \ref{supersub}. Note that the numerically  detected creases in these studies require the introduction of periodic defects. Cao and  Hutchinson \cite{cao2012wrinkles} demonstrate the remarkable sensitivity of  wrinkling  results to  physical imperfections and defects. This is not  surprising, since  it is a general property of the bifurcation theory \cite{cross1993pattern}. \\
 
The case of the self-contacting interface is much more difficult to handle, since analyticity is not preserved on a line (or on a segment) in the plane, so the elasticity equations  are no longer valid. If we approximate the two  Heaviside distributions that mimic the self-contacting interfaces by an analytic function such as $\Phi =-b^2 \sqrt{ Z^2+a^2}$, where $a$ is a tiny quantity, there  is no reason to assume  real contact between the two surfaces, which will remain separated by $2 a$.  Thus,  self-contacting interfaces are intentionally created  like fractures. They can be nucleated by defects and then, they will have a better  chance to be observed in thin samples, i.e.in $2$ dimensions compared to  $3$ dimensions (see Dervaux's thesis \cite{dervaux2011shape}). Nevertheless, such triggered singularities  remain a very attractive area of study, as shown by the  experiment of a deflated cavity localized at a finite distance from the upper boundary \cite{karpitschka2017cusp}. Before the generation of the self-contact, a quasi-singular profile is obtained with  the scaling $x\sim \vert Y\vert^{2/3}$, which is similar to our last profile function $\Phi$  of eq.(\ref{singular}), on the right but  with a different exponent. The curvature at the singularity varies as $\vert Y \vert ^{-1/3}$  like $x^{-1/2}$. This experiment is strongly reminiscent of  the equivalent one realized in viscous flow by Jeong and Moffatt \cite{jeong1992free} with contra-rotating motors. Although the interface behavior recovers the same exponent at some distance from the singularity,  the curvature remains finite and parabolic at the singularity, the only unknown being the radius of curvature at the tip,  which is chosen by the surface tension.
 \\
In conclusion, the observation of a bifurcation occurring before the Biot threshold is possible if at least  $3$ harmonic  modes  are initially coupled in the nonlinear regime. For the quasi-singular profile, the answer depends too much on the mathematical description of the profile. However, here we have presented a way to fully analyze the nature of the bifurcation in the neighborhood of the Biot threshold in order to obtain valuable predictions.

\section{How to escape the Biot threshold?}
\label{escape}
In the previous sections, the existence  of creases occurring at or just after the Biot threshold was examined. There is no difficulty in generating such creases as shown above, using the tools of complex analysis. However, it has been suggested by heuristic arguments \cite{wong2010surface,weiss2013creases,li2012mechanics} that singularities inside elastic samples can induce bifurcation below  the threshold: $J_B$.
Singularities induced by stresses are not forbidden in plane strain or plane stress elasticity, provided that the local elastic energy remains finite even if  the energy density does not. In practice in $2D$, this means that the strains are not more singular than  $R^{1/2}$ and the elastic deformation gradient or the stresses are not more singular than $R^{-1/2}$, where  $R$ represents  the distance to the singular point $R\rightarrow 0$. In linear plain strain elasticity, this is the case  for the fracture tips and also for edge dislocations. The main difference between linear and this work is the fact that linear elasticity does not consider the nucleation of such  defects that exist prior to loading and focuses more on the opening and/or the  displacement of the fracture. There are very few theoretical or experimental investigations about the fracture nucleation \cite{bonn1998delayed,guarino1999failure,ciliberto2001effect,tanne2018crack,kumar2020revisiting,pomeau2002fundamental}. The hope  here is indeed to generate these peculiar structures  by volumetric growth or by compression. The main question we have to solve is the following: is it possible to lower the bifurcation threshold
by considering singularities inside the sample?\label{nobiot}

As already mentioned in eq.(\ref{charact}), the solvability condition to observe periodic solutions implies either $J=J_B$ or $\Re \,[\Phi_Z]=0$, for $X=0$ so $\Phi$ is an even  function of $Z$. Here we avoid here a singularity at the interface $X=0$,  which requires a modification of  the elastic model  with a surface energy \cite{hohlfeld2011unfolding}. A nonlinear  singular solution emerges, but  it does not satisfy the simultaneous  cancellation of normal and shear stress at $X=0$ \cite{hohlfeld2012scale}. So we focus on singularities inside the sample.  An even function of $Z$, that can be represented by  $\Phi(Z)= F(Z)+F(-Z)$, automatically exhibits singularities inside the sample if  convergence  at positive infinity  is required: a holomorphic function, other than a constant, cannot converge at both infinities without singularities. The choice is then  a periodic allowed singular function, eliminating  poles and avoiding  as much as possible extended  branch cuts, always a source of stress. 
\subsection{Singular Profiles below the Biot threshold}
\begin{figure*}	
 \centering 
    \begin{minipage}[b]{1\textwidth} 
    \subfigure[\,\bf{Interface profiles}]{ 
         \includegraphics[width=.41\textwidth]{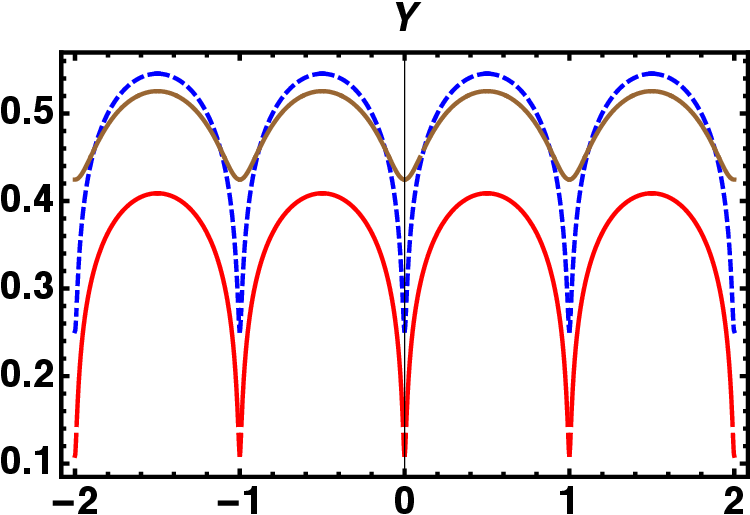} 
          } 
        \hspace{0.3in} 
        \subfigure[\, \bf{Singularities of} $\bm{S_{11}} $]{
            \includegraphics[width=0.43\textwidth]{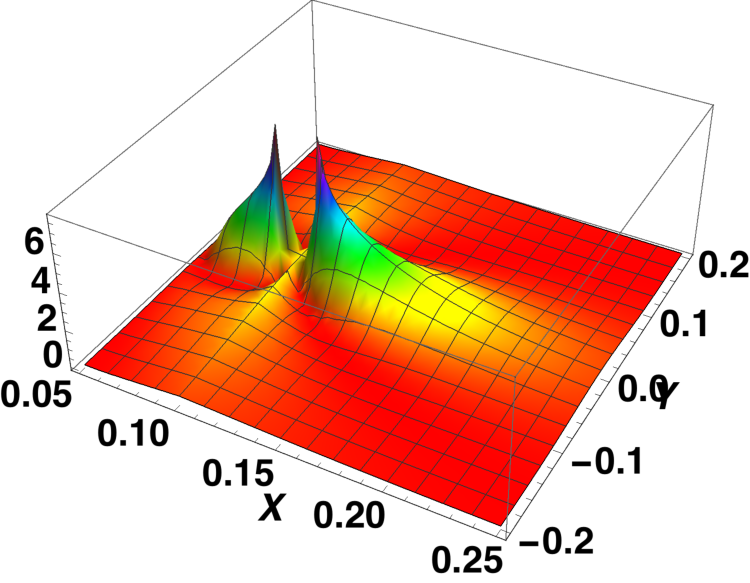} 
        } 
            \end{minipage}
    \caption{On the left, profiles corresponding to the deformation given by eq.(\ref{profsin}). The parameters are $J=3,\epsilon=0.1$. For the blue profile $X_0=0.01$ and  $l_0=0.001$, for the red curve, the same $X_0$ value   and $l_0=0.005$, for the brown profile, $l_0=0.001$ and $X_0=0.1$. Note that $l_0$ has little effect on the shape profile, which depends strongly on the distance of $X_0$ from $X=0$. The difference in height of the averaged surface has no physical meaning. On the right, the stress $ S_{11}$ in logarithmic scale ($Log(1+S_{11}^2$), multiplied by a scaling factor of $0.01$ for  visualization purposes, with $l_0=0.01$ and $X_0=0.1$. Note the two singular zones around $X_0$ and $X_0/J$. These singularities will merge in the boundary layer. The stress  decreases rapidly so that the map is confined  to a limited area of the sample between $0.0<X<0.25$ and $-0.2<Y<0.2$.}
      \label{inner}
    \end{figure*}

In finite elasticity, such as the Neo-Hookean elasticity,  allowed singularities in plane strain must induce a  local finite elastic energy, which in practice means  that $\vert F_Z\vert$ cannot be more singular than $\vert Z-X_0\vert^{-1/2} $, where  the positive constant $X_0$ denotes the central position of the singularity. Larger exponents are allowed since they do not contribute locally to the total elastic energy. Existing branch-cut singularities  must remain limited  in size. However, singular solutions can locally invalidate the hypotheses of the initial elastic model and may require more complexity  in the elastc representation and even in the growth process. As an example, 
   in fast fracture dynamics \cite{bouchbinder2008weakly,bouchbinder20091,li2023crack},  $3$ possible zones for the crack tip  have been identified: first the viscoelastic, then the nonlinear elastic,  and  finally the traditional  linear-elastic zone, which produces the square root singularity \cite{long2020fracture}.  Of course, such a description requires different physical models at different length scales.  Whatever the complexity introduced into the modeling, such as  multiple invariants of finite elasticity, compressibility, plasticity or strain hardening, and eventually  growth variations, it must remain  localized in  a small domain and must be treated as an inner  boundary layer.  Let us fix  this domain around $X_0$ and choose $\Phi$ as:  
\begin{equation}
\label{parity}
 \Phi(Z)=F(Z-X_0)+F(-Z-X_0)\,,
 \end{equation}
where $ F(Z)=\bar F(Z) $. The function $F$ is  periodic with period one, real or its Laurent series has only real coefficients and is convergent for  $ Z \rightarrow \infty$. Calculating the derivative of $\Phi$ for $Z=I Y$, it is easy to see that  $\Re \,[\Phi_Z]=0$ so the normal stress  $S_{11}$ vanishes and there is no need  to cancel ${\cal Q}(J)$, see Eqs.(\ref{stress},\ref{disper},\ref{charact}).  For $F$,  two square root singularities are chosen, located at $X_0 \pm l_0$ and separated by a branch cut. For symmetry reasons, we will fix the branch cut along the $X$ axis as follows:
  \begin{equation}
  \label{derivative}
F_Z=\frac{1}{\sqrt{ \tanh^2 (\pi Z)-\tanh^2( \pi l_0)}}-\cosh(\pi l_0)\,.
\end{equation}
The last term fixes the cancellation of $F_Z$ when $Z\rightarrow \infty$. $l_0$ is a tiny  parameter that specifies the size of the branch cut $ 2 l_0$.
Useful for the following, $F(Z)$ is  the primitive function of eq.(\ref{derivative}): 
\begin{equation}
F(Z)=f(Z)-Z \cosh(\pi l_0) \,,
\end{equation}
where
\begin{equation}
\label{profsin}
f(Z)=\sqrt{ \frac{ \sinh ^2(\pi  Z) -\sinh ^2(\pi l_0)}{\tanh ^2(\pi  Z)-\tanh ^2(\pi l_0)} }\frac{h(Z)}{\pi \cosh(\pi Z)}\,,
%\tanh ^{-1}\left(\frac{\sinh (\pi  Z)}{\sqrt{\sinh ^2(\pi  Z)-\sinh ^2(\pi l_0)}}\right)}
%{\pi \cosh(\pi  Z)  -Z \cosh (\pi l_0)
\end{equation}
and
\begin{equation}
  h(Z)=\tanh ^{-1}\left(\frac{\sinh (\pi  Z)}{\sqrt{\sinh ^2(\pi  Z)-\sinh ^2(\pi l_0)}}\right)\,.
\end{equation}
Several observations must be made at this stage:
\begin{itemize}
\item The choice of $F_Z$ is somewhat arbitrary, but takes into account symmetry arguments and is also  dictated by its simplicity.
\item Less singular functions with the square root replaced by a power law $(w)^{1/2}  \rightarrow (w)^a $, where $a>1/2$, are also possible a priori.
\item As soon as we introduce a singular zone around $X_0$, this automatically produces   another singularity around  $X_0/J$  which will influence  the interface more strongly, at $X=0$.  So we are faced  with  two boundary layers which are easier to treat  independently.  So $J>1$ and $l_0<<<1$.
\item When introducing such profiles, we are faced with a minimal list of  parameters such as $l_0$, $X_0$, $\epsilon$ and $J$. We hope to find $J$ below the Biot threshold $J_B$ and to find constraints on these  parameters. These parameters must be fixed in a consistent way. 
\item Finally, such localized singularities are often found  in periodic viscous flows where the equivalent of our "blob" is   the bubble in the viscous flow \cite{taylor1959note,vasconcelos1993exact,tanveer1995time,crowdy2009multiple}.
Note that in the case of bubbles, there are several families of solutions that depend  on the bubble location and its symmetries.   
\end{itemize}
The contribution of such a function $F(Z)$ on the interface and on the stress accumulated inside the sample are shown in Fig.(\ref{inner}).  
In order to show that such a  scenario can exit, it is necessary  to  establish the existence of semi-singular patches where the stresses are  concentrated. it will also be necessary  to relate the four parameters  $l_0$, $X_0$, $\epsilon$ and $J$ to other physical processes occurring in the inner boundary layer which are neglected in the outer zone. 

 \subsection{Physical origins of the patches}
  Several origins can explain the existence of patches, sites of focusing of the elastic stress. Such focusing locally destroys  the linear expansion of the elastic deformation in $\epsilon$ , making the validity of  the linear expansion questionable. But also, at large strains, it can invalidate the model itself: the choice of  the neo-Hookean elastic energy, the incompressibility hypothesis, the constant volumetric growth. Let us examine each of these possible causes:
  \begin{itemize}
 \item   The neo-Hookean model, very convenient for its simplicity fails to describe a focusing, either too strong or too localized and must be corrected by a more sophisticated hyperelasticity model involving nonlinearities in $I_1$ or other invariant such as $I_2$.
 \item The incompressibility limit is a mathematical limit  that  is not appropriate for large strains. A more physically relevant model may be  a  quasi-incompressible approximation  with a strong coefficient multiplying the invariant $I_3$ in the elastic energy \cite{weiss2013creases}. 
 \item Spatially constant volumetric growth is a naive approximation of  a true biological process. In fact, for growing living species, an excess of compressible stress is known to inhibit the cell proliferation. It is reported as the principle of homeostasis \cite{bernard1957introduction}.
 \item  This phenomenon also occurs in swelling. Everyone knows how to expel water from a wet sponge,  simply by applying a pressure to it.
\end{itemize}
  The complete solution of  the nonlinearities is hopeless, since it is impossible to find a nonlinear solution inside the patch that asymptotically  recovers the expansion given by Eqs.(\ref{derivative},\ref{profsin}). But, we know  similar situations in solid mechanics where  nonlinearities  play an essential role in a localized region around a singular point,  as  in fracture mechanics or dislocation theory, and are responsible for a constant  stress intensity factor. As shown by Barenblatt \cite{barenblatt2006scaling,barenblatt2014flow},  w these nonlinearities, if they  remain localized, do not  prevent the validity of the linear theory, except in the zones of high predicted stress where they soften. In the next section, we relax some of the  limitations of the model, by 
adding  nonlinearities  and compressibility, while keeping the volume growth constant.
 \subsection{Patches as inner boundary layer}
 In the first patch located at  $Z=X_0$, a new  coordinate $U$ such as  $ Z-X_0=\frac{1}{\pi} \tanh^{-1} (\tanh (\pi l_0)U )$ gives the following expansion of eq.(\ref{profsin})  in the limit of small $l_0$
\begin{equation}
\label{approx1}
F(Z)\sim \frac{1}{2 \pi} \left \{ Log \left (\frac{U +\sqrt{U^2-1}}{-U+\sqrt{U^2-1}}\right) +\phi_0 \right\}\,,
\end{equation}
and $\quad x \sim J X + \epsilon J (\Re \,[ F(Z)]+ x_1)$ with  $x_1= \Re \,[\phi_0+F(-X_0)+J \tau_1 \Re \,(F((J-1)X_0)+ F(-(J+1) X_1)]$. This expansion shows that  $F(Z)$ has  two singular points $U\pm 1$ separated by a branch cut. For  $\vert Z-X_0\vert >l_0$, then  $U>1$, so $l_0<Z-X_0<1$, the asymptotic behavior of $F(Z)$ is logarithmic,   which gives  for the outer profile:
 \begin{equation}
 \label{t}
\begin{cases}
 x &\sim J X + \epsilon  \left \{ \frac{J}{2 \pi}  Log \left ( \frac{(X-X_0)^2+Y^2}{ l_0^2}\right ) + x_1 \right\}\,,\\
 y &\sim Y - \frac{\epsilon}{ \pi}  \tan^{-1}\frac{Y}{X-X_0}\,.
 \end{cases}
 \end{equation}
 For the second patch ${\cal S_J}$  located at $X_0/J$, a similar expansion for $Z_1$ with the definition of $U_1=J U_X+I U_Y$ gives a similar result, the only difference  being  the expansion of coordinates $(x,y)$:
  \begin{equation}
 \label{tbis}
 \begin{cases}
 x &\sim J X + \frac{\epsilon J  \tau_1}{2 \pi} \left\{Log \left (\frac{(J X-X_0)^2+Y^2)}{ l_0^2}\right) +x_2\right \} \,,\\
 y &\sim Y - \frac{ J \epsilon \tau_1 }{ \pi}  \tan^{-1}\frac{Y}{J X-X_0}\,.
\end{cases}
\end{equation}
The fact that we have two  independent small  parameters ( $\epsilon$ and $ l_0$ ) suggests a complicated double  boundary layer. An example is also given by fracture mechanics, where a separation of length scales is necessary  and has been estimated in  \cite{long2020fracture,bouchbinder2009}.  Indeed, in soft and highly deformable materials, we cannot neglect nonlinear deformations and the dissipation for fast dynamics even if Hookean elasticity  gives a reasonable picture of  cracks. The separation of length scales, in \cite{long2020fracture} includes  the inner scale (the crack tip) due to  a large amount of dissipation, the intermediate scale due to the nonlinearities of elasticity and finally the outer scale where linear elasticity is allowed. Again,  two different length scales:  $\zeta$ for dissipation and $l$ for nonlinearities coexist and have been  experimentally verified \cite{long2020fracture}. In our case, we can disregard dissipation since either growth or gel swelling are  slow  processes, see Fig.(\ref{cuspexp})(A8,A9) and \cite{dervaux2011shape}. However, as with fractures, additional physical properties are needed in order to fulfill  the neo-Hookean model with incompressibility which  has an inherent lack of parameters and length scales. In the next section,  we  extend the neo-Hookean model by  incorporating  nonlinearities in the elastic energy such as $I_1^2$ and $I_2$ and also  the third invariant  $I_3$, which is  responsible for the compressibility.   
\section{Theoretical evidence for internal singularities} 
\label{evidence}
\subsection{New elastic model for large stresses}
We start by locally assuming a more complex elastic energy
and we focus on the patches located around $X_0$ and $X_0/J$. They are separated by a distance $X_0 (J-1)/J  \sim  2/3 X_0$.  We assume   two boundary layers around each patch: an inner zone of size $l_0$ and an outer zone of size $l_p$ such that  $l_0<l_p<<X_0$, to avoid overlapping  between the two patches. The existence of two boundary layers is  required,  first, to eliminate the square root singularity  and,  second, the necessity to recover the logarithmic asymptotic  before the linear expansion in $\Phi(Z)$.  Besides nonlinearities in the modeling of the hyperelasticity density, the limit of complete  incompressibility can be questionable when the strains become very large. There are  a significant  number of  models in the literature  that treat compressibility, see chapters $6$  and $8$ of \cite{holzapfel2000nonlinear}. Due to the large strains localized in the patches, the best  approach is the compressible model of R. Ogden \cite{ogden1997non,holzapfel2000nonlinear}, which separates the elastic energy density into two parts: $\Psi_{iso}$ and a volumetric part, also called the bulk part $\Psi_{vol}$. In practice however, it is very difficult to find explicit  solutions  that  occur on a scale smaller than  $l_0$. So we focus first  on  the  intermediate regime where $\vert Z-X_0\vert >l_0$  and we restrict on  nonlinearities in $I_1$ and a compressibility penalty treated  as a quadratic expansion. For the elastic energy density: either $\Psi^{(p)}$ (nonlinear model) or $\Psi^{(mr)}$ (the Mooney-Rivlin model) the following expressions are given for geometric invariants which are corrected by the growth according to eq.(\ref{Ione}):
 \begin{equation}
 \label{energsing}
 \begin{cases}
 \Psi^{(p)} &= \tilde I_1+\frac{\omega_p}{J^{2p-1}} \tilde I_1^{2p}+\frac{\kappa}{J}(\tilde I_3-J)^2 \\
 \Psi^{(mr)} &= \tilde I_1+\frac{\omega_{mr}}{J} \tilde I_2 +\frac{\kappa}{J}(\tilde I_3-J )^2\,,
  \end{cases}
  \end{equation}
 where $p$ is a real number greater than $1$,  $\omega_p$ and  $\omega_{mr}$ are  small parameters, positive or negative, representing a first correction to the neo-Hookean elasticity, but $\kappa$ is expected to be a large positive quantity to mimic the quasi-incompressibility.
    \subsection{The intermediate boundary layer analysis}
    \label{intermediateboundarylayer}
For the first patch, we define the rescaled quantities for the space coordinates in the initial configuration:
\begin{equation}
\label{scaling}
\hat X=(X-X_0)/l_p;  \quad \hat Y= Y/l_p\,.
%; \quad \delta \hat x=(x-JX_0)/\epsilon; \quad \delta \hat y= (y-Y_0)/\epsilon
\end{equation}
 A localized patch ${\cal S}$ or $ {\cal S}_J$ requires  $l_p \le  X_0$ and to recover  
 the asymptotics of $F(Z)$, we assume  a solution for the profile function $\epsilon  (x_s(\hat R),\epsilon y_s)$ given by:
 \begin{equation}
\label{repres}
 x_s= \frac{\epsilon}{\pi} f_s(\hat R) +J X_0\quad {\mbox{and}} \quad  y_s \sim  \frac{-\epsilon}{\pi} T\,,
\end{equation} 
where  $\hat R =\sqrt{\hat X^2+\hat Y^2} $   and $T=\arctan (\hat Y/\hat X)$. Assuming a possible correspondence with  eq.(\ref{t}),  the unknown function $f_s$  must  satisfy  $f_s(\hat R) \sim  J Log \hat R$, for  large values of $\hat R$,  but must remain finite for small values of $\hat R$, a property which is not verified  by eq.(\ref{t}). This  gives the following scaling for the invariants:
\begin{equation}
\label{scale}
\tilde I_1= \epsilon^2 /(\pi l_p)^2\,,\quad \tilde I_1^p\sim \epsilon^{4p}/(\pi l_p)^{4p}\,,\quad \tilde I_3\sim \epsilon^2/(\pi l_p)^2. 
\end{equation}
 The size of the boundary layer given by $l_p$ must eliminate arbitrary coefficients as much as possible. Note that $l_p$ is a tiny value, unlike $\kappa$ which can take  large values to represent  quasi-incompressible material. Define:
 \begin{equation}
 \begin{cases}
 \tilde I_1 =\frac{\epsilon^2}{\pi^2 l_p^2} \hat I_1;
 \quad \hat I_1=\frac{1}{\hat R^2} +f_s'(\hat R)^2,
 \\
 \tilde I_3 -J=-J \hat K; \quad \hat K=1+ \vert\omega_p\vert^{-1/p_0} \frac{f_s'(\hat R)}{\hat R },
 \end{cases}
 \end{equation}
 where  $p_0=2 p-1$ and $l_p=\vert \epsilon\vert /\pi\vert\omega_p\vert^{1/(2p_0)} J^{-1/2}$,
then  the elastic energy, for the new elasticity model  given by eq.(\ref{energsing}), is transformed to :
  \begin{equation}
  \label{esing}
      {\cal E}_{s}=\frac{ \epsilon^2}{\pi}\int_0^{\infty}\hat R d\hat R 
     \left( \hat I_1 +\hat  I_1^{2 p}+J \kappa\hat K^2\right).
  \end{equation}
  Variation with respect to $f_s$ leads to a second order E-L equation, which can be integrated once without difficulty, and finally we are faced with a first order nonlinear differential equation:
    \begin{equation}
 \label{compint}
\hat R f_s'(\hat R) \left (1+2 p  \left(f_s'(\hat R)^2+\frac{1}{\hat R^2}\right)^{p_0}\right)\\
+\kappa_0 \frac{f_s'(\hat R)}{\hat R} =C_S,
\end{equation}
where   $\kappa_0= J  \kappa \,\vert \omega_p\vert^{-1/p_0}$  and $p\ge1$. $C_s$ is an arbitrary  integration constant at this stage. In the quadratic case, $p=p_0=1$, the Euler-Lagrange  equation for $\hat x_s$ is easily solved, and the second Euler-Lagrange equation for $\hat y$  is automatically satisfied  once the relation eq.(\ref{repres}) is imposed. Even if an exact  solution can be found, we  focus on the two limits of interest, first for  $\hat R \rightarrow 0$ and then for $\hat R \rightarrow \infty$:
    
\begin{equation}
\label{solp}
{\mbox{For}}\quad \hat R\rightarrow 0\quad   f'_s( \hat R)=\frac{Cs \hat R^{4p-3}}{2 p + \kappa_0 \hat R^{4(p-1)}}\,. 
\end{equation}
Regardless of  the value of $p\ge 1$, the behavior of $f_s$ is a regular function of $\hat R$, as it is required for a physical solution. Note  that $\kappa_0$ only plays a critical role  if $p=1$, which leads to:
\begin{equation}
\label{quadrares}
{\mbox{For}}\quad \hat R\rightarrow 0\quad  f'_s(\hat R)=\frac{ C_S}{2 p+\kappa_0}\hat R\,.
%\quad {\mbox{for}}\quad \hat R\rightarrow \infty\quad \frac{d f_s}{d \hat R}=\frac{C_s}{\pi \hat R}
\end{equation}
 Finally, choosing $C_s=J$  leads to the convenient asymptotic for $f_s$ when $\hat R\rightarrow \infty$ whatever the $p$ values.
 So the outer boundary layer seems to satisfy the physical requirements, see eq.(\ref{t}). Since  we have introduced two boundary layers: an inner core of size $l_0$ and an intermediate zone of size $l_p>l_0$, this  allows to easily perform  the adjustment  in the asymptotics of $F(Z)$ given by eq.(\ref{approx1}).  Another standard way to modify the neo-Hookean model is the Mooney-Rivlin model, which is even simpler:
     \begin{equation}
 \label{mooney}
\hat R \left(\frac{d f_s}{d \hat R}\right) \left(1 +\frac{1}{ \hat R^2}\right)  =-\frac{\kappa_{mr}}{\hat R} \left(\frac{d f_{s}}{d \hat R}\right) +C_s\,,
\end{equation}
with $\kappa_{mr}= J \kappa/\omega_{mr}$  and $l_{mr}=\epsilon/\pi \sqrt{\omega_{mr}/J}$. The solution for $f_s$ is easily found and in physical units, we have:
\begin{equation}
x_s=\frac{J\epsilon }{2 \pi} Log\left\{\frac{ R^2}{l_{mr}^2}+\left(1+ \frac{\kappa_{mr}}{\omega_{mr}}\right ) \right \}\,,
\end{equation}
with no change for $y_s=-\epsilon  T/\pi $. 
\\
 However, in the inner region very close to $X_0$ where $\vert Z-X_0\vert\le l_0$, these solutions given by  eq.(\ref{solp},\ref{mooney})  present a singularity of the strain   in $1/R$  due to the choice of  $y_s$, which limits their validity in this inner zone.  
\\
 Note that, for both models, $x_s$ has the same asymptotic limit for $R>\it{l}_p$ and they are regular for $R->0$, so even if we are not able to find the inner core solution, we can recover the result of eq.(\ref{t}) and the correct asymptotic of $\Phi(Z)$. \\

The same analysis applies to  the upper singularity centered around  $X_0/J$. For the second patch  ${\cal S_J}$, the same strategy is followed,  and it is easy to show that the associated deformations $(\hat x_{S_J}, \hat y_{S_J})$ according  to eq.(\ref{tbis}) satisfy :
\begin{equation}
 \label{bound1}
\begin{cases}  
\hat x_{s_J}=  \frac{1}{2\pi} Log \left (\frac{1}{\pi^2}+ J^2 \hat X^2+\hat Y^2 \right)\\ 
 {\mbox{and}}\\
 \hat y_{s_J}= - \frac{1}{\pi}  \tan^{-1}\frac{\hat Y}{J \hat X}\,.
\end{cases}
\end{equation}
 The same matching between deformations in the second patch is done by 
 $\epsilon  \tau_1 (J x_{s_J},y_{s_J}) $ which gives the same analysis for both patches.  So, both singularities can be treated in the same way.\\
 	\\
		 In summary, once linearized, the primary  incompressible model can have two patches ${\cal S}$ and $ {\cal S}_J $ with a  singularity inside. The actual patch size is $l_p$. Each patch contains an inner core of size $l_0^p$ and an intermediate zone of size $l_p$, corresponding to  two boundary layers with a more regular stress distribution. The main information we obtain for the treatment of the intermediate zone is in fact its size $l_p$ given by  $\epsilon$ and  the first nonlinear correction of the neo-Hookean energy scaled by the constant: $\vert \omega_p\vert$. 
				 \subsection{The inner core }
		 \label{innersection}
		 In the inner core, the strains and stresses are expected to increase again, and perhaps  the specimen  is  strongly modified by nonlinearities  going from stress hardening to plasticity
   \cite{ritchie1973relationship}  Such material transformations have been studied in detail in the fracture mechanics literature \cite{li2023crack,long2020fracture}. Here we try to stay as close as possible to our original model although we agree that any modification of the material structure  is possible and may change some conclusions. Therefore, we keep the compressible hyperelasticity point of view but we reinforce the compressibility analysis  of the material with the Ogden model, where the elastic energy  is decoupled into a purely volumetric elastic response and a dilatational  part \cite{ogden1997non}. As explained in section \ref{Invariant} at the very beginning of the paper, this requires a new definition of the strains according to eq.(\ref{energiecomp}).
   \subsubsection{Rescaling the strains and the invariants}
		We  retain the quadratic compressibility model of the previous paragraph, which is still a function of $I_3$ although other approximations may be more appropriate for $R \sim l_0^p$ (see Chapter $6.5$ of \cite{holzapfel2000nonlinear}). Within  the inner core, we suggest the following description for the current coordinates $( x_c, y_c)$:
		 \begin{equation}
   x_c= \frac{\epsilon}{\pi}  l_0^{s} \bar F(\rho, \theta);\,\, 
              y_c= -\frac{\epsilon l_0^{q}}{\pi} \bar G (\rho, \theta)\,.\\ 
\label{core}
 \end{equation}
with $ \rho=R/l_0^{p}$ and   $\theta=T/l_0^q$. We restrict ourselves to a cubic nonlinearity $I_1^3$ for reason of simplicity  but the method can be applied to any kind  of dilatational  hyperelasticity.   We impose that $p$ and $q$ are positive and that $x_c$ and $y_c$ are  regular for $\rho \rightarrow 0$. For $\rho \rightarrow \infty$, the matched asymptotic analysis required that $x_c$ and $y_c$ coincide with the behavior of $x_s$ and $y_s$ for $R\rightarrow0$. For $y_c$, this simply requires $G \rightarrow -\theta$.  For $x_c$, we first consider $x_s$ from  eq.({\ref{quadrares}), in the neighborhood of  $R\rightarrow 0$: 
\begin{equation}
    x_s=\frac{\epsilon J }{\pi} \frac{R^2}{6 \, l_{3/2}^2}\quad   \mbox{where}\quad  l_{3/2}= \frac{\epsilon}{\pi \sqrt{J}}\vert \omega_{3/2}\vert ^{1/4}.
\end{equation}
Thus, at infinity, $x_c$ must behave as $x_c \sim \rho^2$ and  matching  with $x_s$ gives a more precise result:
\begin{equation}
    x_c=\frac{\epsilon}{\pi} l_0^{s-2 p} R^2\, \Rightarrow\,l_0=  \left( \frac{6 \, l_{3/2}^2}{J}\right) ^{1/(2 p-s)}.
    \end{equation} 
If these conditions are satisfied, the inner coordinates $(x_c,y_c)$ can correctly match  the corresponding ones  $(x_s,y_s)$ in the intermediate zone,  allowing to recover eq.(\ref{bound1}). However, our analysis involves many degrees of freedom  in addition to the unknown material parameters $\kappa$ and $\omega_{3/2}$. Most likely, there will be several possibilities and to limit them, we recapitulate the different constraints. Since the singularity of the strains calculated with $x_s$ and $y_s$ comes from the derivative with respect to $T$, we impose that the corresponding strains for $x_c$ and $y_c$ dominate. After evaluating  the amplitude of the strains, we get:
\begin{equation}
\begin{cases}
    \frac{\partial  x_c}{\partial R}&=\frac{\epsilon}{\pi}l_0^{s-p}\frac{\partial \bar F}{\partial \rho} << \frac{1}{ R} \frac{\partial x_c}{\partial T}\sim  \frac{\epsilon}{\pi}l_0^{s-p-q} \frac{1}{\rho} \frac{\partial \bar F}{\partial \theta},\\
    \\
   \frac{\partial  y_c}{\partial R}&=-\frac{\epsilon  l_0^{q-p}}{\pi  }\frac{\partial \bar G}{\partial \rho} << \frac{1}{ R} \frac{\partial y_c}{\partial T}\sim -\frac{\epsilon}{\pi \rho}  l_0^{-p}\frac{\partial \bar G}{\partial \theta}\,.
\end{cases}
\label{ineq}
\end{equation}
 $ \frac{\partial \bar F}{\partial \rho},\,\,\frac{\partial \bar F}{\partial \theta},\,\,\frac{\partial \bar G}{\partial \rho},\,\, \frac{\partial \bar G}{\partial \theta}$  being  quantities of order one, then  the exponent $q$ is positive.  Now we examine $I_1$, reduced to the shear strains: 
\begin{equation}
\label{simIone}
    I_1\simeq\frac{\epsilon^2}{\pi^2} l_0^{-2 p} \frac{1}{\rho^2}\left(l_0^{2(s-q)}\left(\frac{\partial \bar F}{\partial \theta}\right) ^2+\left(\frac{\partial \bar G}{\partial \theta}\right)^2\right).
\end{equation}
The two terms in $I_1$ are of  different weight. However, only the first one  gives the correct asymptotics for  the Euler-Lagrange equations, such as $x_c\sim \rho^2$.  This imposes $s<q$. 
Then we define $\hat I_1$ and $\hat I_3$
\begin{equation}
\begin{cases}
 I_1= \frac{\epsilon^2}{\pi^2} l_0^{2(s-q- p)}  \hat I_1 ,\quad I_3=\frac{\epsilon^2}{\pi^2} l_0^{s-2p} \hat I_3,\\
\hat I_1=\frac{1}{\rho^2}\left(\frac{\partial \bar F}{\partial \theta}\right) ^2,\\
    \hat I_3=\frac{1}{ \rho^2} \left(\frac{\partial \bar F}{\partial \rho}\frac{\partial \bar G}{\partial \theta}-\frac{\partial \bar F}{\partial \theta}\frac{\partial \bar G}{\partial \rho}\right).
    \end{cases}
\end{equation}
\subsubsection{The energy  density of the inner core}
Although families of possible deformations have been published  for arbitrary elastic energy densities, very few concern compressible materials (\cite{klingbeil1966class,carroll1967controllable,yavari2021universal}). Since we can expect a high compression in the inner core, we modify the bulk energy and adopt the Ogden model \cite{ogden1997non,holzapfel2000nonlinear} of compressible constrained materials, but keep the penalty term as before.
There are  then  two choices, depending on  the  value of $\vartheta= \vert \omega_{3/2}\vert l_0^{2s-4q}/J^2$. In fact  $\vert \omega_{3/2}\vert$   is a tiny amount, but $s<q$ so $\vartheta$ is arbitrary.
\begin{itemize}
\item $\vartheta<<1$,  defining $K_1=\frac{\kappa}{J} \frac{\epsilon^4}{\pi^4}  l_0^{s-4 p+2 q}$
\begin{equation}
\label{hyp1}
    W_c=l_0^{s-2q}\left(\frac{ \hat I_1}{\hat I_3}+ K_1\hat I_3^2 \right).
\end{equation}

\item $\vartheta>>1$, and $K_2=\frac{J \kappa}{\vert \omega_{3/2}\vert}\frac{\epsilon^4}{\pi^4}  l_0^{6 q-4p -s}$
\begin{equation}
\label{hyp2}
    W_c=\frac{l_0^{3s-6q} \vert\omega_{3/2}\vert}{J^2} \left(\pm\left(\frac{ \hat I_1}{\hat I_3}\right)^3+ K_2\hat I_3^2\right).
    \end{equation}
    \end{itemize}
A good way to get very low values of $l_0$ is to choose $p\simeq s/2$.  However, it is sufficient  to  have $2 p< 2+s$. In addition, one must  reduce the elastic energy which is confined in the core.This  means that $2 q-s$ must be as small as possible. In this context, the elastic energy trapped  in the core can be estimated to be around $l_0^{s-2 q+2 p} $, which   means that  $s-2q+2 p>0$ must be positive for the first hypothesis, and for the second, a necessary condition is $3 s-6 q+2p>0$. It is relatively easy to choose good parameters, such as $s=0,p>q$  or $s=0, p>3/2 q$ for the second choice. However, these conditions may  not be sufficient and must  be compared with the elastic energy in the intermediate range.
Note that the nonlinear eigenvalues $K_1$ and $K_2$ are  numbers that are difficult to predict even as an order of magnitude. Whatever the hypothesis, eq.(\ref{hyp1}) or eq.(\ref{hyp2}), the  corresponding asymptotics  perfectly check the overlap with $x_s$ and $y_s$ for $\rho\rightarrow\infty$, and the behavior for $\rho\rightarrow 0$, for eq.(\ref{hyp1}) is $\bar F(\rho,\theta)\sim \rho$ and the same for $\bar G$. The analysis for the second hypothesis is less obvious. This study proves that the matching is possible and the deformation  remains regular, but  due to the degrees of freedom as $p$ and $q$, we do not get any information about $J$ and $\epsilon$, $X_0$ and $l_0$ as a function of the material parameters $\kappa$ and $\omega_{3/2}$.
At this stage, it remains to evaluate how the zero order elastic energy is modified by the localized compression zones. In fact, a bifurcation is possible if the elastic energy  is reduced.

\subsection{ Energy of the patches}
The goal now  is  to evaluate the elastic energy involving  the entire physical plane  (the zero-order and linear expansion in $\epsilon$) and the energy due to the patches, including the inner core and the outer ring of both singular patches.
As shown before, in section \ref{energetic}, the expansion of the elastic field has $3$  contributions in power of  $\vert\epsilon\vert$, see eq.(\ref{Eonetwo}). If there are  no singularities inside the physical plane, the linear term ${\cal E}_1$ vanishes. If there are  singularities, this is  not the case and one has to evaluate  this term. For this, we use the same method of complex analysis as described in  \ref{Ordertwo}:
\begin{equation}
\begin{cases}
{\cal E}_1&=\tau_0\iint_{\cal P} dS \Re \,[\Phi_Z+J \tau_1 \Phi_{Z_1}]\\
&=\frac{\tau_0^2}{2 J^2}\Re \,[\frac{1}{2 I} \iint_{\cal P} dZ d \bar Z \Phi_Z]
\end{cases}
\end{equation}
Without singularities inside the physical plane, the  integration contour called ${\cal \partial P}$, which can be observed in Fig.(\ref{3modester}), first row  on the left, corresponds to the outer boundary of the physical strip and because of the periodicity of $\Phi$ the integration gives no contribution. This is not the case now since the calculation requires to cross the cell in  the middle along paths that go from the point $M_0$  up to the first and second singularities    ${\cal S}$ and  ${\cal S}_J$ and  up along the neighboring path on the right (see Fig.(\ref{3modester}, first row)). So the additive contribution comes from the two circular contours around the singularities. After simple simplifications and taking the size of the singularity radius as  $l_S\sim l_1$, for both   patches, we get:
\begin{equation}
\begin{cases}
\label{firstorder}
{\cal E}_1 &=\frac{\tau_0^2}{2 J^2}\Re\,[\frac{1}{2 I} \oint_{\cal \partial P} dZ \Phi(\bar Z)\\
&=\frac{\tau_0^2}{4 J} \frac{l_s}{\pi} \Re \,\, [ \int_{-\pi}^{\pi}  d T e^{I T}( -I T)]
= \frac{\tau_0^2}{2 J}l_1\,.
\end{cases}
\end{equation}

Thus, the singularities inside the sample give a correction of  the elastic energy  of order  $\epsilon l_1$ which, if $\epsilon<0$, indicates a possible bifurcation: for  negative $\epsilon$, the singular solution is less energetic than the uniform axial  growth.
 However, one must also  evaluate the energy inside the core of the  patches, for both singularities. Evaluating the energy inside the inner core in each singular patch  leads to :
 \begin{equation}
{\cal E}_c\simeq \vert \omega_{3/2}\vert l_{3/2}^\varpi;\quad 
\varpi = \frac{2}{2 p-s} (3s-6q +2p).
\end{equation}
where $\varpi = p (3s-6q +2p)/(2p-s)$.  In order to compare with the energy of the intermediate zone,  $\varpi$ must be as large as  possible. Unfortunately the value $s=2p$ is not compatible with our constraints but $s=0$, $q$ small  and $p=1$ leads to $\varpi\simeq 2$ which is enough for ${\cal E}_c$ to be negligible. This quantity has to  be compared with $\epsilon l_{3/2} =\epsilon^2 \vert \omega_{3/2}\vert ^{1/4}/\pi$. Also  the small amount $\vert\omega_{3/2}\vert^{1/4}>\vert\omega_{3/2}\vert$  justifies the fact that the intermediate singular zone  dominates. 
We conclude that  the dominant energy density  corresponds to the  uniform axial growth corrected by:  
\begin{equation}
\delta {\cal  E}\sim \epsilon \frac{(J^2-1)^2}{2J }\vert \epsilon \vert \vert\omega_{3/2}\vert^{1/4}\,.
\end{equation}
Thus, singularities inside the sample lower the  instability threshold and are at the origin of a bifurcation.  At this stage, there is no restriction on the  values of $J$  except that the distance between the two singularities must be greater than $2 l_1<X_0 (J-1)/J$, which means  that $\delta J=J-1$ must be larger than $2 J l_1/X_0$. This is a necessary condition for our analysis based on the separation of the two patches. Thus, the new bifurcation threshold results from controlled deviations from the neo-Hookean model.  

\section{ Path independent contour integrals}
\label{path}
A fancy and easy way to determine unknown parameters in singular elasto-static fields is to use path-independent integrals \cite{cherepanov1967crack,rice1968mathematical,rice1968path}. They result from Noether's first theorem \cite{noether1918invarianten}  as  recently demonstrated and recalled by J. Rice and collaborators \cite{zhang2017path}. This theoretical method relates the geometric parameters of the singularities to the boundary conditions imposed on the far-field elasticity. It  has been successfully applied to many topics  of elasticity \cite{amar2018creases}, but also to  other physical fields as soon as they are  governed by variational principles. One can think of interfacial potential flows (Darcy or Euler flows, \cite{ben2002exact})and  electrostatic fields (\cite{amar1999void}, \cite{zhang2017path}). It is widely used  in all aspects of solid mechanics such  as fracture \cite{dascalu1994energy}, dislocations. \cite{shi2004j,lubarda2000conservation,ni2008self}, notches \cite{rice1968path}, erosion \cite{meyer2017path}, it is not limited to time-independent formulations \cite{knowles1971class}, nor to linear elasticity although the finite elasticity singularity must be treated  with care \cite{lubarda2000conservation}, especially in the case of  non-quadratic formulations. Nonlinear problems, sometimes time-dependent, are often interpreted in terms of  internal  forces acting on defects present in  materials and  path-independent integrals have also been established in these cases \cite{eshelby1985fundamentals,dascalu1994energy}. It remains, however, that some  applications in nonlinear elasticity have been questioned  \cite{rice1968plane,hutchinson1968singular}, more precisely  for the so-called M integrals. Proofs of the application of this technique are justified  in the Appendix section \ref{JINTPROOF}. Our goal  is to to discover relationships between $l_0$ and $X_0$ and the neo-Hookean parameters  with growth such as $\epsilon$ and $J$. 
\begin{figure*}	
 \centering 
    \begin{minipage}[b]{1\textwidth} 
    \subfigure[\, $\bm{{\cal R}_0=X_0/\vert\epsilon}\vert$ \bf{versus} $\bm{\epsilon}$ ]{ 
         \includegraphics[width=.32\textwidth]{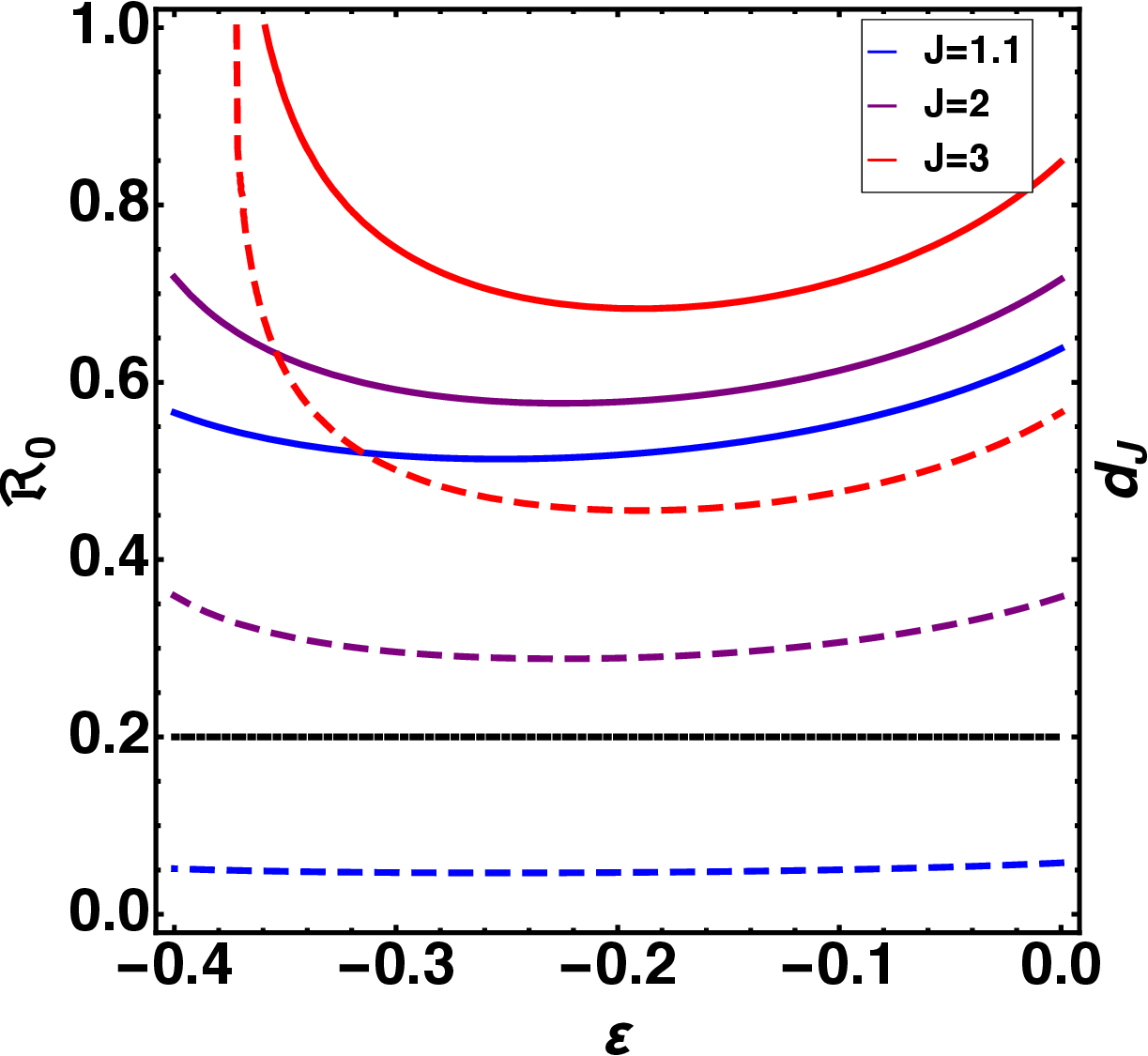} 
          } 
        \hspace{0in} 
        \subfigure[\, \bf{Corrected values of $\bm{{\cal R}_0}$ }  ]{
            \includegraphics[width=0.32\textwidth]{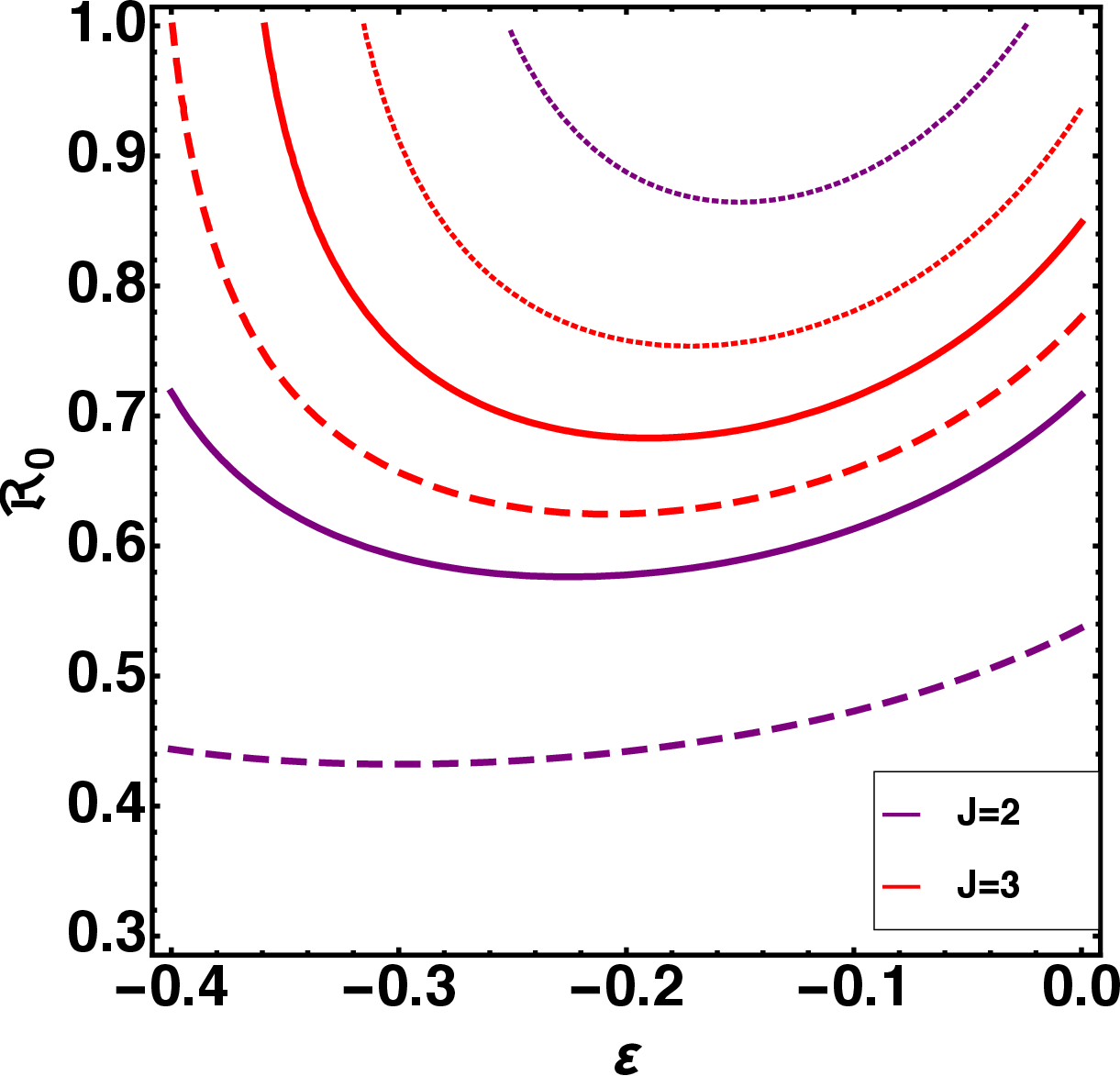} 
        } 
         \subfigure[\, \bf{Growth difference from bottom to top} ]{\includegraphics[width=.29\textwidth]{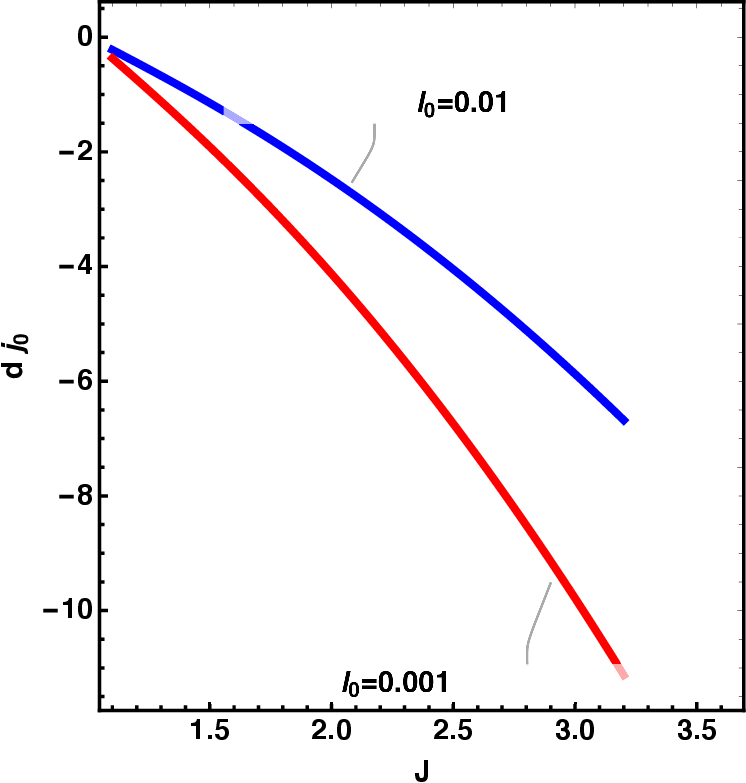}
         }
          \end{minipage}
    \caption{ Panel (a) Continuous lines: the ratio ${\cal R}_0$ between $X_0$ and $\vert\epsilon\vert$  for different  values of $J$, dashed lines: $d_j={\cal R}_0 (J-1)/J$ the distance (divided by $\vert \epsilon\vert$) between $S_J$ and $S$,  which must be greater than
    $2 l_1/\vert \epsilon\vert=2 \sqrt{\vert \omega_1\vert}$ represented by the dot-dashed curve in black. Panel (b) continuous lines: ${\cal R}_0$,dotted and dot-dashed lines, corrected values due to change of  $J$ at large distances: $j_0 \pm  0.1$,  
    with  dot-dashed lines for $+$ and dotted lines for $-$,  see   eq.(\ref{corrX0}).\\ In (c) the quantity $j_0 d$ (jump of the growth values between the bottom and the top of the strip the times the thickness $d$ of the sample) as a function of $J$ required for two sizes of the  singularity, $l_0=0.01$, and $l_0=0.001$.}
      \label{estimate}
    \end{figure*}
    
    \subsection{The J-Integral}
\label{Jim}
This approach is not fully nonlinear because  we perform an incremental expansion. In addition, our sample is pre-streched by growth which destroys the spatial isotropy. Thus, we cannot claim that the J-integral methods are directly applicable. Knowles and Sternberg have demonstrated the  validity of the $J$ integral in fully nonlinear elasticity  and also for incremental deformations but only when the initial state is stress-free,  which is a different case. Therefore, it is important to verify that the J-integral remains  valid for the model described in section \ref{variational}. This is done in Appendix (\ref{JINTPROOF}) and  we define ${\cal J}$ which is  a contour  integral, see Fig. (\ref{3modester}), panel (B) or (C) above:
\begin{equation}
\label{calJ}
 {\cal J}=\oint dS (E {\bf }. {\vec {\bf e}_x}- N_k S_{ik} F_{i1})=0 \,.
\end{equation}
The stress  $S_{ik}$ was introduced in the section \ref{variational}, eq.(\ref{borderstress}) and eq.(\ref{borderstressno}) and the strains are simply given by $F_{ij}=\partial_j x_i$.  Note that the J-integral is a vector so ${\cal J}$ is only one component. Not all of  the J-component can give pertinent information. The contour, shown in Fig.(\ref{3modester}), panel B on top, first goes  from $M$ to $M_0$, the center of  ${\cal C}_0$, then it goes down to avoid the two singularities ${\cal S}_J$ and $ {\cal S}$, it climbs up to $M_0$ to join the point $M_1$, it  continues along $M_1M_2$, then $M_2 M_3$ and finally $M_3 M$. Only the brown paths  can  contribute, the blue paths cancel each other out for reasons of periodicity and symmetry. Focusing on the contour ${\cal C}_0$, which is  $MM_1$ at the  top of the domain, only the energy density contributes, since both normal stress components vanish, a necessary condition for a free boundary. Decomposing  ${\cal J}$ into  ${\cal J}^{(0)}+\epsilon  {\cal J}^{(1)}+\epsilon^2 {\cal J}^{(2)}$, we get for the upper boundary ${\cal C}_0$ :
\begin{equation} 
\label{Jup}
{\cal J}^{(0)}_{{\cal C}_0} =\frac{(J-1)^2}{2};\, {\cal J}^{(2)}_{{\cal C}_0}=\frac{\tau_0^3}{ 8 J^2}\int_{-1/2}^{1/2} dY (\phi_Z)^2\vert_{_{Z=I Y}}\,.\\
\end{equation}
and $\quad {\cal J}^{(1)}_{{\cal C}_0}=0 $. The last integral is  difficult to evaluate exactly, but  in the limit of small $l_0$, it gives: 
\begin{equation}
{\cal J}^{(2)}_{{\cal C}_0}\sim -\frac{\tau_0^3}{2 J^2} (\coth (2\pi X_0)-1)\,.
\end{equation}
If we  now focus on the singularities ${\cal S}_J$ and $ {\cal S}$, the vertical brown contours have no contribution, so only the singularities ${\cal S}$ and ${\cal S_J}$  play  a role. By defining  a small radius around each singularity: $R=((X-X_0)^2+Y^2)^{1/2}$ for $l_0<R<1$, one can approximate $\phi_Z$ close to  ${\cal C_S}$ by:
\begin{equation}
\label{firstpatch}
\phi_Z\sim \frac{ e^{-I T}}{\pi R}-\coth (2 \pi X_0)\,,
\end{equation}
and in the neighborhood  of ${\cal C_{S_J}}$ by:
\begin{equation}
\label{secondpatch}
\phi_Z \sim \frac{1}{\pi R (J \cos(T)+I \sin(T))} -\coth (2 \pi X_0)\,,
\end{equation}
and derive the contributions of these singularities to ${\cal J}$, first  for ${\cal S}$:
\begin{equation}
{\cal J}^{(1)}_{{\cal S}}=2 \tau_0\quad {\cal J}^{(2)}_{{\cal S}}\sim 2 \tau_0\coth (2 \pi X_0)\,,
\end{equation}
and  for  ${\cal S}_J$:
\begin{equation}
{\cal J}^{(1)}_{{\cal S_J}}=2 J \tau_0\tau_1; \quad J^{(2)}_{{\cal S_J}}=- 2  J^2 \tau_0\tau_1^2  \coth (2 \pi X_0)  \,,
\end{equation}
  Finally, after adding the integral contribution  ${\cal J}$ at   ${\cal C}_0$, around the singularities ${\cal S}$ and ${\cal S_J}$, and on the two trajectories between $M_0$ and the singularities and on the two  vertical lines $M_1M_2$ and $M_3M$, and after simplifications, we determine the value of   ${\cal J}_P= {\cal J}_{{\cal C}_0}+{\cal J}_{{\cal S}}+{\cal J}_{{\cal S}_J}={\cal J}_{{\cal C}_1} $, where ${\cal J}_{{\cal C}_1} $ is restricted to the horizontal lower segment. If at $+\infty$, the volumetric growth is kept at  $J$, then ${\cal J}_{{\cal C}_1}={\cal J}^{(0)}_{{\cal C}_0}$, 
  which  eliminates  the zero order and results in: 
 \begin{equation}
\label{partialJ}
{\cal J}= -\frac{\epsilon\tau_0(J-1)^2}{J}
\left \{ 1 +\frac{\epsilon(1+J)^2}{2 J}\left(\frac{2}{ \tanh(2 \pi X_0)} -1\right )\right \}
\,.
\end{equation}
This evaluation  is correct   for $X_0>l_0$, which results in   $X_0\sim -\epsilon (1+J)^2/(2 \pi J)$. Again, the negative sign of $\epsilon$ is confirmed. The numerical values of the $X_0$ solution of eq.(\ref{partialJ})  are shown in Fig.(\ref{estimate}),  which shows  the ratio ${\cal R}_0=X_0/\vert \epsilon\vert$  in panel (a) with continuous lines while the distance of both singularities also divided by $\vert \epsilon\vert$:  $d_J=X_0 (J-1)/J$ is represented by dashed curves. Both sets of curves show a decrease at small  $\epsilon$ and then an increase, so we  can deduce that $X_0 \sim -\epsilon$. Since $d_J$ must be greater than $2 l_p=2\vert\epsilon\vert \vert\omega_p\vert ^{1/2p}$, a threshold value for  $J$  depending on   $\vert \sqrt{\vert\omega_1\vert}$ can be proposed: as an example, from Fig.(\ref{estimate}), the value $J+1.1$ is obviously too low. Our analysis assumes that the growth conditions are maintained at infinity and that the sample is  infinite, which is not true in real experiments. If the sample has a finite depth $d\ge 1$, we know that the  elastic deformations decay exponentially from a distance to the interface of the order of the wavelength, our approach remains valid near the interface, but we must consider a substrate that may alter our estimate of  ${\cal J}_{{\cal C}_1}$. In addition,  the growth law may change   away from the interface. These two points will be  explored below. 
\subsubsection{ Constant growth and finite size effects}
 We now assume that our sample has a finite size $d$ and it  is attached on a solid substrate. For $X=d$, the growing material cannot penetrate the substrate but can slide freely on it. The deformation $\Phi(Z)$, estimated from the top of the layer, decreases exponentially as $\Phi(Z)\sim -2 X_0 e^{-2\pi Z}$  when $\vert Z\vert>>1$. We need to  adjust  this deformation near the substrate when $Z\sim d$. Following the eq.(\ref{expansion})  at a distance $d$, the profile function can be represented  by:
\begin{equation}
\begin{cases}
     x&= JX-d_1+\epsilon_1 \cos(2 \pi Y)
     \left(e^{-2 \pi X}-\tilde \tau e^{-2 \pi J X}\right),\\
   y&=Y-\epsilon_1\sin(2 \pi Y)
     \left(e^{-2 \pi X}-J \tilde \tau e^{-2 \pi J X}\right).
     \end{cases}
     \end{equation}
  where $d_1=d (J-1)$, $\epsilon_1=-2 X_0 \epsilon$  and $\tilde \tau=-e^{2 \pi d (J-1)}$. $\epsilon_1$ results from the matching with the lower expansion and is of the order of $\epsilon^2$.   ${\cal J}_{{\cal C}_1}$ is easy to compute and read: ${\cal J}_{{\cal C}_1}=(J-1)^2/2\{1-(4\epsilon\pi X_0 e^{-2\pi d} )^2(J+1)\}$. Obviously, once $d$ is of the order of the wavelength and even larger, this correction becomes negligible: for $d=1$, $e^{-4\pi d}=3.46 10^{-6}$, for $d=2.5$, $e^{-4\pi d}=2 10^{-14}$. Note that a sliding substrate allows an  easy estimation of finite size effects. Clamped conditions, as discussed later  in the section \ref{Finitesize} are much more difficult to fit to our singular deformation mode.
  \subsubsection{Inhomogeneous volumetric growth}
  If the growth becomes slightly inhomogeneous at large distances,  becoming $\tilde J=J+\epsilon j_0$ at the bottom of the sample, then  estimating  ${\cal J}_{{\cal C}_1}=(1-J)^2/2 +\epsilon (J-1)j_0$ will change  the  $X_0$ value into :
  \begin{equation}
  \label{corrX0}
        X_0 \sim - \epsilon \frac{(J-1)^2 (1 + J)^3}{2 \pi J (1-J-J^2 +J^3+ j_0 J)}\,.
  \end{equation}
  This estimate for $X_0$ is given for small $\epsilon$, see Fig. (\ref{estimate})(b). Increasing the volumetric growth at the bottom ($j_0<0$) also increases the value of $X_0$. However, such an estimate is valid for a change in volumetric growth localized  only at the bottom. 
%(mathetica cor2)
 
\subsection{The M-Integral}
\label{Mint}

Despite debates about the validity of the M-integrals in finite elasticity, let us now consider these  integrals which  have the advantage of explicitly  introducing the size of the elastic samples. Unlike  the $J$  and $L$ integrals, proved valid  by Knowles and  Sternberg \cite{knowles1971class}, the M-integral technique turns out not to be  always applicable for arbitrary energy densities. Nevertheless, when applicable, it remains a very useful tool for demonstrating properties of nonlinear fields 
such as the creeping closure, for example \cite{meyer2017path}. As before for the $J$ integral, it is better to convince ourselves that a path-independent integral ${\cal M}$ can be constructed,  and this is realized in the Appendix section  \ref{JINTPROOF}. For our modeling, the definition of ${\cal M}$ follows: 
\begin{eqnarray}
 \label{calM}
&{\cal M} = \oint ds \left(E-\frac{(J-1)^2}{2} \right) \vec {\bf X}. \vec {\bf N}-  S_{jk}U_{ji} X_i.N_k {\nonumber}\\
&- (J+1)\left (U_x\cdot N_X+U_y\cdot N_Y
 \right ) =0\,.
\end{eqnarray}
where $U_X=x-J X$ and $U_y=y-Y$, the equation (\ref{calM}) being valid up to $O(\epsilon^3)$. As before for ${\cal J}$,  ${\cal M}$ results from $4$ contributions along  the horizontal axis $X=0$,  the two patches ${\cal C_S}$ and ${\cal C_{S_J}}$, and the far field. The vertical lines do not contribute as before. Considering the upper boundary where $X=0$, only the third term in eq.(\ref{calM})  of order $\vert\epsilon\vert$ contributes. The contribution of the two patches is of order  ${\cal J} X_0$ for the first two terms of eq.(\ref{calM}). Each term, either ${\cal J}$ or $X_0$, is  of order $\vert \epsilon \vert$ and the result will be neglected. For the last term, it  is of order $\vert \epsilon\vert  l_1$, which is  even smaller than $\vert \epsilon\vert X_0$, so the patches make a subdominant contribution. Consequently, the only way to compensate ${\cal M}_{X=0}$ is  to close the contour at a finite distance $d$ as done before and to assume a slight difference in the volumetric growth. Let us first evaluate  ${\cal M}_{X=0}$ for a very small value of $l_0$

\begin{equation}
 \begin{cases}
  {\cal M}_{X=0}&=-(J+1) \int_{-1/2}^{1/2}(x-J X)dY \\
  &= -J (J+1)(1+\tau_1)\epsilon\int_{-1/2}^{1/2} dY I_f\,,
\end{cases}
\end{equation}
with $I_f=F(I Y-X_0)+F(-I Y-X_0)$. A careful analysis of the integral of $I_f$ gives:  $$ -2 X_0 + 2 Log(\sinh(\pi l_0))/\pi$$ for small values of $l_0$, so the last term dominates which finally leads   to:
%\begin{equation}
%I_f\sim \frac{1}{\pi} \tanh^{-1} \left (\frac{\sinh(\pi (i Y-X0))}{\sqrt{ \sinh(\pi ( %iY-X_0))^2-\sinh(\pi l_0)^2}}\right)-\frac{1}{\pi}\tanh^{-1} \left (\frac{\sinh(\pi (i Y+X0))}{\sqrt{ %\sinh(\pi ( iY+X_0))^2-\sinh(\pi l_0)^2}}\right)
% \end{equation}
\begin{equation}
  {\cal M}_{X=0}\sim  -\frac{1}{J\pi}(J-1)(J+1)^2 \epsilon Log(\sinh(\pi l_0))\,.
\end{equation}
Now let us evaluate  the contribution of the lower boundary at $X=d$\, where $\tilde J=J+\epsilon j_0$. The first and last terms contribute giving : $(J-1)d \epsilon j_0  -(J+1) d \epsilon j_0$ and then:
\begin{equation}
\label{MM}
{\cal M}_{X=d}= -2 d \epsilon j_0 \quad \mbox{so }\quad l_0\sim \frac{1}{\pi}e^{\left(\frac{2\pi d J j_0 }{(J-1)(J+1)^2}\right)}\,.
\end{equation} 
Since $l_0$ is a tiny quantity, the model is validated if $j_0<0$, that is, if the volumetric growth is greater at the bottom  than at the top. Noticing that values of $l_0$  of order $10^{-2}$ or $10^{-3}$ require both a negative jump value $j_0$ and a finite thickness for the sample  $d$, a graph representing the product of $j_0 \times d$, see Fig.(\ref{estimate})(c), shows that this product must be of the order of several units for suitable $l_0$ values, except for very low growth values: $J-1\sim 0.1$. In other words, thin shells $d\sim 1$ are more likely to reach low values of $l_0$.\\
\\

    In conclusion, our solution is based  on the determination of  $\Phi(Z)$,  which involves  2 parameters $ X_0$ and $l_0$. $X_0/J$ gives the position of the closest singularity of our doublet from the top  while $X_0$ indicates the position of the second singularity. $l_0$is a parameter that determines the outer solutions.  Since these two  constants are  relevant to the outer solution, they are   automatically eliminated in the boundary layer treatment which concerns the inner solutions as shown by Eq$(111)$, and so are not detected at this level.  To capture them, the path-independent integral treatment  is appropriate,  since this fancy technique  introduces the boundary conditions at the level of the whole domain.  In  fracture theory, the $J$ integral relates  the singular stress at the fracture tip  to the dimensions of the specimen while the M-integral gives  access to more complex singular fields and in particular to interfacial fractures. Here, the $X_0$ value is determined  by the $J$ integral, and can be slightly modified by growth inhomogeneity at large distances from the interface. The balance of the ${\cal M}$  integral is dominated by the two horizontal  boundaries: above  and below when the volumetric growth varies at both ends. This is due to the fact that the  ${\cal M}$ integrals associated with the singularities are subdominant, being of order $\epsilon^2$. They are evaluated in section \ref{JINTPROOF}. Obviously, introducing  growth heterogeneity  at the bottom  is the best way to fix $l_0$. 
One may wonder whether our results concerning for either $ \cal J$ or $\cal {M}$   remain valid when we add the growth heterogeneity. In fact, the initial axial growth makes the elastic model anisotropic : therefore, we check the validity of these approaches in section.(\ref{JINTPROOF}). In addition, we add local heterogeneity at the bottom.
 It  can  be shown that the method, which is  valid at order $\epsilon^2$ for constant volumetric growth,  remains valid only at order $\epsilon$, in the case of heterogeneity. This  is also the reason  why we assume that the growth jump  is localized at the bottom. 
 Finally, at this stage, $X_0$ and $l_0$  are completely determined by $J$ and $j_0$ and $\epsilon$. Since  $J$ is given  by the nonlinearity of the hyperelastic model, the only unknown is $\epsilon$. \\
Thus, in order to conveniently treat the two boundary layers, the neo-Hookean model must be  modified. A weak compressibility is required, as well as  a nonlinearity  of the elastic energy in $I_1^p$ and finally a variation of the growth in the far field.  Surprisingly, a case studied  by   Pandurangi {\it et al.} \cite{pandurangi2022nucleation} for a semi-infinite sample, consists in an elastic energy modeling that also includes compressibility and a  quadratic energy in  $I_2$. However, they also introduce a  graded material property in  the vertical direction, while our choice consists in a  graded growth in this direction. We can conclude that, although the two approaches are different, the physics of creases requires going  beyond the simple incompressible  neo-Hookean hyperelasticity. 
 %   solvabilitynewessay 
    \begin{figure*}	
 \centering 
    \begin{minipage}[b]{1\textwidth} 
    \subfigure[\,\bf{Critical Threshold} $\bm{J_d}$ \bf{versus the thickness}]{ 
         \includegraphics[width=.32\textwidth]{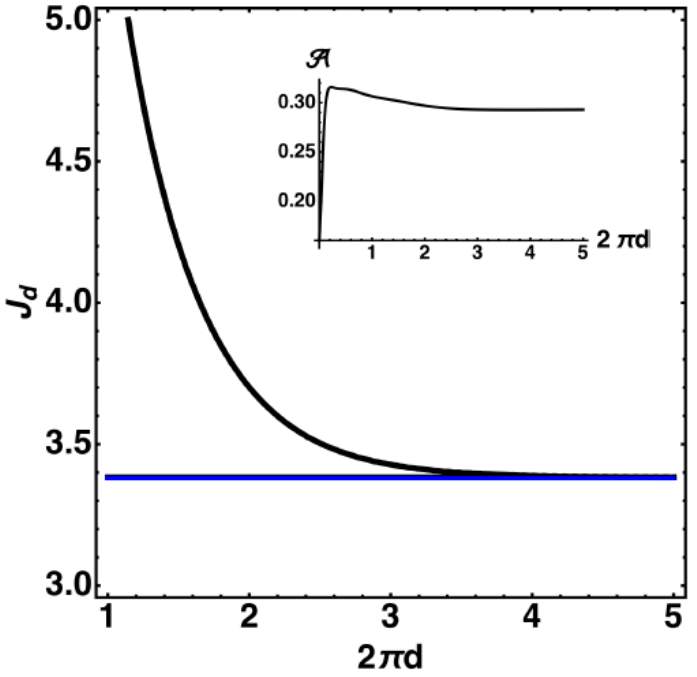} 
          } 
        \hspace{0in} 
        \subfigure[\, \bf{Singular Normalized Profile} ]{
            \includegraphics[width=0.3\textwidth]{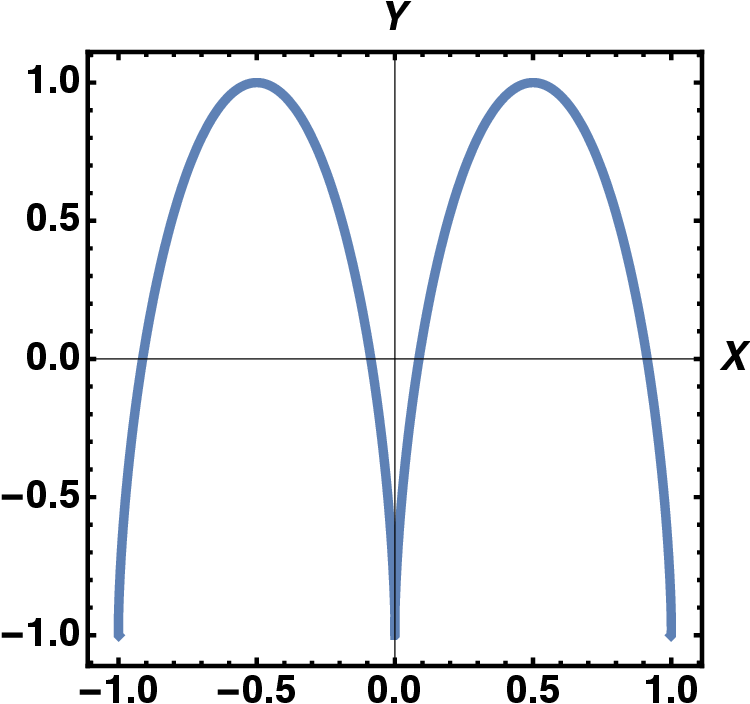} 
        } 
        \hspace{0in} 
        \subfigure[\, \bf{Regular and Nonregular Profiles}]{ 
                    \includegraphics[width=0.3\textwidth]{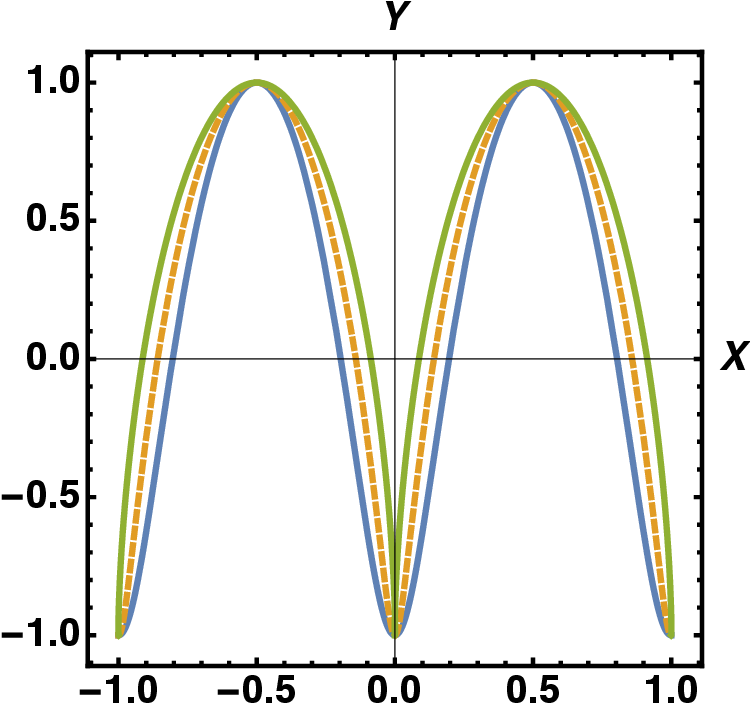} 
        } 
    \end{minipage}
    \caption{ The selected threshold $J_d $ as a function of  $2 \pi d $, where $d$ is the ratio between of thickness to pattern wavelength. For a thin film, the threshold increases dramatically as  $J_d\sim 4.93347/(2 \pi \Lambda d)$. When the thickness is of order or greater than   $\Lambda/\pi$,  $J_d$ reaches the asymptotic limit $J_d \sim J_{\infty}=J_b$, so  $3.3829$ represented by a solid blue line in (a).  In the inset, the critical amplitude  ${\cal A}$ defined in eq.(\ref{finite})  (in units of the wavelength for observing a crease with a single harmonic mode). A plateau of order $0.3$, is reached very quickly. In b) a normalized pattern  is plotted for a value of  ${\cal A}$ at the critical value $0.3$  where $d=t_3/\pi$ and $J_c$ the threshold value. The amplitude $x$ is divided by ${\cal A}$ for normalization. A cusp can be observed for $Y=0$, which repeats periodically for $ Y=n \pi$. In c) superposition of the profiles for increasing  amplitudes for ${\cal A}= 0.1$ in blue, $ {\cal A} =0.2$ and ${\cal A}= 0.3$.}
    \label{disp}
    \end{figure*}
\section{Finite-size effects or the buckling of layers}
\label{Finitesize}
One may wonder whether the degeneracy of the solutions presented above is not  due  either to the simplicity of the neo-Hookean model  or to the fact that the initial geometry is too  simple. It is obvious that a length scale is missing in our 
formulation, since  we  arbitrarily set the wavelength to unity. Consider the case of a gel layer whose height is given by the  parameter $d$. In order to keep as much as possible the same definitions and equations given in the previous section, we continue to use the wavelength as the unit of length. This situation was also  considered by Biot in 1963, but with different boundary conditions \cite{biot1963surface} and with a different point of view: the limit of small height $d$ compared to the wavelength  where the analogy with  Euler's buckling becomes more explicit \cite{biot1963surface}. Sinusoidal modes have been  found and the dispersion relation given the wavelength as a function of $d$ has been  given numerically. In this section, our aim is to revisit his results  with different boundary conditions at the substrate and considering $d>1$.
For a single layer, there is no need to change the main equations, just remember that boundary conditions have to be applied on both horizontal sides $X=0$ and now $X=d$ and during the growth, the layer can be free or glued at the bottom. In the first case, and unlike the second case, symmetry allows us to choose half of the sample with boundary conditions for $X=d/2$, which is now  the bottom. In any case, we will have strain conditions at the bottom and free stress conditions at the top: $X=0$. These two cases are physically similar and will only differ in numerical values. The two sets of boundary conditions to be applied either at the  top or at the bottom  are different in nature, and due to the finite extension of the layer, divergent solutions at infinity  are now relevant.  They were not allowed  and eliminated in the previous section.  The  non-symmetric case  adapts  more easily to bilayers and is therefore of more interest. This is especially true when the second layer is stiffer than the upper layer 
 (see \cite{amar2017mimicking}) 
 although  the case of a soft substrate is  more considered in the literature \cite{jiang2007finite,audoly2008buckling,sultan2008buckling,cao2012buckling}. Under growth, the description of the new positions $(x,y)$ can  follow the same perturbation  scheme  as before, but  must now include $2$ holomorphic functions: $\Phi_e(Z)$, (an even function of $Z$)  and $\Phi_o(Z)$
 (an odd  function of $Z$). 
 Then, by defining $\tilde Z=Z-d=X+I Y-d $ and 
$\tilde Z_1=Z_1-J d=J(X-d)+I Y$,
 the Euler-Lagrange equation associated with the incompressibility condition gives the following results for the deformations:
\begin{equation}
    \begin{cases}
x=J X -(J-1) d+J \epsilon \Re \,[F_1(\tilde Z)+F_2(\tilde Z_1)] \,,\\
y=Y- \epsilon \Im \,[F_1(\tilde Z)+J F_2(\tilde Z_1)]\,.\\ 
\end{cases}
\end{equation}
where
\begin{equation}
\begin{cases}
\label{prof}
F_1 = \Phi_e(\tilde Z )+a_1 \Phi_o(\tilde Z )\,,\\
F_2=b_1  \Phi_e (\tilde Z_1)+b_2 \Phi_o (\tilde Z_1)\,.
\end{cases}
\end{equation}
With this definition, the incompressibility condition  valid everywhere in the sample:
  \begin{equation}
  \frac{ \partial (x-J X)}{\partial X}+J  \frac{ \partial (y-Y)}{\partial Y} \quad  \forall \quad  X  \quad {\mbox{and}} \quad  Y\,,
 \end{equation}
is automatically checked  at first order in $\epsilon$. 
The boundary conditions of anchoring to the solid substrate impose: $x=d$ and $y=Y$ for $X=d$. Since $\Re \, \,[\Phi_o(I Y)]$ and $\Im \,[\Phi_e(I Y)]$ vanish independently of  the $Y$ values, the anchorage then imposes   $b_1=-1$ and $b_2=- a_1/J$.  It remains to apply the free stress conditions involving  $S_{11}$ and $S_{21}$ on the upper boundary (X=0) which must be verified for arbitrary $Y$ values.  Let us  limit  ourselves to  the harmonic modes.

 \subsection{Selection of a unique harmonic mode}
Selecting $\Phi_e(Z)=\cosh(2 \pi Z)$ and $\Phi_o(Z)=\sinh(2 \pi Z)$, to represent the current $(x,y)$ coordinates, the first order incremental  correction in $\epsilon$ becomes: $\delta x=J \cos(2 \pi Y) f_1(\tilde X,X_1)$  and $\delta y=\sin( 2 \pi Y) f_2(\tilde X, \tilde X_1)$, where $\tilde X=2 \pi (X-d)$ and $\tilde X_1=2 \pi J (X-d)$ and
\begin{equation} 
\begin{cases}%\begin{eqnarray}
f_1&=\cosh\tilde X-\cosh\tilde X_1+a_1 (\sinh\tilde X-\frac{ \sinh\tilde X_1}{J}) \,,\nonumber \\
f_2&= a_1 (\cosh \tilde X_1 - \cosh\tilde X)+J \sinh\tilde X_1 -\sinh\tilde X  \,.
\end{cases}
%\end{eqnarray}
\end{equation}
Now, considering the cancellation of the normal $S_{11}$ and of the shear stress $S_{21}$ at  the top of the strip, we  derive  the value of the coefficient $a_1$ in a first step:
  \begin{equation}
a_1= \frac{J (2 J \sinh(\tilde d) - (1 + J^2) \sinh(J\tilde d)}{
 2 J^2 \cosh(\tilde d ) - (1 + J^2) \cosh(J \tilde d)}\,,
\end{equation}
where $\tilde d=2 \pi d$. The  dispersion relation ${\cal D}$ gives the new threshold $J_{d}$ as a function of the ratio of wavenumber to thickness, so $2 \pi d$  is  the solution of a transcendental equation:
\begin{eqnarray}
{\cal D}&=-4 J_d^2(1+J_d^2) +  (1 +2 J_d^2 + 5 J_d^4 )\cosh(\tilde d) \cosh(J_d \tilde d) \,,\nonumber\\
&-J_d(1 +6 J_d^2 + J_d^4 )  \sinh(\tilde d) \sinh(J_d \tilde d)=0\,,
\label{ddispers}
\end{eqnarray}
 
\subsection{Nonlinearity and creasing above threshold for  growing layer}
Why  focus on a single harmonic mode? Each harmonic mode will correspond to $\cosh(2 \pi m Z)$ and $\sinh(2 \pi m Z)$ and  will have a different  threshold given by $J_{m d}$ as opposed  to  the  unique threshold independent of $m$ for an  infinite thickness, see section (\ref{nonlinearcoupling}). Thus,  we cannot simply combine different modes and evaluate the nonlinearities.  In fact,  nonlinear profiles do not result from  a single mode and traditional techniques become more difficult. Other asymptotic techniques consist of  the coupling mode approach of classical bifurcation theory such as  the Landau formalism of the amplitude equations  \cite{crawford1991introduction,cross1993pattern,meron2015nonlinear,charru2011hydrodynamic} 
but it remains complex or even impossible to use them with partial differential equations with  boundary conditions. One method, different from the present one, concerns the use of a  nonlinear stream function  introduced in \cite{amar2010swelling}, which exactly treats the incompressibility and reports the nonlinearities  on the elastic energy \cite{amar2013anisotropic,jia2013theoretical,amar2017mimicking}. This method, valid only in $2D$ geometry  or when the elastic deformations are reduced to a two-dimensional  space, has allowed to demonstrate  a third order of expansion: \cite{amar2010swelling} :
\begin{equation}
\label{invert}
 \begin{cases}
 x=J X -(J-1)d + \epsilon J \cos ( 2 \pi y)f_1\,, \\
Y=y -\epsilon \sin( 2 \pi y) f_2\,.
\end{cases}
\end{equation}
The parameter $\epsilon$ being predicted at the third order. Of course, this prediction depends on the size of the layer $d$. This formulation assumes that all inversion formulas  can be achieved, which is obviously  not the case when creases occur at the interface. They appear when $ \partial Y/\partial y$ vanishes for $X=0$  according to the theorem of the implicit function, (\cite{crawford1991introduction}), which gives the critical value ${\cal A}$, which is equal to $\epsilon J f_1$ of the deformation at the cusp position (see  inset in Fig.(\ref{disp}), panel (a)) and eq.(\ref{invert}). It can be noticed  that the required  amplitude saturates to a finite value around 0.3 for values $d$ of the sample width of the order of the wavelength or more, and that very thin samples, very easily  exhibit cusps as they grow,  although their threshold $J_d$ is obtained for a higher value. In Fig.(\ref{disp}), panel (b) we plot the profile of the cusp function over one period. It is divided by  ${\cal A}$ for normalization so that the amplitude varies between $-1$ and $1$.  To evaluate whether the amplitude ${\cal A}$ can be obtained in practice requires an  analytical treatment of the nonlinearities (which is not reported here, see Ref.(\cite{amar2017mimicking}). It approximates  the amplitude of the wavy regular pattern above the bifurcation threshold $J_d$  as  
\begin{equation}
\label{finite}
x \sim -(J-1)d\pm 0.537 \sqrt{J-J_d} \cos( 2 \pi Y)
\end{equation}
where the numerical number $0.537$ is analytically predicted  and  the zero order in eq.(\ref{finite}) indicates the increase in height due to growth, which appears negative due to our choice of  coordinate system.  The nonlinear treatment assumes a regular wave pattern and does not assume a priori  singularities as cusps. This  estimation must be compared with the amplitude ${\cal A}$ and results in  a growth parameter inducing a possible crease  given by ${\cal A}\sim 0.3=0.537 \sqrt{J-J_d}$, which implies a distance  from the threshold approximately equal to $0.3$, or  $10\%$ of the Biot threshold. Thus, for a  thickness of the order of the pattern wavelength: $d/\Lambda\sim 1$, creases appear rather quickly at the interface  once the threshold value is exceeded.
Although  nonlinearities can be responsible for  creases, they always appear above the Biot threshold and not below.

\subsection{Conclusion}
We have shown  that the sinusoidal Biot  profile is not the only solution that occurs in a growing hyperelastic sample.  Restricting to  the simplest configuration of a semi-infinite two-dimensional  neo-Hookean  sample, growing with an isotropic constant growth rate $J$, we show that other candidates are possible solutions that appear  exactly  at the same threshold.  Among them, quasi-singular solutions with a periodic array of cusps can be found, at  the Biot threshold. Nonlinearities can be evaluated by classical nonlinear treatments, supercritical bifurcations are rather common, but also subcritical bifurcations can appear slightly below the Biot threshold when several harmonics are coupled. This explains the diversity of experimental observations independent of the elasticity model, since this diversity occurs at the level of the simplest growth formalism. Independent of these patterns, which are  always related to the Biot threshold,  it has been suggested that patterns can occur well below the Biot threshold if local singularities also occur within the material.  We consider this conjecture and show that it can be  the source of new families of solutions.  In this case, tiny linear singularities occur through pairs near  (but not at ) the interface where the compressive elastic  field is concentrated. The high level of stress generated  requires a slight local modification of the elastic model. Relevant parameters such as the linear extension of the singularities and their positions are recovered by path-independent integrals. In addition, this study proposes a threshold value for the volumetric growth  below the Biot threshold determined by the nonlinearities above the neo-Hookean approach.
\section{Acknowledgements}
I  would like to thank Linda Cummings, Darren Crowdy and Saleh Tanveer for insightful discussions during the  programme "Complex analysis: techniques, applications and computations" (Fall 2019, July 2023) of  the Isaac Newton Institute of Mathematical Sciences, Cambridge,  for its support and hospitality. I acknowledge the support of the ANR (Agence Nationale de la Recherche) under the contract MecaTiss (ANR-17-CE30-0007) and the contract EpiMorph (ANR-2018-CE13-0008).

\section{Appendix}
\label{supplement}
\setcounter{equation}{0}
\renewcommand{\theequation}{S\arabic{equation}}

\subsection{Nonlinear elasticity at first order: stress and energy expansion }
\label{appendexp} 
This appendix is written with  the elastic fields given by a complex function. It is written in the initial frame of coordinates and the stress corresponds to the Piola stress tensor\cite{ogden1997non}.

\begin{eqnarray}
x&=&J X+\epsilon J \Re \, \,[\Phi(Z)+ \tau_1\Phi(Z_1)]\,, {\nonumber}\\
y&=& Y-\epsilon \Im \,[\Phi(Z)+ \tau_1 J\Phi(Z_1)]\,.
\end{eqnarray}
In this work, we restrict to $\Phi=\bar \Phi$ or $\Phi$ function having a real expansion in $Z$. The Jacobian in $2 D$ is given by $I_3=x_X \cdot y_Y-x_Y\cdot y_X-J$.  At linear order it reads:

\begin{eqnarray}
\label{sim}
x_X&=&J +  \epsilon J \Re \, \,[\Phi_Z+J \tau_1 \Phi_{Z_1}] \,, {\nonumber}\\
y_Y&=& Y- \epsilon \Re \, \,[\Phi_Z+ J \tau_1  \Phi_{Z_1}] \,, {\nonumber}\\
x_Y&=&  -\epsilon J  \Im \,[\Phi_Z+ \tau_1 \Phi_{Z_1}] \,,{\nonumber}\\
y_X&=&-I \epsilon \Im \,[\Phi_Z+J^2 \tau_1  \Phi_{Z_1}]\,.
\end{eqnarray}

With this choice for the deformation field, we can verify that at linear order in $\epsilon$,  $I_3=0$ as required.

The Euler-Lagrange equation for the Neo-Hookean model are  given by:
\begin{equation}
\label{Laplace}
\Delta x =x_{XX}+x_{YY}=Q_X;\,\,\Delta y =y_{XX}+y_{YY}=J Q_Y\,.
\end{equation}
Obviously, only the term depending on $Z_1$ and $\bar Z_1$ are concerned by these equations which implies that  the Lagrange parameter $Q$ is only a function of   $Z_1$ and
 $\bar Z_1$. Both equations of eq.(\ref{Laplace}) determines $Q$ as:
 \begin{equation}
 \label{Lagrange}
 Q=J+\frac{1}{2} \epsilon \tau_0\tau_1 \left\{ \Phi_{Z_1}+ \tilde \Phi_{Z_1}\right\}
 \end{equation}
 It remains to check that the boundary verifies the  cancellation of  the shear $S_{21}$ and  normal stress $S_{11}$  for the free surface $X=0$.
Using  eq. (\ref{sim}), we first obtain  the components of the Piola stress tensor. For the diagonal elements:\\
$S_{11}  =  x_X-Q y_Y$ and for $S_{22}  =  y_Y-Q x_X$: 
\begin{equation}
\label{stressbis}
\begin{cases}
S_{11}&=\epsilon \Re \,[ 2 J \Phi_{ Z}+ (1+J^2) \tau_1 \Phi_{ Z_1}] \,,\\
S_{22}&= -\tau_0-\epsilon \Re \,[ (1+J^2)\Phi_{ Z}+ 2 J^3 \tau_1 \Phi_{ Z_1}] \,.
\end{cases}
\end{equation}
\\
For  the off-diagonal elements, we have:\\
\\
$S_{21}  =  y_X+Q x_Y$ and  $S_{12}  =  x_Y+Q y_X$
which reads:
\begin{equation}
\begin{cases}
S_{21} & =- \epsilon  (1+J^2) \Im \,[\Phi_{ Z}+ 2 J^2 \tau_1 \Phi_{ Z_1}],\\
S_{12} & 
=- \epsilon \Im \,[ 2 J \Phi_{ Z}+ (1+J^2) \tau_1 \Phi_{ Z_1}].
\end{cases}
\end{equation}
\\
Note that, for $X=0$ so for $Z=I Y$,
 the normal stresses on the top stress become:
 \begin{equation}
 \label{sxx}
 S_{11}=\frac{1}{2} (2 J+(1+J^2) \tau_1) (\Phi_{ Z}+ \bar \Phi_{\bar Z})\,,
\end{equation}
 \begin{equation}
 \label{syx}
 S_{21}=\frac{I}{2} (1+J^2+ 2 J^2  \tau_1) (\Phi_{ Z}- \bar \Phi_{\bar Z})\,.
\end{equation}
 The boundary conditions impose $S_{11}=S_{21}=0$  but If $\Phi(Z)$ is an even function of $Z$, the derivative  $\Phi'(Z)$ is odd, so  $S_{11}$ vanishes automatically, and we only need to choose $\tau_1=-(1+J^2)/(2 J^2)$ to cancel $S_{21}$. If $\Phi(Z)$ is an odd function of $Z$, then $S_{21}$ cancels  automatically   and $S_{11}=0$ imposes  $ \tau_2=-2 J/(1+J^2)$. Our choice in this manuscript  corresponds to the first case. The Biot solution $\Phi(Z)=e^{-2 \pi Z}$ has no parity so both normal stress components  must vanish at the top of the strip $X=0$, which explains the existence of the threshold $J=J_B$. Regardless of the choice of $\tau_1$ or $\tau_2$, the threshold value $J_B$ is identical.
\\
\subsection{Expansion of the elastic and capillary energy density}
\label{develop}
Expansion of the  the elastic energy density  $E$ given by eq.(\ref{energy})  at third order of the parameter $\epsilon$: 
 $E=E_0+ \epsilon E_1+\epsilon^2 E_2+\epsilon^3 E_3$, reads:
\begin{equation}
\label{densityel}
  \begin{cases}
E_0 =\frac{1}{2}(J-1)^2\,,\\
\\
E_1 = \tau_0 \Re \,[\Phi_Z+J \tau_1 \Phi_{Z_1}],\\
\\
E_2= \frac{1}{2}(3 J^2+1)( \vert \Phi_Z \vert^2+ J^2 \tau_1^2 \vert\Phi_{Z_1}\vert^2)\\ 
+\frac{J \tau_1}{2} \Re \,[(J+1)^3\Phi_Z  \Phi_{\bar Z_1}-(J-1)^3\Phi_{Z}  \Phi_{ {Z_1}}])\,,\\
\\
E_3/(J\tau_0\tau_1)=\Re \,[\Phi_{Z_1}] \times (\vert \Phi_Z\vert^2 +J^2 \tau_1^2   \vert\Phi_{Z_1}\vert^2\\
+\frac{\tau_1}{2}\Re \,[(J+1)^2\Phi_Z \Phi_{\bar Z_1}-(J-1)^2\Phi_Z \Phi_{Z_1}])\,.
\end{cases}
\end{equation}
 
Note that if $J=1$  (no growth) all the coefficients $E_i$ vanish  whatever the function $\Phi(Z)$. 
\begin{figure*}
 \centering 
    \begin{minipage}[b]{1\textwidth}  
    \subfigure[\,\,\,$\bm{e_3}$ \bf{coefficient}]{ 
         \includegraphics[width=.235\textwidth]{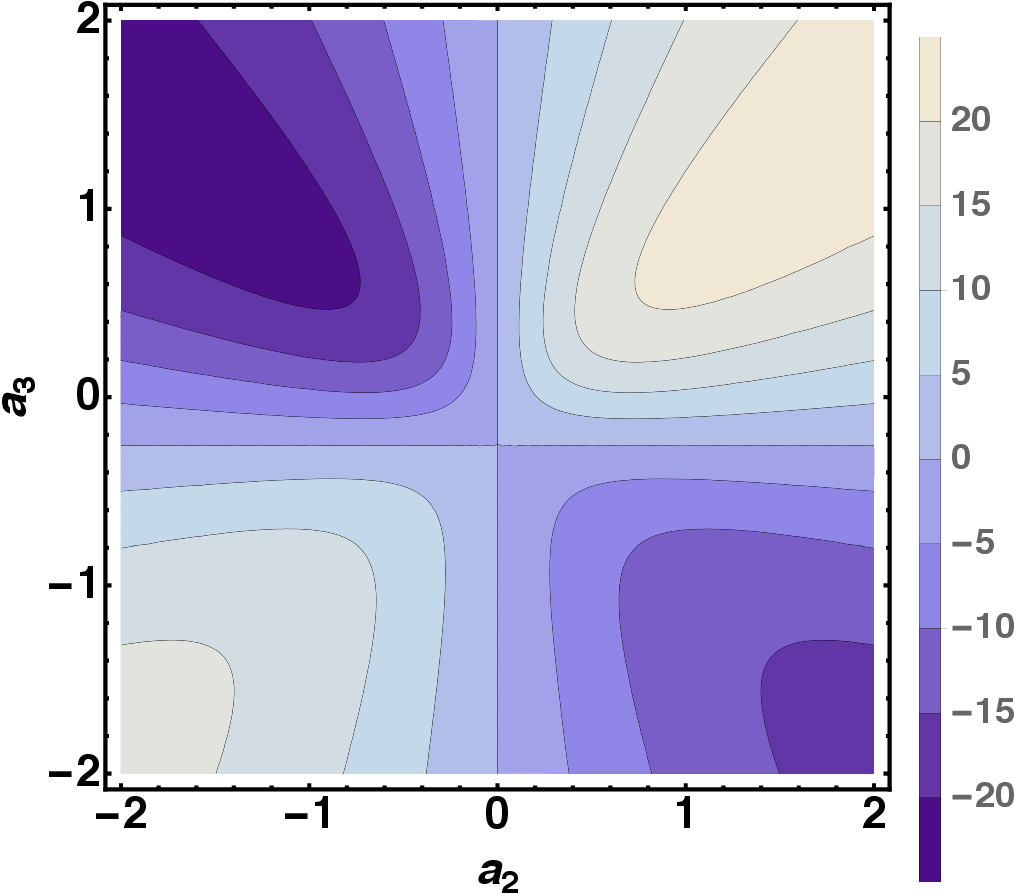} 
          } 
        \hspace{0in} 
        \subfigure[\,\,\, $\bm{g_2}$ \bf{coeffient}]{
            \includegraphics[width=0.235\textwidth]{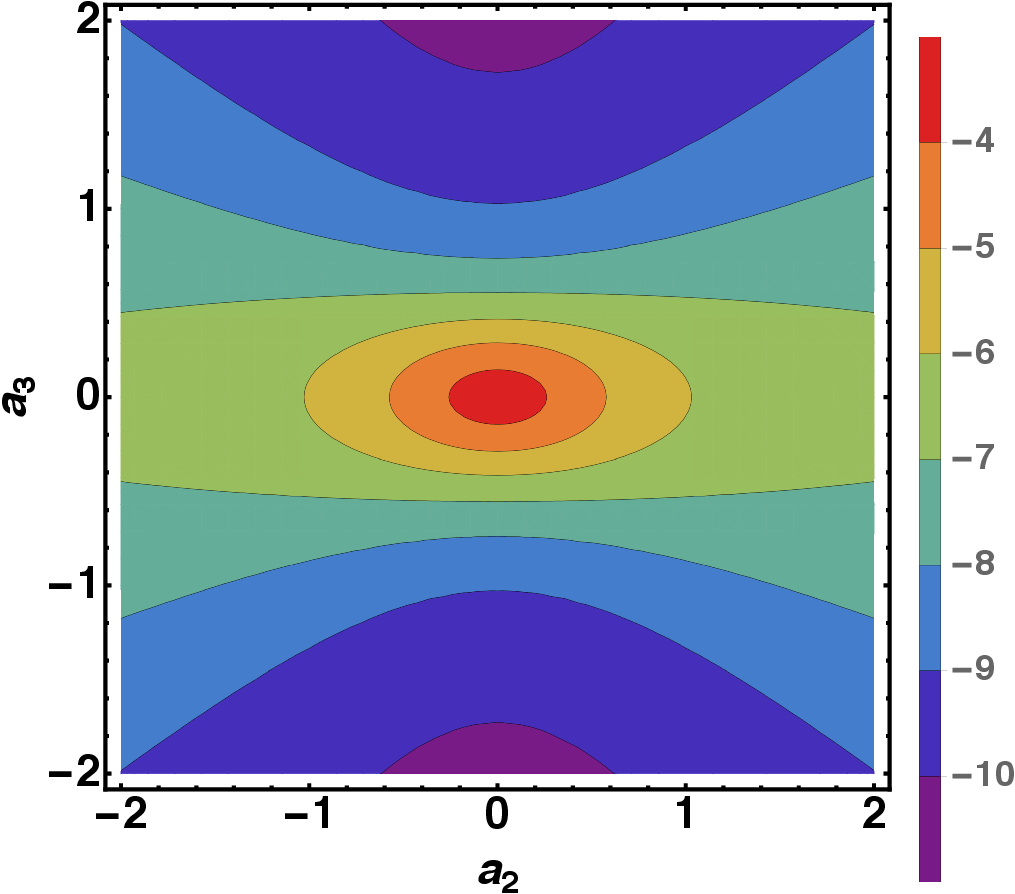} 
        } 
  \hspace{0in}   
    \subfigure[\,\,\,$\bm{g_3}$ \bf{coefficient}]{ 
         \includegraphics[width=.23\textwidth]{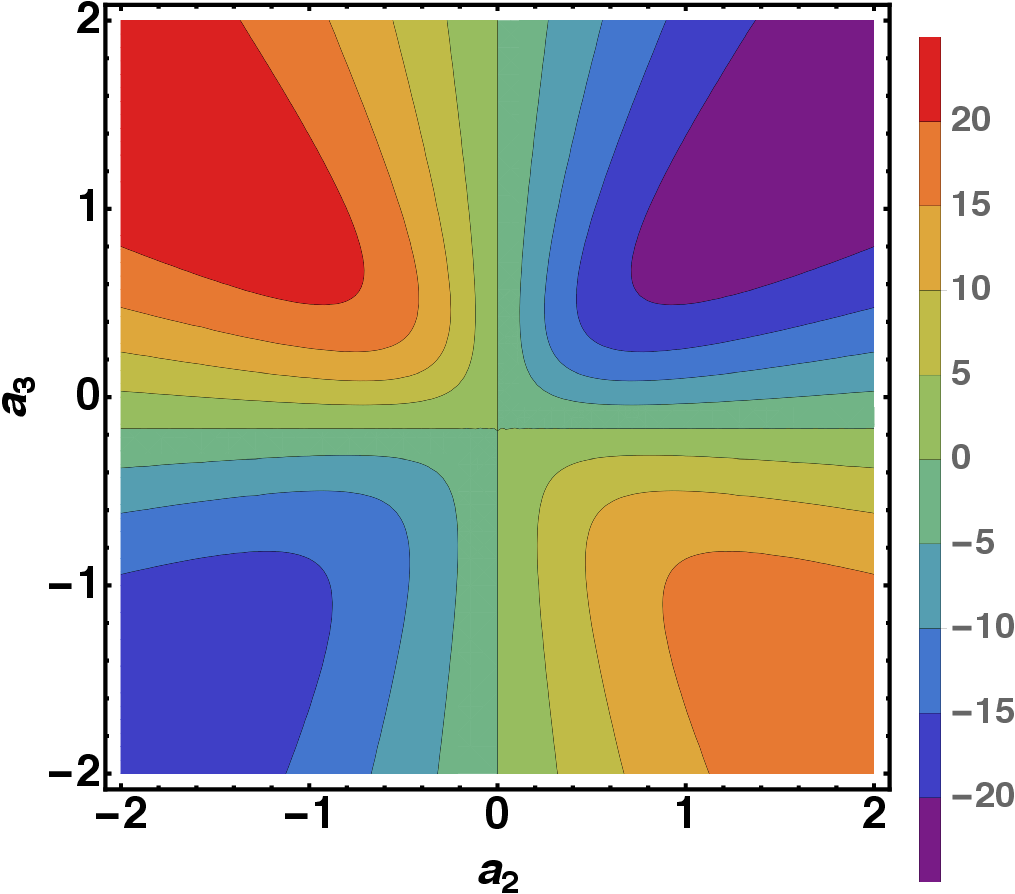} 
          } 
      \hspace{0in} 
        \subfigure[\,\,\, $\bm{g_4}$ \bf{coefficient}]{
          \includegraphics[width=0.23\textwidth]{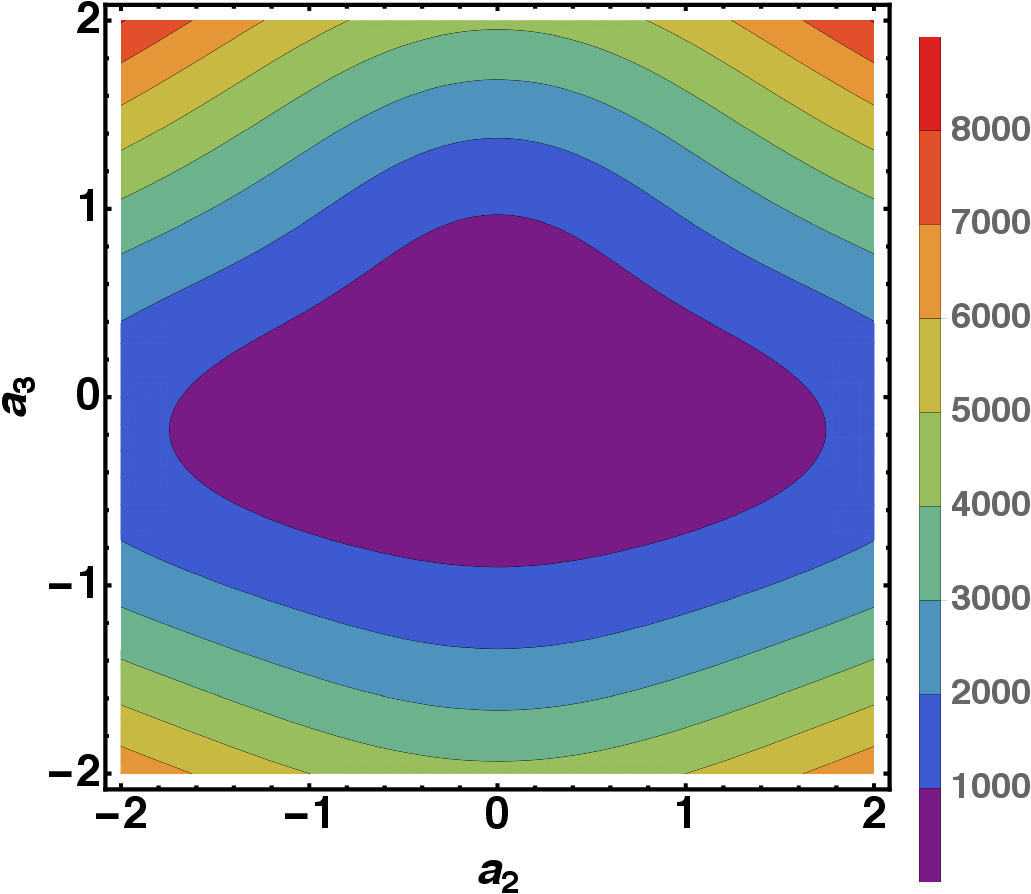} 
       } 
          \hspace{10in} 
          \end{minipage}
   \hspace{10in} 
   \subfigure[\, \bf{Profile for an outer Logarithmic singularity} ]{ 
            \includegraphics[width=0.33\textwidth]{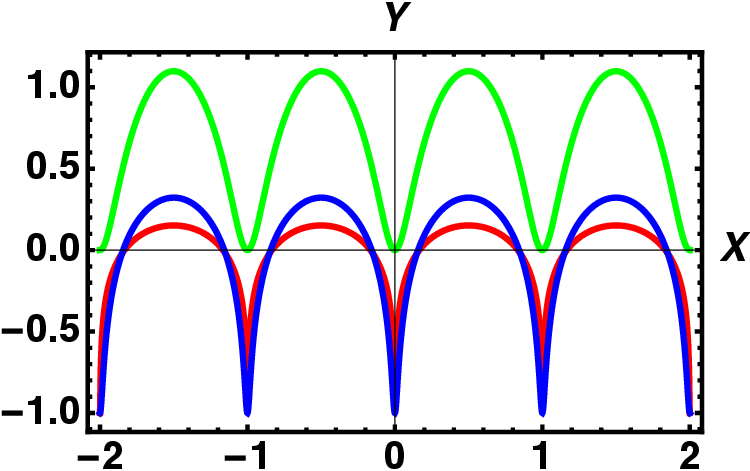} 
        } 
        \hspace{0in} 
        \subfigure[\,\bf{Real Part}]{ 
            \includegraphics[width=0.3\textwidth]{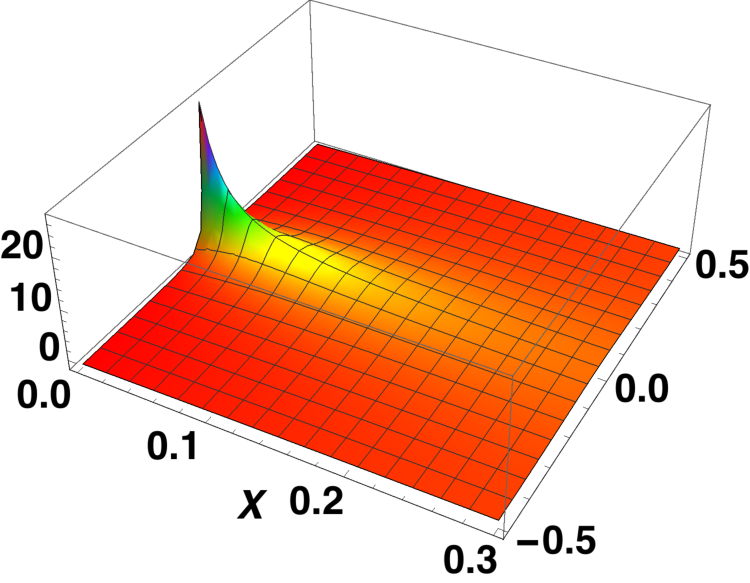} 
        } 
        \hspace{0in} 
        \subfigure[\,\bf{Imaginary part}]{ 
            \includegraphics[width=0.3\textwidth]{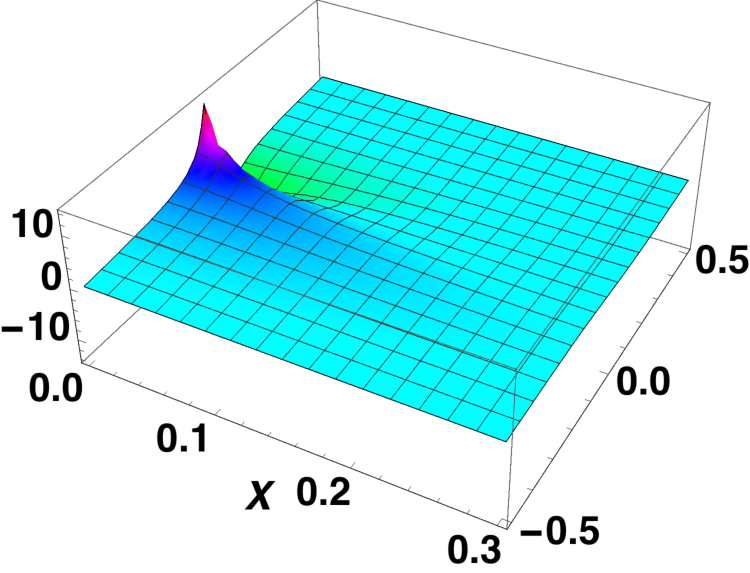} 
       }
\hspace{10in} 
\subfigure[\,\bf{Profile for an outer square root singularity}]{ 
            \includegraphics[width=0.33\textwidth]{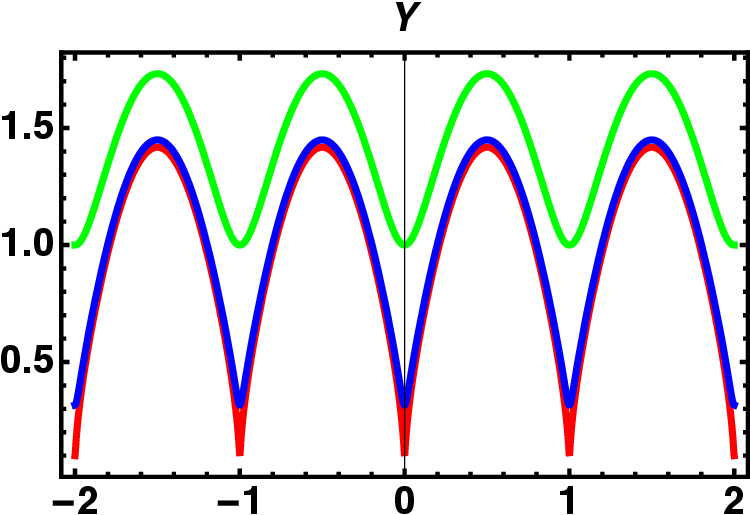} 
        } 
        \hspace{0in} 
        \subfigure[\,\bf{Real Part}  ]{ 
            \includegraphics[width=0.3\textwidth]{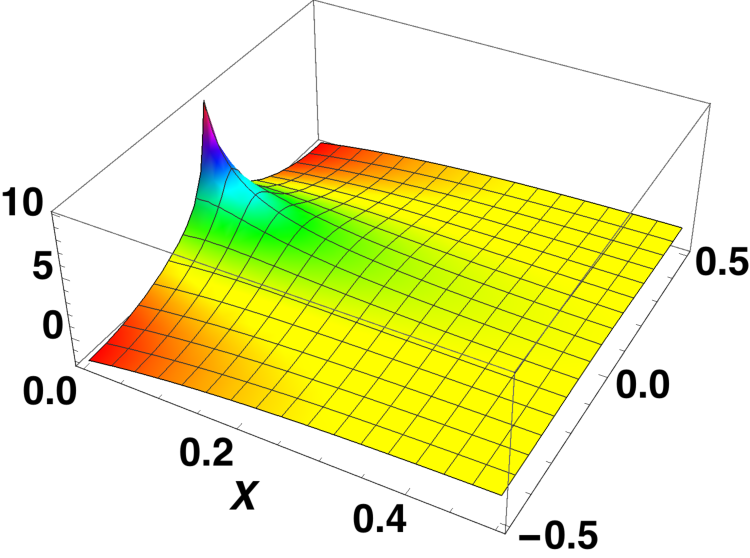} 
        } 
        \hspace{0in} 
        \subfigure[\,\bf{Imaginary part}]{ 
            \includegraphics[width=0.3\textwidth]{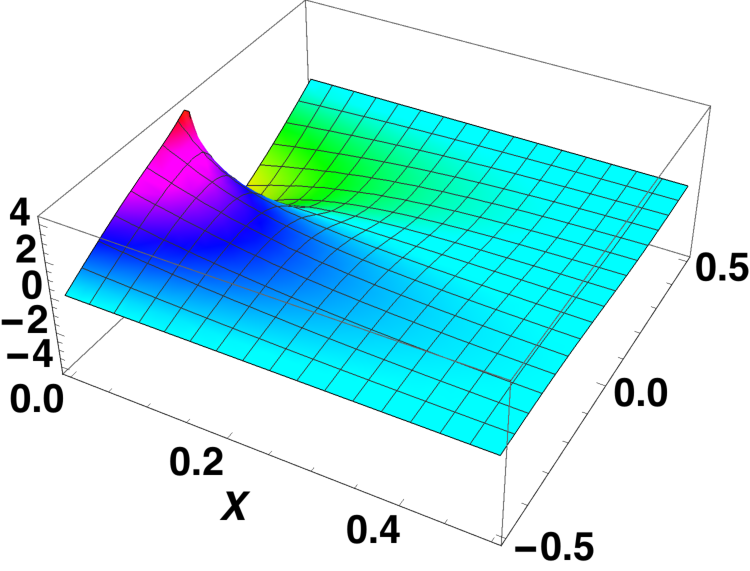} 
        } 
    \caption{In (a) and (b) density plots of the coefficients $e_3$ and $g_2$ entering the expansion of the free energy eq.(\ref{normalform}) as a function of $a_2$ and $a_3$. The numerical values can be estimated from  the legend to the right of each panel. In (c) and (d) Density plots of the coefficients $g_3$ and $g_4$ entering the expansion of the free energy eq.(\ref{normalform}) as a function of $a_2$ and $a_3$.\\
   In   (e) the interface profile for a logarithmic singularity according to   eq.(\ref{singular}) with the parameter $a=0.01$ for the red curve, $a=0.1$ for the blue curve, $a=1.$ for the green curve. For $a=1$ the definition is changed and the plot only concerns $\Re \,[\Phi(\zeta_a)]$, while for $a=0.01$ or $a=0.1$, it is given by eq.(\ref{singular}). In the second row, in  (h), the interface profile for a square root  singularity according to   eq.(\ref{singular}) with the parameter $a=0.01$ for the red curve, $a=0.1$ for the blue curve, $a=1.$ for the green curve.  In (f,i),  the real part of the  $\Phi(Z)$ derivative contributes to the stress, see $S_{11}$, eq.(\ref{stressbis}). In (i,j)  the imaginary part is shown, it goes  into the formula for  $S_{21}$.  One wavelength  ${\mathcal{P}} $  is shown and the rainbow colors indicate the sharp variations near  the origin $\mathcal{O}$. Note also that the scale for $X$ varies from $0$ to $0.3$ for (f,g), and from $0$ to $0.5$, for (i,j). Although quite singular near $\mathcal{O}$, the stress remains finite and goes to zero very quickly. } 
      \label{EC3EC4}
    \end{figure*}
    
    The capillary energy density evaluated at $X=0$ so for $Z=I Y$ can be expanded up to forth order:
    \begin{equation}
  {\cal E}_{c}=\gamma_0 \epsilon (E_{1c}+\epsilon E_{2c}+\epsilon^2 E_{3c}+\epsilon^3 E_{4c})\,,
  \end{equation} 
where   $\gamma_0 $ is the ratio between the capillary energy and the shear modulus multiplied by the wavelength and
where:
\begin{equation}
    \label{densitycap}
  \begin{cases}  
E_{1c}=\frac{(J-1)^2}{2 J} \Re \,[\Phi_Z] \,,\\
\\
E_{2c}=\frac{\tau_0^2}{8 J^2} (\Im \, [\Phi_Z ])^2\,, \\
\\
E_{3c}=-\frac{\tau_0^2 (J-1)^2}{16J^3} \Im \, [\Phi_Z ]^2 \Re \,[\Phi_Z] \,,\\
\\
E_{4c}=\frac{\tau_0^2(J-1)^2}{128 J^4}\Im \, [\Phi_Z]^2 \times\\
  \{(5-6J+5 J^2) \Re \, [\Phi_Z]^2 -(J+1)^2 \vert\Phi_Z\vert^2 \}\,.
    \end{cases}
   \end{equation}
 \\
 \\
 
 \subsection{Evaluation of the total energy for a single mode, double and triple  mode}
\label{evaluation}
We consider first a  single mode: $\zeta=e^{-2 \pi Z}$. in this case, only $E_{2c}$ and $E_{4c}$ and $E_{6c}$  contribute to the capillary energy $E_{cs}$. After integrating  on the top interface and  defining the following quantities:
\begin{equation}
\label{coefec}
\begin{cases}
\alpha_c=\left(\frac{(J-1)\pi}{2 J}\right)^2; E_{s0}= \gamma_0 \alpha_c \epsilon^2 (1+J)^2,\\
\\
{\cal Q}_{1c}=\frac{1}{4}( J^2-14 J+1),\\ 
\\
Q_{2c}=\frac{1}{4}(1 - 12 J + 102 J^2 - 12 J^3 + J^4).
\end{cases}
\end{equation}
it reads:
\begin{equation}
 {\cal E}_{cs} =E_{s0}\left(1+ \epsilon^2 \alpha_c {\cal Q}_{1c}+ \epsilon^4 \alpha_c^2 {\cal Q}_{2c}\right)
\label{onemode}
\end{equation}
If the base state includes other harmonics such as $\zeta^2$ and $\zeta^3$   with decreasing amplitudes as in section \ref{supersub}, $\phi[Z]=\zeta +\epsilon B_2 \zeta^2-\epsilon^2 B_3 \zeta^3 $,  the capillary energy includes  ${\cal E}_{cs}$ but also  additive terms in $\epsilon^4$ and $\epsilon^6$:
\begin{equation}
 {\cal E}_{c}={\cal E}_{cs}+E_{s0} \epsilon^2 \left(e_{1c}+
+\epsilon^2 e_{2c}\right).
\end{equation}
\begin{equation}
    \begin{cases}
e_{1c}=B_2(4 B_2 + \pi (J-1)^2/J),\\
e_{2c}=9 B_3^2 -\frac{3\alpha_c}{\pi} B_3 (8 J  B_2 +\pi(1 + J)^2 )\\
+4\alpha_c Q_{1c} e_{1c}.
\label{B2cap}
\end{cases}
\end{equation}

We now consider the 3 mode coupling  with $\Phi(Z)= \zeta + a_2 \zeta^2 +a_3 \zeta^3$ of the section \ref{threemodes}. The goal is to evaluate the weight of each term in the expansion of the elastic energy, eq.(\ref{3modes}), and to compare it with the capillary energy. 
Writing the total energy: the elastic plus the capillary energy, we obtain for ${\cal E}_{t}= -E_f\epsilon^2 {\cal \tilde E}_t$ (see eq.(\ref{3modes})) where: 
\begin{equation}
\label{normalform}
\begin{cases}
E_f=\pi {\cal Q}_2 (1+2 a_2^2 +3 a_3^2)\,,\\
  {\cal \tilde E}_{t}= \delta J + 2 \gamma_0g_2+(e_3+2 \gamma_0 g_3)\epsilon+2 \gamma_0 g_4 \epsilon^2. 
\end{cases}
\end{equation}
where ${\cal Q}_2$ has been given in eq.(\ref{calQ}), $e_3$  in eq.(\ref{bifone}) and $e_4=0$:
\begin{equation}
\label{normalform1}
e_3=\pi^2 a_2 {\cal Q}_3 (1+a_3 {\cal Q}_{33})/E_f;\,g_i=-E_{ic}/E_f\\
\end{equation}
We first define:
\begin{equation}
 f_g=-\frac{ (-1 + J)^2 (1 + J)}{(1 + 2 a2^2 + 
    3 a3^2) (3 J^2 -6 J -1)}
\end{equation}
\begin{equation}
    \begin{cases}
        g_2=(1+4 a_2^2+9 a_3^2)\frac{J \pi f_g}{ (J-1)^2}, \\
 g_3= a_2(1+6 a_3)\pi^2 f_g, \\
 g_4=\frac{ \pi^3 f_g}{4 J} \{
  3 a_3 (1 + J)^2 +Q_{1c}\times(1 + 36 a_3^2   \\
  + 81 a_3^4 
      +16 a_2^2 (1 +a_2^2+3 a_3 + 9 a_3^2)  \}
    \end{cases}
    \end{equation}
$E_f$ is positive and all polynomials  must be evaluated for $J=J_B$. The order of magnitude of each coefficient  as $e_3, g_2, g_3, g_4$  is   function of the two  parameters $a_2$ and $a_3$, once $J=J_B$ is imposed and is represented by density plots (see Fig.(\ref{EC3EC4})).  More specifically, Fig.(\ref{EC3EC4}(a) gives the order of magnitude of $e_3$  while panel (b) gives the $g_2$ coefficient as function of the same quantities. $g_2$ is negative and so is responsible for  shifting  the threshold towards higher values, a similar result was found in \cite{amar2010swelling}.  Fig.(\ref{EC3EC4}c) and  Fig.(\ref{EC3EC4}d) are dedicated respectively to $g_3$ and $g_4$ respectively. Note that all $g_i$ must be multiplied by the capillary number $\gamma_0$ and only $g_4$ appears alone in the asymptotics eq.(\ref{normalform}).
   \subsection{Profiles and Cartography of the stress}
 \label{appenprofile}
 The profiles and the stress of both a logarithmic singularity and of  a  square root singularity, localized outside the elastic sample, are displayed in 
 Fig. (\ref{EC3EC4}). 
Panels (e,f,g) are devoted  to $\Phi(Z) = -Log 1+a-e^{-2  \pi  Z}/Log(a)$ for $a=0.01, 0.1$ and  to $\Phi(Z) = Log (2-e^{-2 I \pi  Y})$ for $a=1$. This corresponds to singularities in the physical plane located to a distance $d_a=-0.00158,-0.01516,-0.110318$, outside the elastic sample. The  panels (h,i,j) concern the square root singularity:$ \sqrt{e^{-2  \pi  Z}- 1 - a}$.
 An indication of the normal stress $S_{11}$,  which includes the real part of   $\Phi_Z$, and the shear stress $S_{21}$, which includes the imaginary part are also shown in panels (f,i) and (g,j) of 
 Fig.(\ref{EC3EC4}). The stress contributes significantly at the boundary, but decreases very rapidly.
 $S_{11}$ and $S_{21}$ are given by eq.(\ref{stressbis}).

\subsection{ Weakly nonlinear analysis for quasi-singular profiles}
\label{quasisingular}
These profiles have been discussed in section (\ref{quasisinguar}. A full exhaustive study cannot be achieved and it iis not certain that the method of weakly nonlinear analysis will converge for these choices of quasi-singularity. The evaluation of the integrals is not easy in $X$ and $Y$ geometry and  and we choose a particular mode:  $\phi(Z)=-(Log(1+a -e^{-2 \pi Z}) -Log(1+a))/Log(a)$. Obviously, an expansion in sinusoidal modes: $\zeta^p$ have little chance to converge quickly so we have to modify our strategy.
The algebra is more complex even with formal mathematical software and we cannot  find in handbooks of integrals the result for the previously defined integrals that go into the energy expansion. Thus, the results cannot  be obtained  completely  analytically without approximations. It is also  not possible, except for $L_3$, to integrate in the complex plane of $\zeta$ or the physical plane $\Omega$ due to the juxtaposition of $Z$ and $Z_1$. 
Let us give an estimate. A first integration on $Y$ induces  the evaluation of two kinds of integrals, ${\cal L}_{m,n}$ :

\begin{equation}
\label{true}
{\cal L}_{m,n}=
 \int_{0}^{\infty}\frac{e^{-2 \pi (m+n) u} du}{\left(\alpha^2 -e^{-2 \pi m u}\right) \left(\alpha^2 -e^{-2 \pi n u}\right)}\,,
  \end{equation}
   where $m$ and $n$ are positive integers,$\alpha=a+1$. Using Watson's lemma \cite{fokas1997complex}  and the fact that the denominator is singular  near $X=0$ for vanishing $a$ values, once the parameter $l_m$ is introduced with  $l_m=a(2+a)/(2m\pi)$, we approximate the denominator by linear expansion  giving  
\begin{equation}
{\cal L}_{m,n} \sim \int_{0}^{\infty} \frac{du}{p_a} e^{-2 \pi (m+n) u} 
\left( \frac{1 }{l_m+ u}-\frac{1}{l_n+ u}\right) \,,\\
\end{equation}
with $p_a=2 \pi a (2+a) (m-n)$.  The two last integrals can be computed explicitly with  the function exponential integrals  $(E_i)$  and the limit   $l_m->0$  finally gives  the following result:
 \begin{equation}
 \label{approxlmn}
{\cal L}_{m,n} \sim \frac{1 }{ 4 \pi a} l_{m,n}+O(a)\,\, \mbox{with}\,\,  l_{m,n}=\frac{Log(m/n)}{m-n} \,.
\end{equation}
For $m=n$, $l_{n,n}=1/n$. Then, taking into account only the contribution in $ a^{-1}$ and defining $m_a=-2\pi^2/(a Log(a)^3)$, we get:
\begin{equation}
\label{approx2}
\begin{cases}
L_1&\sim m_a l_{2,1+J}=\frac{m_a}{1-J}Log\frac{2}{1+J}\,,\\
L_2&= m_a l_{2J,2J}=\frac{m_a}{2J}\,,\\
L_4&\sim \frac{m_a}{2} l_{2J,1+J}=\frac{m_a}{2(J-1)}Log\frac{2J}{1+J}\,,\\
L_3 &\sim \frac{m_a}{2}(l_{1+J,1+J}+l_{2J,1+J})\\
& \sim\frac{m_a}{2}\left (\frac{1}{1+J}+\frac{Log (2J/(1+J))}{J-1}\right)\,.\\ 
\end{cases}
\end{equation}
A comparison between numerical values of integrals (see eq.(\ref{true}) and the estimate given by eq.(\ref{approxlmn}) is correct for $a\sim 0.01$, but smaller values are necessary for eq.(\ref{approx2}). This treatment can always be done numerically, the advantage here is to find the scaling of ${\cal E}_3$.
For the logarithmic choice, we finally derive:
\begin{equation}
\label{Eonefive}
\begin{cases}
\Pi_1 \sim  4 J^2 (J+1)^2 Log \left(\frac{2}{J+1}\right)\\
 +\left(J^2+1\right) \left(4 J^2 Log (J)+(J-1)^2\right)\,,
  \end{cases} 
  \end{equation} 
and the third order correction gives:
\begin{equation}
\label{energquasi}
    {\cal E}_3=-\frac{ (J+1)}{8J^2} \Pi_1 \tau_1 m_a \simeq -\frac{38.38}{a Log(a)^3}\,.
\end{equation}

%%%%%%%%%%%%%%%%%%%%%%%%%%%%%%%%%%%%%%%%%%%%w<
\subsection{Path-independent integrals}
\label{JINTPROOF}
It may seem pointless to demonstrate that conservation laws remain valid in  incremental models of  finite elasticity   after the pioneering work  of Knowles and Sternberg \cite{knowles1971class}, but there are some slight differences with our approach. Fist, we are concerned with growth so our initial state is an anisotropic  (axially pre-stretched)  state. Second, our demonstration is achieved at the second order in $\epsilon$ and not at first order. For this reason, if we can predict the result for the $J$ integral, it is less obvious for the $M$ integral, which is not believed  to be valid in nonlinear elasticity. 
The strategy  to prove  the existence of path-independent integrals is simple: it consists in relating a scalar or a vector to a vector or a tensor which is  divergence free. In $2$ dimensions, we shall  demonstrate that it is the case for $\cal J$ and $\cal M$. First let us begin by ${\cal J}$
  \begin{equation}
{\cal  J}_i=\iint dS \left \{\frac{\partial  (E \delta_{ik})}{\partial X_k}- \frac{\partial (S_{jk}F_{ji})}{\partial X_k} \right\}\,.
\end{equation}
where we use the Einstein convention for repeated indices and $dS=dXdY$ and the brackets represent ${\cal T}_i$ The index $i$ reminds us that the $J$ integral is indeed a vector but here only the component $\vec {\bf e}_X$ is important. $\bf F$ is the gradient of the deformation tensor, whose components are: $F_{ij}=\partial x_i/\partial X_j$. It can be replaced by $\bf U$ so that  $U_{ij}=F_{ij}$ if $i \ne j$ and $U_{11}=F_{11}- J$ and $U_{22}=F_{22}-1$. We want  to show that   ${\cal T}_i$ is a divergence. In $2$ dimensions, we have :
\begin{equation}
\label{Juno}
{\cal T}_1= \frac{\partial  E}{\partial X}- \frac{\partial (S_{11} U_{11}+S_{21} U_{21})}{\partial X}- \frac{\partial (S_{12} U_{11}+S_{22} U_{21})}{\partial Y}\,.
\end{equation}
where $E$ has been given in eq.(\ref{energy}), $ \bf S$ is the Piola stress tensor  already defined  in eq.(\ref{stress}). At equilibrium, the divergence of the Piola stress tensor cancels:
\begin{equation}
\frac{\partial S_{11}}{\partial X}+\frac{\partial  S_{12}}{\partial Y}=0  \quad  \frac{\partial S_{21}}{\partial X}+\frac{\partial S_{22}}{\partial Y}=0\,.
\end{equation}
\begin{equation}
\label{calJ1bis}
\begin{cases}
{\cal T}_1&=  \frac{\partial  E}{\partial X}-\left(S_{11} \frac{\partial^2  x}{\partial X^2}+S_{21} \frac{\partial^2  y}{\partial X^2}\right)\\
\\
&-\left( S_{12}  \frac{\partial^2  x}{\partial X \partial  Y}+S_{22}  \frac{\partial^2 y}{\partial X \partial Y} \right ).
\end{cases}
\end{equation}
We recall the Piola stress components, given in section {\ref{variational}, Eqs.(\ref{borderstress},\ref{borderstressno}), that we evaluate at linear order in $\epsilon$}:
\begin{equation}
\label{stressnew}
\begin{cases}
S_{11}=\frac{\partial x}{\partial X}-Q\frac{\partial y}{\partial Y}\quad S_{12}=\frac{\partial x}{\partial Y}+Q\frac{\partial y}{\partial X}\,,\\
\\
S_{21}=\frac{\partial y}{\partial X}+Q\frac{\partial x}{\partial Y}\quad S_{22}=\frac{\partial y}{\partial Y}-Q\frac{\partial x}{\partial X}\,.
\end{cases}
\end{equation}
From eq. (\ref{calJ1bis}) it is easy to show that all terms of the neo-Hookean part are eliminated by the stress contribution of the same equation and that terms proportional to $Q$ also cancel each other, a relation which is always true even without expansion in $\epsilon$. So  ${\cal  T}_1$ vanishes and  the same result is obtained for ${\cal T}_2$. In our case,  ${\cal J}_1$  and an  ${\cal J}_2$ vanishes rigorously on a closed contour,  and  so also up to second order.
Although  ${\cal J}_2$ vanishes identically  in our geometry because of the chosen contour, we write its generic form, which will be useful to establish  the $\cal M$ integral:
\begin{equation}
\label{Jdue}
\begin{cases}
 {\cal T}_2= \frac{\partial  E}{\partial Y}\\ -\frac{\partial (S_{11} U_{12}+S_{21} U_{22})}{\partial X}
 - \frac{\partial (S_{12} U_{12}+S_{22}U_{22})}{\partial Y}\,.
\end{cases}
\end{equation}
Let us consider now ${\cal  M}=\iint dS  N$
\begin{equation}
N=\frac{\partial  \delta E X_i}{\partial X_i}-  \frac{\partial \left(S_{jk}U_{ji} X_i \right )}{\partial X_k} -\tau_0/J \frac{\partial U_X}{\partial X}\,.
\end{equation}
where $\delta E=E-(J-1)^2/2$,  $U_i$ is the displacement:  $U_{1}=x-J X$ and $U_2=y-Y$. The last term can be replaced by $-(1+J) Div(U)=-(1+J)(\partial_X U_X+\partial_Y U_Y$.  This result, which is  easy to demonstrate with mathematica software, turns out to be less obvious to show analytically and differs from the result obtained by Knowles and Sternberg  for incremental elasticity. The reason is due to growth  since $J^2-1$ different from  from zero, in our case. Nevertheless, our definition of $M$ satisfies the criterion for  defining  a path independent integral similar to the classical  $M$ integral of linear elasticity up to order $\epsilon^2$.
%Taking into account that we have already demonstrated  ${\cal T}_1={\cal T}_2=0$, then we get
 After some algebra with this definition, which is slightly different   from the one given by \cite{knowles1971class}, we can  construct a  path-independent integral, valid up to second order for a pre-stretched  sample. Compared to eq.(3.36) of \cite{knowles1971class}'s equation, the only difference comes from the last term of  the previous equation. Transforming  ${\cal M}$ into a closed contour integral: ${\cal M}=
 \oint ds {\it m}=0$
 \begin{equation}
 \label{Mintegral}
 \begin{cases}
{\it m}= \left(E-\frac{(J-1)^2}{2} \right) \vec {\bf X}. \vec {\bf N}
- S_{jk}U_{ji} X_i.N_k\\
- (J+1)\left \{(x-J X)\cdot N_X+(y-Y)\cdot N_Y
 \right \} \,.
\end{cases}
\end{equation}
This result is true up to $O(\epsilon^3)$.
Then the horizontal boundary at the top  contributes to the order of $\epsilon$. We have already discussed  the case of  two boundaries in the main text section (\ref{Mint}). For completeness we will also  consider the patches here, at least to confirm our approach.
Considering now the $2$ patches where the function $F$ or $\Phi$  has singularities in $X=X_0$ and $X=X_0/J$, we can then consider $3$ contributions.  We start with  a separation into  three contributions for   ${\cal M}_{{\cal J}_S}$ into:
${\cal M}^{(1)}+{\cal M}^{(2)}+ {\cal M}^{(3)}$ which will be done for the two patches one after the other.
Around the patch located at $X=X_0$, the contour being a circle with radius $R$ greater than $l_0$ but less than $X_0$, 
definined by 
\begin{equation}
\begin{cases}
m_1&=E-\frac{(J-1)^2}{2} -S_{11}U_{11}-S_{21} U_{21},\\
m_3&=E-\frac{(J-1)^2}{2} -S_{12}U_{12}-S_{22}U_{22},\\
m_2&=S_{12}U_{11}+S_{22} U_{22},\\
m_4&=S_{11}U_{12}+S_{21}U_{22}.
 \end{cases}
\end{equation}
Note that each $m_i$ is of order $\epsilon^2$
\begin{equation}
{\cal M}_{{\cal J}_S}=R  \int_{-\pi}^{\pi} dT (X_0+R \cos {T})
\left\{ \cos{T} m_1-m_2 \sin{T} \right \}\,.
\nonumber
\end{equation}
This relation can eventually be simplified to give:
\begin{equation}
{\cal M}^{(1)}_{{\cal J}_S}=X_0  {\cal J_S} +R^2  \int_{-\pi}^{\pi} dT 
\cos{T}(\cos{T} m_1 -\sin{T} m_2)\,.
\nonumber
\end{equation}
Defining in the same way:
\begin{equation}
{\cal M}^{(2)}_{{\cal J}_S}= R^2  \int_{-\pi}^{\pi} dT \sin{T}\left \{\sin{T} m_3 - \cos{T} m_4 \right \},
\nonumber \end{equation}
 and finally: 
\begin{equation}
{\cal M}^{(3)}_{{\cal J}_S}= \tau_0 R  \int^{\pi}_{-\pi} dT \sin{T} (y-Y)\,.
\nonumber\end{equation}
So the leading order for ${\cal M}_{{\cal J}_S}$ is $X_0 J_S$ so $\epsilon^2$.
For the second patch which is  around $X=X_0/J$, the difference comes from the fact that $X_0$ has to be changed to $X_0/J$ and of course each integral is different due to the local behavior of the function $\Phi$ in each patch given by Eqs.(\ref{firstpatch},\ref{secondpatch}). From this expansion, we can  deduce that $ {\cal M}^{(3)}_{\cal S} $ and $ {\cal M}^{(3)}_{\cal S_J}$ is negligible for $R\rightarrow 0$ compared to the other contributions. Also it is clear that $ X_0  {\cal J_S} $ is  of the order of $ X_0  {\cal J_S} \sim X_0 \epsilon\sim \epsilon^2$ as  the integrals in $R^2$. Then
we conclude that only the upper boundary and lower boundary contribute to the $\cal M$ integral  and the inner singularities contribute only to the order of$\epsilon^2$.

%\bibliographystyle{ieeetr}
%\bibliography{biblio}

%%%%%%%%%%%%%%%%%%%%%%%%%%%%%%%%%%%%%%%%%%%%%%%%%%%%%%%%%%%%%%%%%%%%%%%%%%%

\end{document}